# Surface-induced magnetism of the solids with impurities and vacancies

***A. N. Morozovska*[1,2*], *E.A. Eliseev*[1], *M.D. Glinchuk*[1] *and R. Blinc*[3]**

[1]Institute for Problems of Materials Science, National Academy of Sciences of Ukraine,
Krjijanovskogo 3, 03142 Kiev, Ukraine,

[2] V. Lashkarev Institute of Semiconductor Physics, National Academy of Sciences of Ukraine,
prospect Nauki 41, 03028 Kiev, Ukraine

[3]Jožef Stefan Institute, P. O. Box 3000, 1001 Ljubljana, Slovenia

Using the quantum-mechanical approach combined with the image charge method we calculated the lowest energy levels of the impurities and neutral vacancies with two electrons or holes located in the vicinity of flat surface of different solids. Unexpectedly we obtained that the magnetic triplet state is the ground state of the impurities and neutral vacancies in the vicinity of surface, while the nonmagnetic singlet is the ground state in the bulk for e.g. He atom, $Li^{+}$, $Be^{++}$, etc. ions. The energy difference between the lowest triplet and singlet states strongly depends on the electron (hole) effective mass μ, dielectric permittivity of the solid $\varepsilon_2$ and the distance from the surface $z_0$. For $z_0 = 0$ and defect charge $|Z| = 2$ the energy difference is more than several hundreds of Kelvins at $\mu = (0.5 - 1)m_e$ and $\varepsilon_2 = 2\text{-}10$, more than several tens of Kelvins at $\mu = (0.1 - 0.2)m_e$ and $\varepsilon_2 = 5\text{-}10$, and not more than several Kelvins at $\mu < 0.1m_e$ and $\varepsilon_2 > 15$ ($m_e$ is the mass of a free electron). Pair interaction of the identical surface defects (two doubly charged impurities or vacancies with two electrons or holes) reveals the ferromagnetic spin state with the maximal exchange energy at the definite distance between the defects (~5-25 nm). We estimated the critical concentration of surface defects and transition temperature of ferromagnetic long-range order appearance in the framework of percolation and mean field theories, and RKKY approach for semiconductors like ZnO. We obtained that the nonmagnetic singlet state is the lowest one for a molecule with two electrons formed by a pair of identical surface impurities (like surface hydrogen), while its next state with deep enough negative energy minimum is the magnetic triplet. The metastable magnetic triplet state appeared for such molecule at the surface indicates the possibility of metastable orto-states of the hydrogen-like molecules, while they are absent in the bulk of material. The two series of spectral lines are expected due to the coexistence of orto- and para-states of the molecules at the surface. We hope that obtained results could provide an alternative mechanism of the room temperature ferromagnetism observed in $TiO_2$, $HfO_2$, and $In_2O_3$ thin films with contribution of the oxygen vacancies. We expect that both anion and cation vacancies near the flat surface act as magnetic defects because of their triplet ground state and Hund's rule. The theoretical forecasts are waiting for experimental justification allowing for the number of the defects in the vicinity of surface is much larger than in the bulk of as-grown samples.

[*] Corresponding author: morozo@i.com.ua

## 1. Introduction

In recent years the room-temperature magnetism in the nonmagnetic systems without doping by 3d and 4f magnetic ions has been observed and/or theoretically predicted (see e.g. [1, 2, 3] and refs. therein).

Numerous experiments revealed the ***ferromagnetic properties*** of the ***nanomaterials***, which are nonmagnetic in the bulk. For instance, remarkable room-temperature ferromagnetism was observed in undoped $TiO_2$, $HfO_2$, and $In_2O_3$ thin films with extrapolated Curie temperatures far in excess of 400 K [4, 5]. Magnetization of $TiO_2$ and $HfO_2$ films strongly decreases after 4 h annealing in oxygen and eventually disappears for 10 h annealing. Thus the authors [4, 5] concluded that the ***oxygen vacancies*** are the main source of the magnetism in $TiO_2$ and $HfO_2$ thin films.

Striking phenomena such as the observation of room-temperature ferromagnetism in spherical nanoparticles (size 7–30 nm) of nonmagnetic oxides such as $CeO_2$, $Al_2O_3$, ZnO, $In_2O_3$, and $SnO_2$ have been reported [6]. These studies show that ferromagnetism is associated only with the nanoparticles, because the corresponding bulk samples are diamagnetic. It was experimentally demonstrated that MgO nanocrystalline powders reveal room-temperature ferromagnetism, while MgO bulk exhibits diamagnetism [7]. The vacuum annealing of MgO nanocrystalline powders reduces ferromagnetism. The authors conclude that the ferromagnetism possibly originates from Mg ***vacancies*** at the surface and near the surfaces of nanograins. Large concentrations of Mg vacancies at the surfaces of nanograins can lead to magnetization via magnetic percolation [7].

An example of magnetization induced by nonmagnetic impurities in a nonmagnetic host one can find in Ref. [8], where Herng et al. reported about ferromagnetic properties of carbon-doped ZnO nanoneedles. It was found that the ferromagnetism of the ZnO:C nanoneedles were stable in ambient air over a period of 1 year and annealing temperature up to 100°C. Thus the authors assumed that ferromagnetism in the ZnO:C nanoneedles could be attributed to C substitution on the O site.

Magnetic properties of vacancies and the possible formation of a magnetic phase in the bulk samples were examined in a number of studies [9, 10, 11, 12, 13, 14]. The magnetic states of cation neutral vacancies with different spins were observed by EPR only in irradiated samples: ZnO [9], MgO [10, 11], GaP [12]. In the latter case the analysis of EPR spectra (performed under the uniaxial strain) has shown that Ga vacancy spin was 3/2. Data for Zn vacancy were taken after a brief exposure to 325 nm laser light. Used irradiation was the following: neutron irradiation [10, 11], 1.5 MeV electron irradiation [10], 2.5 MeV [13] or 3 MeV *in situ* [14]. It might be supposed that the electron or neutron irradiation is required to increase the vacancies concentration, although their equilibrium concentration $10^{18}$ $cm^{-3}$ [2], that is much higher than the sensitivity of commercially available radiospectrometers ($10^{10}$ $cm^{-3}$). So, the absence of EPR data for as grown samples seems to be strange. On the other hand the aforementioned external stimulus can induce new paramagnetic centers (see e.g. [9]) and change the charge state of vacancies. The experiments [7] have shown that as grown bulk MgO presents diamagnetism. The role of irradiation in the appearance of ferromagnetic order e.g. in MgO may be related to the fact, that the concentration of Mg vacancies appeared to be too small to achieve percolation [7] or ferromagnetic state is not stable in the bulk, since the calculated energy difference between ferro- and antiferromagnetic states (28 meV) is smaller than the thermal energy at room temperature [1]. In any case the larger concentration of magnetic defects, the better for magnetization observation.

The defects (impurities and vacancies) concentration increases near the sample surface, in particular allowing for the strong lowering of their formation energies [1, 15, 16]. Density functional calculations show that the energy of vacancy formation on the surface is lower than in the bulk on about 3 eV for GaN [16] and 0.28 eV for MgO [1]. Therefore, the native vacancies should be present largely in the surface layer (more generally, in the vicinity of surface), and the ferromagnetism observed in as grown nonmagnetic solids should arise primarily from the ***surface defects***. Because of this it is not a surprise that the most of experimental results about ferromagnetism induced by nonmagnetic defects in nonmagnetic hosts were obtained in nanosystems, where the influence of surface is strong.

The theoretical calculations had confirmed that ferromagnetism could be the intrinsic property of non-magnetic hosts with non-magnetic defects. The origin of this phenomenon is the strong spin polarization of 2p shell of light atoms from the second raw of the periodic table, for which the Hund's energy is close to that of transition metal atoms (see e.g. the review [2]). The ***first principles calculations*** [1, 3, 17, 18, 19, 20, 21, 22] show that such effect of the ions in the nearest neighbor of cation vacancies is the main reason of the vacancies magnetism. In particular it was shown that Mg vacancies can induce local magnetic moments in MgO, contrary to O vacancies, irrespectively of their concentration [1], local magnetic moments originated from the 2p-orbitals of the nearest O atoms.

The main conclusion about a magnetic state for a neutral cation vacancy, but not for an anion vacancy, was obtained for Zn*A* (*A* = S, Se, Te) [3], CaO [17, 18], $TiO_2$ [19], GaN or BN [21] and $HfO_2$ [22]. Maca et al [20] demonstrated that the K impurity leads to a robust induced magnetic moment on the surrounding O atoms in the cubic $ZrO_2$ host, while Ca impurity leads to a nonmagnetic ground state.

The calculations devoted to surface magnetism are not numerous. In particular the occurrence of spin polarization at the ideal, oxygen-ended surfaces MgO, $ZrO_2$, $Al_2O_3$ was explored from the *ab initio* calculations performed by Gallego et al [23].

Therefore, in accordance with the numerous first-principles studies [1, 3, 17-22] the origin of the magnetism and related properties in otherwise nonmagnetic materials is ***cation defect only***, i.e.

induced by the magnetic ground state of the ***neutral cation vacancies***. Thus, it is obvious that some experimental results, namely the fact that oxygen annealing suppresses ferromagnetism in $HfO_2$ and $TiO_2$ films [4], i.e. ***anion oxygen vacancies*** play the main role in ferromagnetism appearance, seems in a disagreement with the aforementioned first principles calculations. In the most cases the influence of surface was taken into account mainly by the decrease of the coordination number of ions surrounded the vacancy. As a result the properties of the subsurface layers (e.g. the formation energy of Mg vacancies in MgO) appeared to be practically the same as in the bulk [1]. To our mind these questionable results originate from the supposition that localized sp-states are responsible for magnetic moments formation (see e.g. [21]). However, it is obvious from the symmetry consideration that s-states cannot exist in the vicinity of surface. Quantum mechanics calculations performed earlier [24] confirmed the pure p-type of one-electron impurity ground state at the surface, which was s-type in the bulk, so that some mixture of p- and s-states exists in the subsurface layers. As a matter of fact this leads us to the idea to check if the abovementioned change of the electronic structure in the vicinity of surface could lead to the appearance of high spin states of the two-electrons (holes) impurities and vacancies, which can lead to the magnetization of the surface and subsurface layers. Such possibility could exist even without any irradiation or other external stimulus, because the concentration of defects in the surface region can be much higher than in the bulk as we discussed earlier.

The ***quantum-mechanical*** calculations presented in the paper show that the ground state of the impurities like He, $Li^{+}$, $Be^{2+}$ etc, cation and anion vacancies in the binary solids is indeed the triplet one (spin $\Sigma=1$) at the surface and under the surface. To check if the defect spins could be ordered ferromagnetically or antiferromagnetically we performed the calculations of the exchange interaction of the defect pair and obtained that the ferromagnetic ordering between the nearest neighbors is energetically preferable. The estimations have shown the possibility of ferromagnetic long-range order appearance in the vicinity of surface.

The calculations of the energetic levels of the impurity hydrogen-like molecule on the surface lead to the interesting result about orto- and para-state of the molecule, because both ground singlet and excited triplet states have deep negative energy minima at the surface for some distance between the impurity atoms. This has to manifest itself in the peculiarities of the molecules spectral lines similar to the ones observed for He atoms long ago [25].

The paper contents are the following. After the Introduction we described the model of calculations and grounded the adopted approximations (e.g. the effective mass approach) [Section 2]. Next parts of the paper include the calculations of the single defect with two localized electrons (or holes) at the surface [Section 3] and in subsurface layers [Section 4]. Then we calculated the exchange energy of the surface defect pair and estimated the conditions of the ferromagnetic long-range order appearance between the considered magnetic defects [Section 5.1]. The energy levels of the two surface impurity atoms coupled into the hydrogen-like molecule are calculated in the Section 5.2. Then we briefly discuss the obtained results and possibility of their application to nanosystems [Section 6].

## 2. Model assumptions

### *2.1. Direct variational method for Schrödinger equation solution*

To find out the wave functions of localized electrons or holes we will solve Schrödinger equation. To take into account the surface influence we applied the image charges method. This approach, proposed in the pioneer paper [24], was successfully used for the ground state calculations of the one-electron impurity center located near the flat surface. For the application of the image charges method the model of ***continuous media*** characterized by ***dielectric permittivity*** and the ***effective mass approximation*** will be used for the wave functions calculations similarly to [24]. These two characteristics could be enough for the description of the principal differences between the solid and its ambient as well as between electrons and holes in dielectrics and various semiconductors.

The variational method for Schrödinger equation solution was applied for ***p-type*** of the electron trial wave function of the defect at the surface, while it transforms into s-state for the defect in the bulk. The choice was specified by the strong anisotropy of the Hamiltonian at the surface, so that the spherical symmetry of the s-type function is forbidden from the symmetry consideration. For the case of sufficiently high barrier between the solid and vacuum the hydrogen-like p-state wave function with zero value at the surface was taken as the trial function with two variational parameters. Due to the dependence of the variational parameters on the defect distance from the solid surface, the p-type wave function reveals the correct transition to the s-type function for the defect in the bulk of the solid and successfully describes the most important physical properties of solids related to the surface influence [24].

Contrary to the one-electron problem solved in [24], we consider two electrons (or holes) localized near ***defect*** located in the point $\mathbf{r}_0 = (0,0,z_0)$ under the solid body flat ***surface*** z=0 in the framework of above-described approach. It is obvious that in such a case the possibility to have symmetric (with zero spins) or antisymmetric (nonzero spin) ground state open the way of defects magnetic state (spin $\Sigma$=1) and magnetic long-range order induced by the defects under the favorable conditions. It is obvious that the realization of such possibility depends on the on the defect wave function type. In particular, for s-type functions only a symmetric singlet state is possible without any magnetization, while the magnetization is possible for p-type triplet ground state. The search of the latter possibility realization conditions is the main goal of our research.

For the better understanding of the applied model and material parameters choice we have to add some remarks in this and next subsections.

The image charges method (as continuum media approach) requires the conception of the media dielectric permittivity $\varepsilon(\omega)$ ($\omega$ is the frequency) to describe the defect Coulomb potential $U_d(\mathbf{r}) \sim -\frac{Ze}{4\pi\varepsilon_0\varepsilon r}$, where $Z$ is the defect effective charge in the units of the electron charge $e$>0. The introduction of the material dielectric permittivity $\varepsilon(\omega)$ has a solid background only when the characteristic size of the carrier localization is several times larger than the lattice constant [26]. Since the defect is immovable the static dielectric permittivity should be used, even when the carrier-defect binding energy frequency $\omega = E_b/\hbar$ is high enough for the noticeable dispersion $\varepsilon(\omega)$ [26]. Note, that a defect effective charge $Z$ may be negative or positive, but the only interactions that can lead to the localized state are "positively charged defect – electron" or "negatively charged defect – hole", both interaction energies are proportional to $Z$ absolute value, i.e. $-\frac{|Z|e^2}{4\pi\varepsilon_0\varepsilon r}$. Note, that the description of the defect center by Coulomb potential in continuous media takes into account its charge, but completely excludes other individual features of the defect related to its chemical nature.

### *2.2. Effective mass approximation for p-states*

Allowing for the fact that effective mass approach originates from the possibility to approximate electrons or holes dispersion law by the parabolic law, let us discuss its applicability to the considered problem in more details.

The characteristic size $r_d$ of the carrier localization at the shallow defect (considered hereinafter) is typically much larger than the lattice constant $a$ [26]. For the case only the wave vectors small in comparison with the inverse lattice constant play significant role in the localized carrier wave function expansion on the plane waves. The assumption allows one to use the effective mass approximation similarly to Ref.[24].

At the first glance the condition $r_d >> a$ of the effective mass approximation validity is strict only for the carriers wave-functions of the s-states. However for the p-states, which are zero at the defect site, the corresponding wave functions are not sensitive to the concrete short-range peculiarities of the defect potential. As a result the effective mass approximation typically describes ***p-states*** (we are interested in) much better than s-states (see Ref [26] and **Appendix A**).

Effective mass approximation for the localized electrons (or holes) is adopted in the assumption of the non-degenerated conductive (or valence) band structure (see e.g. [27]). Actually, the Schrödinger-Vanjie equation for the electron is $(\varepsilon_C(-i\nabla_n) + V(\mathbf{r}_n))\psi(\mathbf{r}_n) = E\psi(\mathbf{r}_n)$, where the electron energy near the extremum of the conductive band is $\varepsilon_C(\mathbf{k}_n) \approx \varepsilon_C(0) + \frac{\hbar^2 k_n^2}{2\mu_n}$ ($\mu_n$ is the electron effective mass, $\nabla_n$ is the gradient operator acting on the electron coordinates $\mathbf{r}_n$). Similarly, Schrödinger-Vanjie equation for the hole is $(-\varepsilon_V(+i\nabla_p)+V(\mathbf{r}_p))\psi(\mathbf{r}_p)=E\psi(\mathbf{r}_p)$, where the hole energy near the extremum of valence band is $\varepsilon_V(\mathbf{k}_p)\approx\varepsilon_V(0)-\frac{\hbar^2 k_p^2}{2\mu_p}$ ($\mu_p$ is the hole effective mass, $\nabla_p$ is the gradient operator acting on the hole coordinates $\mathbf{r}_p$). The band gap is $\varepsilon_G=\varepsilon_C(0)-\varepsilon_V(0)$. Note, that $\mu_n$ is positive near the bottom of conductive band and negative near its edge, while $\mu_p$ is positive near the top of valence band. Then Schrödinger-Vanjie equation can be conveniently represented in the same form: $\left(-\frac{\hbar^2\Delta_i^2}{2\mu_i}+V(\mathbf{r}_i)\right)\psi(\mathbf{r}_i)=E\psi(\mathbf{r}_i)$, $i=n$ for electrons and $i=p$ for holes respectively.

Since the effective mass value depends on the degree of ionicity and covalence, these important characteristics of the material can be taken into account in our approach. For dielectric materials $\mu$ value is close to the mass of a free electron $m_e$. For semiconductors $\mu$ value is determined by the concrete band structure, in particular the condition $\mu << m_e$ is typical for light weakly localized carriers (e.g. free electrons or holes). The condition $\mu > m_e$ corresponds to the heavy carriers strongly localized at the atom sites.

Since the characteristic sizes $\alpha$ of the one-fermion wave-functions $\varphi_{nlm}\sim\exp(-\alpha r)$ should be positive for their square integrability, it is easy to check directly, that they could not satisfy one-particle Schrödinger equation with Coulomb potential, like $\left(-\frac{\hbar^2\Delta_i^2}{2\mu_i}-\frac{|Z|e^2}{4\pi\varepsilon_0\varepsilon r_i}\right)\varphi_{nlm}=E\varphi_{nlm}$ for negative effective mass $\mu$ and any positive $\alpha$. This fact leads to the result that carriers with negative effective mass cannot create the stable localized states for chosen form of the wave function. Thus hereinafter we regard $\mu$ positive, which is true for the electrons with energies near the bottom of conductive band and holes near the top of valence band [27].

### *2.3. Model of surface defects in the continuum media approach*

Impurity atoms, cation or anion vacancies are considered as defects placed in a perfect host lattice in the continuum media approach (see **Figs. 1**). We consider several cases:

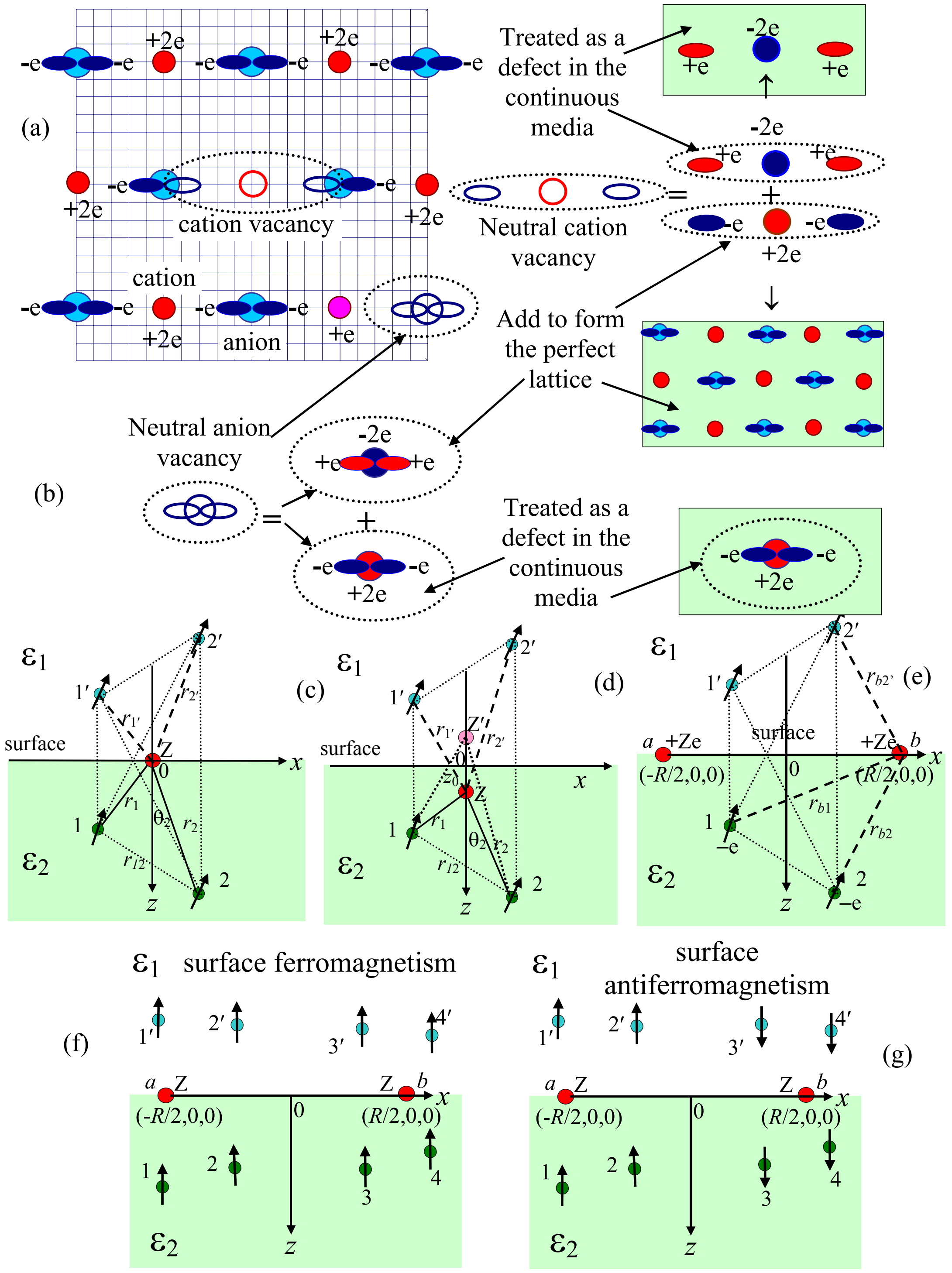

**Fig. 1.** Schematics of the neutral cation (a) and anion (b) vacancies in the continuous media approach. (c,d) Problem geometry for the defect at the surface (c) and at the distance $z_0$ under the surface (d). Two carriers (electrons or holes) 1 and 2 (shown by green circles with arrows) are localized near the defect with effective charge $Ze$ (shown by red circle). The coordinate origin is placed at the defect site. Carrier image charges are shown as 1′ and 2′, defect image $Z'$ appears in the case (d). (e) Impurity molecule with two electrons (1 and 2) at the surface. (f,g) Ferromagnetism (f) and antiferromagnetism (g) for a pair of identical surface defects with four shared carriers.

(1) neutral defects with valency ±2e, which already captured two carriers (electrons or holes), e.g. neutral vacancy, filled donor or acceptor;
(2) the situation, when defect charge ±Ze is not fully compensated by the two electrons or holes allowing for the definite degree of the host lattice covalence;
(3) the hydrogen-like molecules at the surface (two identical atoms with two electrons);
(4) the pair of identical surface defects with shared four electrons, which may appear when the concentration of the defects increases.

For the case of the cation vacancy the cation atom should be added to form the perfect host lattice. Its two electrons should be localized at the nearest anions. As a result a negatively charged defect (-2e) with two holes appears in the perfect continuous media (see **Figs. 1a**). The situation is vise versa for an anion vacancy: it can be modeled as a positively charged defect (+2e) with two electrons in the continuous media (see **Figs. 1b**).

The continuum media around impurity atoms or impurity molecules with two electrons is introduced in the straightforward way [see e.g. **Figs. 1 e**].

We used the image charges method to account the surface influence on the defects electric field [28, 29]. The geometry of the image charges induced by the surface defect with two electrons is shown in **Fig. 1c-e.**

### *2.4 Nearest neighbor approximation for the interaction between defects*

To find out the type of magnetic coupling (ferro- or antiferro-), only the interaction between the spins of the nearest defects will be calculated (see schematics in **Figs. 1 f,g**). Under negligibly small overlap integrals of the electron wave functions localized at the nearest defect sites, the exchange contribution to the energy is mainly determined by the two-fermions exchange integral $J$ [30]. For the ferromagnetic triplet state to be the ground state the condition $J > 0$ is necessary. The condition $J < 0$ is necessary for the antiferromagnetic order, but sometimes not enough. Really, recent atomistic calculations should be considered as the manifestation of strong Dzyaloshinskii-Moriya interaction in nanosystems [31, 32, 33]. Actually, the symmetry lowering near the surface could strongly increase the symmetry related Dzyaloshinskii-Moriya vector value and change its direction [32]. Based on these results one can expect weak surface-induced ferromagnetism even in the case $J < 0$, while the corresponding bulk material may be antiferromagnetic. Allowing for the defect concentration strongly increases near the sample surface, in particular it concerns the neutral vacancies in solids as it was shown both theoretically [1, 15, 16] and experimentally [34], we will pay attention mainly to the interaction inside the defect pair at the surface.

## 3. Energy levels of a single surface defect with two electrons

### *3.1. The problem statement for a surface defect with two electrons*

The geometry of the surface defect with two electrons is shown in **Fig. 1c.** Two-electron (holes) wave function obeys the Paul principle, and thus its coordinate part could be either symmetric or antisymmetric:

$$\varphi_{S,A}(\mathbf{r}_1,\mathbf{r}_2)=\frac{1}{\sqrt{2}}\left(\varphi_m(\mathbf{r}_1)\varphi_n(\mathbf{r}_2)\pm\varphi_m(\mathbf{r}_2)\varphi_n(\mathbf{r}_1)\right). \tag{1}$$

Hereinafter $\mathbf{r}_i=(x_i,y_i,z_i)$ are the coordinates of the electrons (holes) 1 and 2. Their images coordinates are $\mathbf{r}_i^{/}=(x_i,y_i,-z_i)$. The surface defect (impurity or vacancy) is located in the point $O=(0,0,0)$. Letters *m* and *n* in Eq.(1) correspond to the different wave functions, since for *m*=*n* only symmetric wave function exists. Each function can correspond to the set of four quantum numbers.

Schrödinger equation for coordinate part has the form:

$$\left(-\frac{\hbar^2}{2\mu}(\Delta_1+\Delta_2)+V(\mathbf{r}_1,\mathbf{r}_2)\right)\varphi_{S,A}(\mathbf{r}_1,\mathbf{r}_2)=E\varphi_{S,A}(\mathbf{r}_1,\mathbf{r}_2) \tag{2}$$

The boundary conditions $\varphi(z_1=0)=\varphi(z_2=0)=0$ is used, as argued in the Section 2.1. The condition corresponds to the infinitely high barrier at the interface solid-ambient medium (e.g. vacuum, atmosphere or dielectric soft matter). It becomes enough rigorous for the case of narrow-gap semiconductors and approximate for insulators.

There are several contributions into the electrostatic energy *V* of the carriers interaction with defect and polarized half-space (image charges), namely:

$$V(\mathbf{r}_1,\mathbf{r}_2)=V_e(\mathbf{r}_1)+V_e(\mathbf{r}_2)+V_{12}(\mathbf{r}_1,\mathbf{r}_2) \tag{3}$$

1) $V_e$ is the sum of carriers (electron or hole) interaction with the surface defect and the carrier interaction with its own charge image, which has the form

$$V_e(\mathbf{r}_i)=-\frac{e^2}{2\pi\varepsilon_0}\left(\frac{|Z|}{(\varepsilon_1+\varepsilon_2)r_i}-\frac{\varepsilon_2-\varepsilon_1}{\varepsilon_2+\varepsilon_1}\frac{1}{4\varepsilon_2 z_i}\right) \tag{4}$$

Here $\varepsilon_{1,2}$ are the static dielectric permittivity values respectively of the media 1,2; *Z* is the defect effective charge in the units of the electron charge *e*>0. In Eq.(4) we used that $|\mathbf{r}_i|=|\mathbf{r}_i^{/}|$ for the considered surface defect.

2) The electron-electron (or hole-hole) interaction energy and the interaction of electrons (holes) with their images are:

$$V_{12}(\mathbf{r}_1,\mathbf{r}_2)=\frac{e^2}{4\pi\varepsilon_0\varepsilon_2}\left(\frac{1}{r_{12}}+\frac{\varepsilon_2-\varepsilon_1}{\varepsilon_2+\varepsilon_1}\frac{1}{r_{12^{/}}}\right). \tag{5}$$

Here we introduced the conventional designation $r_{12}=|\mathbf{r}_1-\mathbf{r}_2|$ and used that $r_{12'}=|\mathbf{r}_1-\mathbf{r}_2'|\equiv|\mathbf{r}_2-\mathbf{r}_1'|$ for the considered surface defect (see **Appendix A** for mathematical details).

### *3.2. Approximate analytical solution obtained by direct variational method*

We are looking for the ground state of the system by direct variational method using the wave functions localized near the ***surface defect***.

As argued in the Section 2.1, the coordinate dependence of the one-electron (or hole) wave functions was chosen in the form of hydrogen-like atom eigen functions: $\varphi_{nlm}(\mathbf{r})\sim R_{nl}(\alpha|\mathbf{r}-\mathbf{r}_0|)Y_{lm}(\theta,\varphi)$, where $R_{nl}(r)$ are the radial functions and $Y_{lm}(\theta,\varphi)$ are spherical harmonics, $z=r\cos(\theta)$.

For the case $z_0=0$ the lowest hydrogen-like atom wave function that satisfies the boundary condition $\varphi_{nlm}(x,y,z=0)=0$ is 2p$_z$ state, $\varphi_{210}(\mathbf{r}_i)$, that evident form is:

$$\varphi_{210}(\mathbf{r}_i)=\sqrt{\frac{2\alpha^5}{\pi}}\exp(-\alpha r_i)z_i, \qquad (6)$$

The next excited state is 3p$_z$, $\varphi_{310}(\mathbf{r})$, that evident form is:

$$\varphi_{310}(\mathbf{r}_i)=\sqrt{\frac{4\beta^5}{3\pi}}z_i\exp(-\beta r_i)(2-\beta r_i). \qquad (7)$$

Here $i=1,2$. Parameters $\alpha$ and $\beta$ will be determined by direct variational method from the energy minimum, at that the electron-electron (or hole-hole) Coulomb interaction and all interactions with the image charges are considered as perturbations.

Note that the functions (6)-(7) are normalized in the half-space z>0, but not orthogonal at arbitrary $\alpha$ and $\beta$.

Two-fermion wave function obeys the Paul principle [25]:

$$\psi_{22}(\mathbf{r}_1,\mathbf{r}_2)=\varphi_{210}(\mathbf{r}_1)\varphi_{210}(\mathbf{r}_2)\chi_A(s_1,s_2), \quad \text{(singlet, } \Sigma=0) \qquad (8a)$$

$$\psi_{23}^S(\mathbf{r}_1,\mathbf{r}_2)=\frac{1}{\sqrt{2}}(\varphi_{210}(\mathbf{r}_1)\varphi_{310}(\mathbf{r}_2)+\varphi_{210}(\mathbf{r}_2)\varphi_{310}(\mathbf{r}_1))\chi_A(s_1,s_2), \quad \text{(singlet, } \Sigma=0) \qquad (8b)$$

$$\psi_{23}^T(\mathbf{r}_1,\mathbf{r}_2)=\frac{1}{\sqrt{2}}(\varphi_{210}(\mathbf{r}_1)\varphi_{310}(\mathbf{r}_2)-\varphi_{210}(\mathbf{r}_2)\varphi_{310}(\mathbf{r}_1))\hat{\chi}_S(s_1,s_2), \quad \text{(triplet, } \Sigma=1) \qquad (8c)$$

$$\psi_{33}(\mathbf{r}_1,\mathbf{r}_2)=\varphi_{310}(\mathbf{r}_1)\varphi_{310}(\mathbf{r}_2)\chi_A(s_1,s_2). \quad \text{(singlet, } \Sigma=0) \qquad (8d)$$

$\chi_A(s_1,s_2)$ is antisymmetric scalar spin function and $\hat{\chi}_S(s_1,s_2)$ is three symmetric spin Pauly matrix (normal spinors), $\Sigma$ is the full spin.

Introducing dimensionless coordinates $\tilde{\mathbf{r}} = \frac{\mathbf{r}}{a_B}$, renormalized on the surface-perturbed effective Bohr radius $a_B^* = (\varepsilon_1 + \varepsilon_2)\frac{2\pi\varepsilon_0\hbar^2}{|Z\mu|e^2}$, dimensionless Hamiltonian $\tilde{H} = \frac{a_B^{*2}|\mu|}{\hbar^2}H$ and energy $\tilde{E} = \frac{a_B^{*2}|\mu|}{\hbar^2}E$ we rewrite the Schrödinger equation (2) in the form:

$$\left(-\frac{1}{2}\left(\tilde{\Delta}_1 + \tilde{\Delta}_2\right) + \tilde{V}_e(\tilde{\mathbf{r}}_1) + \tilde{V}_e(\tilde{\mathbf{r}}_2) + \tilde{V}_{12}(\tilde{\mathbf{r}}_1, \tilde{\mathbf{r}}_2)\right)\psi_{mn}(\tilde{\mathbf{r}}_1, \tilde{\mathbf{r}}_2) = \tilde{E}_{mn}\psi_{mn}(\tilde{\mathbf{r}}_1, \tilde{\mathbf{r}}_2), \tag{9a}$$

$$\tilde{V}_e(\tilde{\mathbf{r}}_i) = -\frac{1}{\tilde{r}_i} + \frac{\varepsilon_1 + \varepsilon_2}{2\varepsilon_2|Z|}\frac{\zeta}{2\tilde{z}_i}, \quad \tilde{V}_{12}(\tilde{\mathbf{r}}_1, \tilde{\mathbf{r}}_2) = \frac{\varepsilon_1 + \varepsilon_2}{2\varepsilon_2|Z|}\left(\frac{1}{\tilde{r}_{12}} + \frac{\zeta}{2}\left(\frac{1}{\tilde{r}_{1'2}} + \frac{1}{\tilde{r}_{12'}}\right)\right). \tag{9b}$$

Here $\zeta = \frac{\varepsilon_2 - \varepsilon_1}{\varepsilon_2 + \varepsilon_1}$. Note, that the image charges method typically used for dielectrics, can be modified into the integral representation for semiconductors with arbitrary screening length [35], but the differences are not essential for the current study, since we consider only the real dielectric permittivity of the ionic host.

Hereinafter we introduced the dimensionless variational parameters $\tilde{\alpha} = \alpha a_B^*$ and $\tilde{\beta} = \beta a_B^*$ normalized on the effective Bohr radius $a_B^*$.

Eigen values of the coordinate functions (8) was obtained in the framework of the conventional perturbation theory (see [25]):

$$\tilde{E}_{mn} = \frac{\int\limits_{z_1>0} d\tilde{\mathbf{r}}_1 \int\limits_{z_2>0} d\tilde{\mathbf{r}}_2 \psi^*_{mn}(\tilde{\mathbf{r}}_1, \tilde{\mathbf{r}}_2)\left(-\frac{1}{2}\left(\tilde{\Delta}_1 + \tilde{\Delta}_2\right) + \tilde{V}_e(\tilde{\mathbf{r}}_1) + \tilde{V}_e(\tilde{\mathbf{r}}_2) + \tilde{V}_{12}(\tilde{\mathbf{r}}_1, \tilde{\mathbf{r}}_2)\right)\psi_{mn}(\tilde{\mathbf{r}}_1, \tilde{\mathbf{r}}_2)}{\int\limits_{z_1>0} d\tilde{\mathbf{r}}_1 \int\limits_{z_2>0} d\tilde{\mathbf{r}}_2 \left|\psi^*_{mn}(\tilde{\mathbf{r}}_1, \tilde{\mathbf{r}}_2)\right|^2}. \tag{10}$$

In **Appendix B** we derived the singlet-states energies as

$$\tilde{E}_{22} = 2A_{22} + C_{22}, \qquad \tilde{E}_{33} = 2A_{33} + C_{33}, \tag{11a}$$

$$\tilde{E}^S_{23} = \frac{A_{22} + A_{33} + (A_{23} + A_{32})S_{23} + C_{23} + J_{23}}{1 + S_{23}^2}, \tag{11b}$$

and triplet-state energy as:

$$\tilde{E}^T_{23} = \frac{A_{22} + A_{33} - (A_{32} + A_{23})S_{23} + C_{23} - J_{23}}{1 - S_{23}^2}. \tag{11c}$$

The overlap integrals $S_{mn}$, matrix elements $A_{mn}$, Coulomb $C_{mn}$ and exchange $J_{mn}$ integrals have the following form:

$$S_{mn} = \int_{z_1>0} \varphi_{m10}(\tilde{\mathbf{r}}_1)\varphi_{n10}(\tilde{\mathbf{r}}_1)d\tilde{\mathbf{r}}_1 = 32\sqrt{\frac{2\tilde{\alpha}^5\tilde{\beta}^5}{3}}\frac{(2\tilde{\alpha}-3\tilde{\beta})}{(\tilde{\alpha}+\tilde{\beta})^6} \quad (m=2,\ n=3) \tag{12a}$$

$$A_{mn} = \int_{z_1>0} \varphi_{m10}(\tilde{\mathbf{r}}_1)\left(-\frac{\tilde{\Delta}_1}{2}+\tilde{V}_e(\tilde{\mathbf{r}}_1)\right)\varphi_{n10}(\tilde{\mathbf{r}}_1)d\tilde{\mathbf{r}}_1 \tag{12b}$$

Calculation gives $A_{22} = -\frac{\tilde{\alpha}(1-\tilde{\alpha})}{2}+\frac{\zeta}{2|Z|(1+\zeta)}\frac{3\tilde{\alpha}}{4}$, $A_{33} = -\frac{\tilde{\beta}(2-3\tilde{\beta})}{6}+\frac{\zeta}{2|Z|(1+\zeta)}\frac{\tilde{\beta}}{2}$ and

$$A_{23} = A_{32} = 16\sqrt{\frac{2\tilde{\alpha}^5\tilde{\beta}^5}{3}}\frac{\left((2\tilde{\alpha}-3\tilde{\beta})\tilde{\alpha}\tilde{\beta}-\tilde{\alpha}^2+\tilde{\beta}^2\right)}{(\tilde{\alpha}+\tilde{\beta})^6}+\frac{\zeta}{|Z|(1+\zeta)}\frac{4\sqrt{6\tilde{\alpha}^5\tilde{\beta}^5}(\tilde{\alpha}-\tilde{\beta})}{(\tilde{\alpha}+\tilde{\beta})^5}.$$

$$C_{mn} = \int_{z_2>0}\int_{z_1>0}\tilde{V}_{12}(\tilde{\mathbf{r}}_1,\tilde{\mathbf{r}}_2)\varphi_{m10}^2(\tilde{\mathbf{r}}_1)\varphi_{n10}^2(\tilde{\mathbf{r}}_2)d\tilde{\mathbf{r}}_1 d\tilde{\mathbf{r}}_2\,, \tag{12c}$$

$$J_{mn} = \int_{z_2>0}\int_{z_1>0}\tilde{V}_{12}(\tilde{\mathbf{r}}_1,\tilde{\mathbf{r}}_2)\varphi_{m10}(\tilde{\mathbf{r}}_1)\varphi_{m10}(\tilde{\mathbf{r}}_2)\varphi_{n10}(\tilde{\mathbf{r}}_1)\varphi_{n10}(\tilde{\mathbf{r}}_2)d\tilde{\mathbf{r}}_1 d\tilde{\mathbf{r}}_2\,. \tag{12d}$$

Here $m$, $n$ = 2, 3.

In **Appendix B** we had shown that

$$C_{22}(\tilde{\alpha}) = \sum_{n=0}^{\infty}\frac{1+(-1)^n\zeta}{|Z|(1+\zeta)}I_n C_{22}^n(\tilde{\alpha}), \qquad C_{33}(\tilde{\beta}) = \sum_{n=0}^{\infty}\frac{1+(-1)^n\zeta}{|Z|(1+\zeta)}I_n C_{33}^n(\tilde{\beta}), \tag{13a}$$

$$C_{23}(\tilde{\alpha},\tilde{\beta}) = \sum_{n=0}^{\infty}\frac{1+(-1)^n\zeta}{|Z|(1+\zeta)}I_n C_{23}^n(\tilde{\alpha},\tilde{\beta}), \qquad J_{23}(\tilde{\alpha},\tilde{\beta}) = \sum_{n=0}^{\infty}\frac{1+(-1)^n\zeta}{|Z|(1+\zeta)}I_n J_{23}^n(\tilde{\alpha},\tilde{\beta}). \tag{13b}$$

Here the coefficients $I_n$ originated from the integration over the spherical angles, while other coefficients originated from the integration over the radii (see Appendix B).

In accordance with the basis of quantum mechanics wave functions, which correspond to the different eigen values of energy, should be orthogonal. Note, that the orthogonality of the singlet wave functions $\psi_{22}(\mathbf{r}_1,\mathbf{r}_2)$ and $\psi_{33}(\mathbf{r}_1,\mathbf{r}_2)$ leads to the condition $S_{32} = \int_{z_1>0}\varphi_{310}(\tilde{\mathbf{r}}_1)\varphi_{210}(\tilde{\mathbf{r}}_1)d\tilde{\mathbf{r}}_1 = 0$ (i.e. to orthogonality of the one-electron wave functions $\varphi_{210}(\mathbf{r})$ and $\varphi_{310}(\mathbf{r})$). Also we should demand the orthogonality of the singlet $\psi_{23}^S(\mathbf{r}_1,\mathbf{r}_2)$ to the singlet $\psi_{22}(\mathbf{r}_1,\mathbf{r}_2)$ and $\psi_{33}(\mathbf{r}_1,\mathbf{r}_2)$. This again leads to the condition $S_{32} = 0$ that demands the constraint $2\tilde{\alpha}-3\tilde{\beta} = 0$ for all singlet levels. Triplet $\psi_{23}^T(\mathbf{r}_1,\mathbf{r}_2)$ is already orthogonal to all singlet wave functions at arbitrary $\tilde{\alpha}$ and $\tilde{\beta}$, both due to antisymmetric coordinate part and symmetric spin functions, thus for the function we have no need to impose the condition $S_{32} = 0$ and thus no constraints like $2\tilde{\alpha}-3\tilde{\beta} = 0$ exist for the variational parameters $\tilde{\alpha}$ and $\tilde{\beta}$ in the triplet state.

Results of numerical minimization of the energies (11) are presented in the **Table 1** and in **Figs. 2**. For comparison we used the condition $S_{32} = 0$ (i.e. $\tilde{\beta} = 2\tilde{\alpha}/3$) for the energy $\tilde{E}_{23}^{T}$ calculation and obtained that even under the condition it remains the lowest one, though its energy becomes slightly higher than listed in the **Table 1**, namely –0.130366 for $\varepsilon_2 = 3$ and –0.11763 for $\varepsilon_2 = 10$.

**Table 1.** Lowest energy levels of the surface defect and corresponding values of the variational parameters $\tilde{\alpha}$ and $\tilde{\beta}$.

| Energy level in atomic units ($\hbar^2/(a_B^{*2}\lvert\mu\rvert)$) | $\varepsilon_2$=3 | | | $\varepsilon_2$=10 | | |
|---|---|---|---|---|---|---|
| | Energy value | $\tilde{\alpha}$ | $\tilde{\beta}$ | Energy value | $\tilde{\alpha}$ | $\tilde{\beta}$ |
| $\tilde{E}_{33}$ $(\Sigma = 0)$ | -0.0484583 | 0.4210 | 0.2807 $\tilde{\beta} = 2\tilde{\alpha}/3$ | -0.0445196 | 0.400 | 0.267 $\tilde{\beta} = 2\tilde{\alpha}/3$ |
| $\tilde{E}_{22}$ $(\Sigma = 0)$ | -0.0990634 | 0.4210 | 0.2807 $\tilde{\beta} = 2\tilde{\alpha}/3$ | -0.0914374 | 0.400 | 0.267 $\tilde{\beta} = 2\tilde{\alpha}/3$ |
| $\tilde{E}_{23}^{S}$ $(\Sigma = 0)$ | -0.128015 | 0.4210 | 0.2807 $\tilde{\beta} = 2\tilde{\alpha}/3$ | –0.115652 | 0.400 | 0.267 $\tilde{\beta} = 2\tilde{\alpha}/3$ |
| $\tilde{E}_{23}^{T}$ $(\Sigma = 1)$ | -0.132799 | 0.4683 | 0.2614 | –0.11982 | 0.445 | 0.249 |

It is seen from the **Table 1** data (calculated in dimensionless units $\hbar^2/(a_B^{*2}\lvert\mu\rvert)$) and **Figs. 2a,b** curves (calculated in eV and Kelvins) that the triplet state $\tilde{E}_{23}^{T}$ is the ground one at the surface in entire range of dielectric permittivity $\varepsilon_2$~2-20 and defect charge $\lvert Z\rvert$~1-3. The singlet state $\tilde{E}_{23}^{S}$ is the first excited state. Next singlet states ($\tilde{E}_{22}$ and $\tilde{E}_{33}$) are well separated (~0.2–0.5 eV at $\mu=m_e$) from the lowest states $\tilde{E}_{23}^{T}$ and $\tilde{E}_{23}^{S}$ for the defect charge $\lvert Z\rvert \leq 2$ [**Figs. 2a**]. For the case $\lvert Z\rvert$>2 and fixed $\varepsilon_2$~10 the difference singlet state energy $\tilde{E}_{22}$ tends to the energies $\tilde{E}_{23}^{S,T}$, while the energy gap between $\tilde{E}_{33}$ and other lower levels ($\tilde{E}_{22}$ and $\tilde{E}_{23}^{S,T}$) only increases up to 0.6 eV at $\mu=m_e$ [**Figs. 2b**].

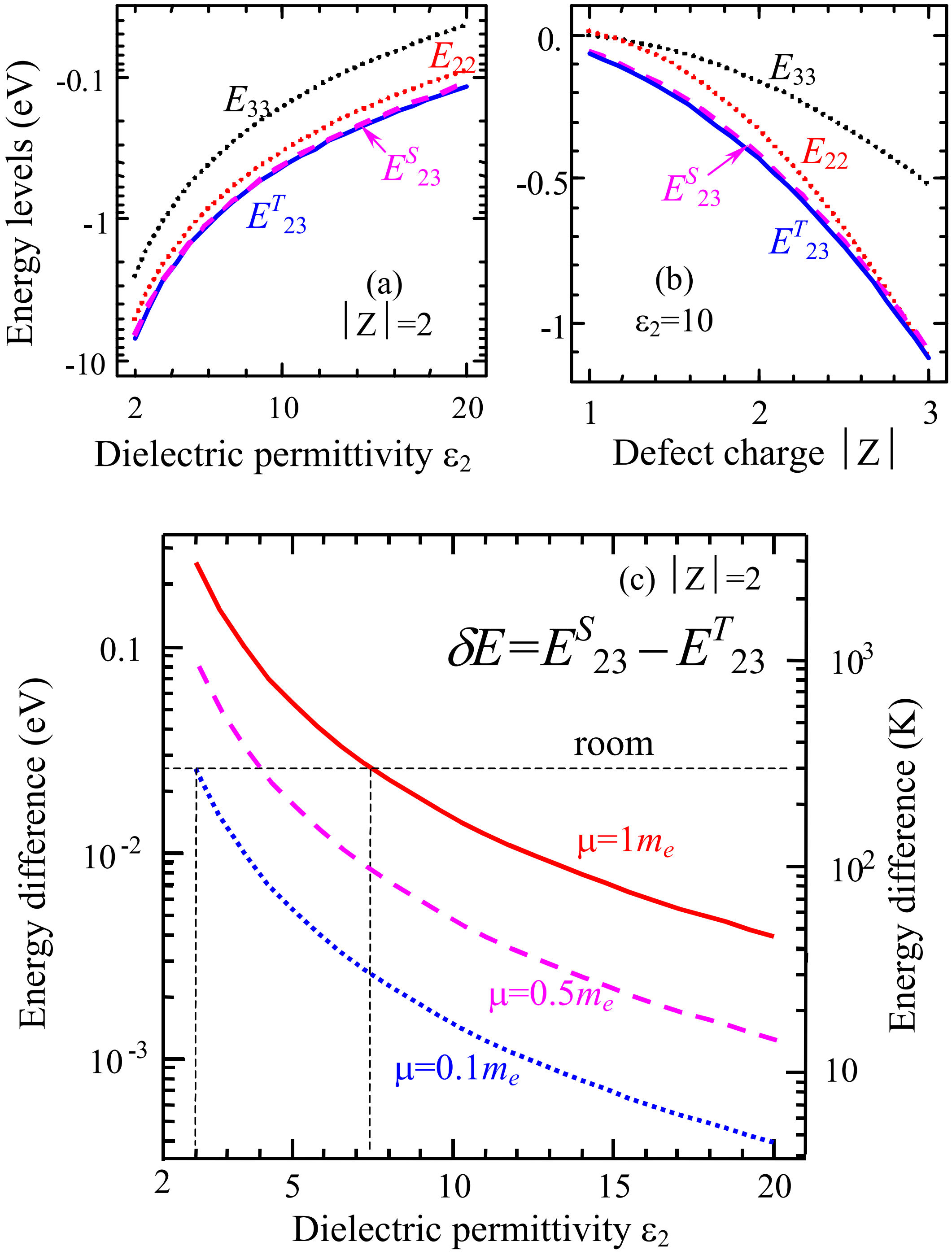

**Fig. 2**. (a) The dependence of energy levels (in eV) on the dielectric permittivity $\varepsilon_2$ calculated for $\varepsilon_1$=1, |Z|=2, $\mu=m_e$. (b) The dependence of energy levels (in eV) on defect charge |Z| calculated for $\varepsilon_1$=1, $\varepsilon_2$=10. (c) The difference between the lowest levels $\delta\tilde{E} = \tilde{E}_{23}^{S} - \tilde{E}_{23}^{T}$ in eV (left scale) and Kelvins (right scale) calculated for |Z|=2, $\mu=m_e$ (solid curve), $\mu=0.5m_e$ (dashed curve) and $\mu=0.1m_e$ (dotted curve).

The energy difference $\delta\tilde{E} = \tilde{E}_{23}^{S} - \tilde{E}_{23}^{T}$ strongly depends on dielectric permittivity $\varepsilon_2$ at fixed other parameters [**Fig. 2c**]. For the case $\mu=m_e$ it is giant (~3000 K) at $\varepsilon_2$=3, more than several hundreds of Kelvins at $\varepsilon_2$=5-10 and about several tens of Kelvins at $\varepsilon_2$>20. For the case $\mu=0.1m_e$ it is about 300 K at $\varepsilon_2$=3, and only about 30-10 K at $\varepsilon_2$=5-10. Generally, the scaling on effective mass exists, namely all energy levels linearly depend on μ value, e.g. the levels splitting $\delta\tilde{E} \sim \mu$ (compared solid, dashed and dotted curves in **Figs. 2c**).

It is seen from the **Figs. 2** that the ground triplet state could be occupied (and correspondingly the higher singlet is empty) even at room temperatures. The triplet state $2p_z3p_z$ should become the magnetic one ($\Sigma_z = \pm 1$) allowing for the Hund's rule that orient two fermions (electrons or holes) spins in the same direction. The situation may lead to the surface-induced magnetism in nonmagnetic solids.

Note, that relatively high values of effective mass μ ($\mu>0.5m_e$), moderate permittivity $\varepsilon_2$<5-10 and $|Z|<2$ seem favorable for the energy levels splitting increase and thus for the magnetic state appearance. To demonstrate this, the dependence of energy levels on the effective mass μ is shown in **Figs. 3**. The difference between the energy levels linearly increases with the effective mass increase at fixed other parameters as shown in **Fig. 3a**. The difference between the lowest levels $\delta\tilde{E} = \tilde{E}^S_{23} - \tilde{E}^T_{23}$ decreases with dielectric permittivity $\varepsilon_2$ increase as shown in **Fig. 3b**. The difference $\delta\tilde{E}$ reaches noticeable values (e.g. 0.1 eV) for $\mu > m_e$ and $\varepsilon_2$<10.

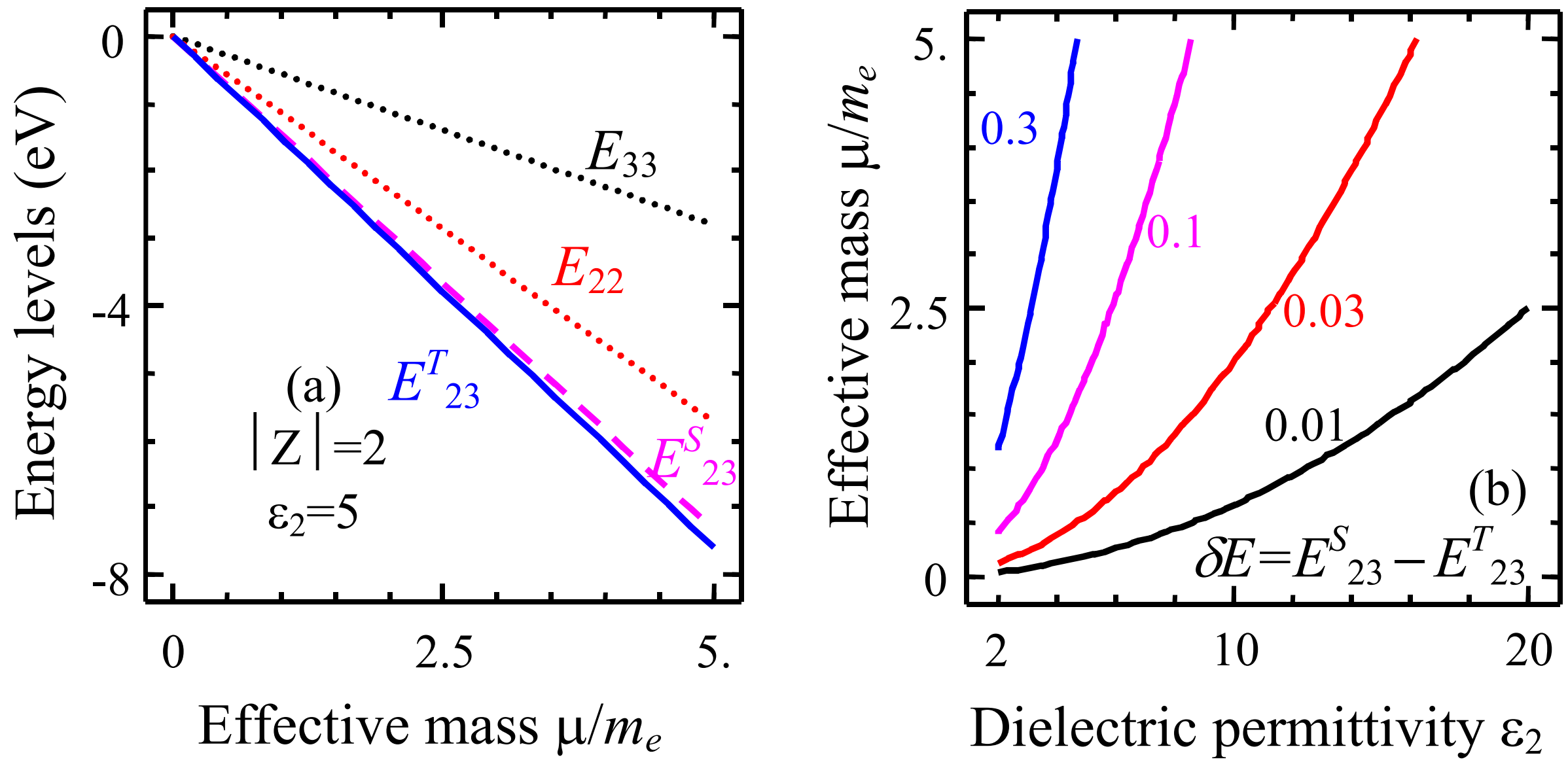

**Fig. 3**. (a) The dependence of energy levels $E$ on effective electron mass calculated for $\varepsilon_1$=1, $\varepsilon_2$=5, $|Z|$=2. (b) Contours of constant values of difference between the lowest levels $\delta\tilde{E} = \tilde{E}^S_{23} - \tilde{E}^T_{23}$ (in eV, marked near the curves) as a function of dielectric permittivity and effective mass calculated for $\varepsilon_1$=1 and $|Z|$=2.

Let us check the validity of the effective mass approximation we used for the energy calculations. Since the values of inverse variational parameters, $\alpha^{-1}$ and $\beta^{-1}$, determine the wave function $\psi_{mn}(\mathbf{r}_1, \mathbf{r}_2)$ radius of state, their dependences on the effective mass and dielectric permittivity contain quantitative information about the validity of the effective mass approximation we used. As it was argued in the subsection 2.2 and **Appendix A**, the model assumptions require the strong inequalities $\frac{1}{\alpha} >> \frac{a}{4\pi}$ and $\frac{1}{\beta} >> \frac{a}{3\pi}$ to be valid. For the lattice constant $a$~0.5 nm, the inequalities $\frac{1}{\alpha} >> 0.04$ nm and $\frac{1}{\beta} >> 0.05$ nm should be valid. The dependences of the variational parameters $\alpha^{-1}$ and $\beta^{-1}$ on the on effective mass μ and dielectric permittivity $\varepsilon_2$ are shown in **Figs. 4**. It is seen from the **Figs. 4a** that the required inequalities $\frac{1}{\alpha} > 0.5$ nm and $\frac{1}{\beta} > 0.5$ nm are valid for light effective masses $\mu < (0.7-1)m_e$ and $\varepsilon_2 \geq 5$ (the situation becomes only better with dielectric permittivity $\varepsilon_2$ increase as shown in **Fig. 4b**). Thus all our analytical results obtained in the effective mass approximation are self-consistent in the parameters range $\mu < (0.7-1)m_e$ and $\varepsilon_2 \geq 5$.

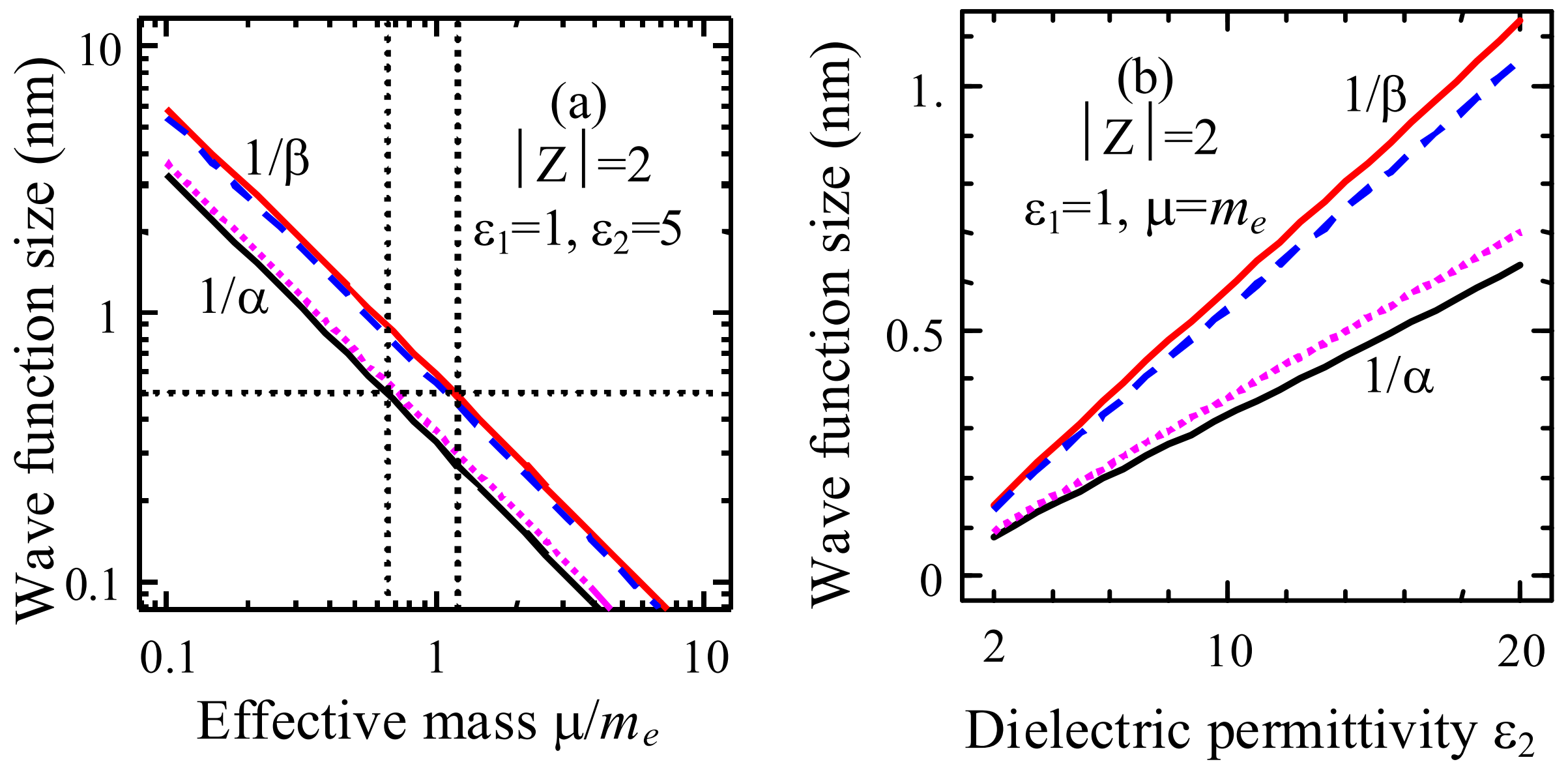

**Fig. 4**. (a) Dependence of the inverse variational parameter $\alpha^{-1}$ (lower curves for triplets and singlets) and $\beta^{-1}$ (upper curves for triplets and singlets) on effective mass μ calculated for $\varepsilon_1$=1, $\varepsilon_2$=5, $|Z|$=2. (b) Dependence of the inverse variational parameter $\alpha^{-1}$ (lower curves for triplets and singlets) and $\beta^{-1}$ (upper curves for triplets and singlets) on the dielectric permittivity $\varepsilon_2$ calculated for $\varepsilon_1$=1 and $|Z|$=2, $\mu=m_e$.

## 4. Energy levels of the defect with two electrons localized under the surface

### *4.1. The problem statement for a defect with two electrons located under the surface*

Analyzing results of the Section 3 the natural question arises: to which levels the functions (6) and (7) correspond in the bulk of material; and (more important) how far from the surface the magnetic triplet state remained the lowest one.

Qualitatively, the trial $2p_z$ and $3p_z$ wave functions (6) and (7) used at the surface $z_0 = 0$ should be gradually transformed into the spherically-symmetric 1s and 2s functions in the limit $z_0 \to \infty$, as anticipated for the lowest states of the hydrogen-like atoms in the bulk. One could expect that the magnetic triplet state should remain the lowest one until the p-type contribution is dominant in the wave functions.

To answer the questions quantitatively let us consider the defect (impurity or vacancy) located in the point $\mathbf{r}_0 = \{0,0,z_0\}$ under the surface. Its image coordinates are $\mathbf{r}_0^{/} = \{0,0,-z_0\} = -\mathbf{r}_0$. Radius-vector $\mathbf{r}_i = (x_i, y_i, z_i)$ is the coordinates of the electrons (holes) 1 and 2. Their images coordinates are $\mathbf{r}_i^{/} = (x_i, y_i, -z_i)$. The geometry of the calculations is shown in **Fig. 1d**.

Schrödinger equation for the two-fermions wave-function coordinate part has the form of Eq. (2) along with Eqs.(9). In dimensionless coordinates $\tilde{\mathbf{r}}$ and $\tilde{\mathbf{r}}_0$, introduced the subsection 4.1, the only different electrostatic energy contribution $\tilde{V}_e(\tilde{\mathbf{r}}_i)$ acquires the form: $\tilde{V}_e(\tilde{\mathbf{r}}_i) = -\frac{\varepsilon_1+\varepsilon_2}{2\varepsilon_2}\left(\frac{1}{|\tilde{\mathbf{r}}_i-\tilde{\mathbf{r}}_0|} + \frac{\varepsilon_2-\varepsilon_1}{\varepsilon_1+\varepsilon_2}\frac{1}{|\tilde{\mathbf{r}}_i+\tilde{\mathbf{r}}_0|}\right) + \zeta\frac{\varepsilon_1+\varepsilon_2}{2\varepsilon_2|Z|}\frac{1}{2\tilde{z}_i}$ (see **Appendix C** for details).

### *4.2. Approximate solution obtained by direct variational method*

Possible two-fermions wave functions are given by Eqs.(8). For the considered case $z_0 \neq 0$ the lowest one-particle wave function, which transforms to Eq.(6) at $z_0 = 0$, has the form:

$$\varphi_{210}(\mathbf{r}) = A(\alpha, z_0)\exp(-\alpha|\mathbf{r}-\mathbf{r}_0|)\cdot z,$$
$$A(\alpha, z_0) = \sqrt{\frac{4\alpha^5}{\pi\left(4\alpha^2 z_0^2 + 4 - (2+\alpha z_0)\exp(-2\alpha z_0)\right)}}. \tag{15}$$

Normalization constant $A(\alpha, z_0)$ is proportional to $1/z_0$ at $z_0 \to \infty$. Thus the wave function $\varphi_{210}(\mathbf{r}) \to \exp(-\alpha|\mathbf{r}-\mathbf{r}_0|)\frac{z}{z_0} \approx \exp(-\alpha|\mathbf{r}-\mathbf{r}_0|)$ at $z_0 \to \infty$ with $|z - z_0| \sim \alpha^{-1}$, so that it tends to the 1s wave function in the bulk (see also **Fig. 5**).

Excited one-electron wave function, which transforms to Eq.(7) at $z_0 = 0$, has the form:

$$\varphi_{310}(\mathbf{r}) = B(\beta, z_0)\cdot z\left(b(z_0) - \beta|\mathbf{r}-\mathbf{r}_0|\right)\exp\left(-\beta|\mathbf{r}-\mathbf{r}_0|\right),$$

$$B(\beta, z_0) = \sqrt{\frac{4\beta^5}{\pi}}\begin{pmatrix} 30 - 20b + 4b^2 + 4\beta^2 z_0^2\left(3 - 3b + b^2\right) + \\ \exp(-2\beta z_0)\left(2b\left(5 - 11\beta z_0 - 5\beta^2 z_0^2 - \beta^3 z_0^3\right) - 15 - b^2(2 + \beta z_0)\right)\end{pmatrix}^{-\frac{1}{2}} \tag{16}$$

Normalization constant $B \sim 1/z_0$ at $z_0 \to \infty$.

The variational parameter $b(z_0)$ should provide the orthogonality of the functions (16) and (15). Actually, since the wave functions, which correspond to the different eigen values of energy, should be orthogonal, the condition $\int_{z_1>0} \varphi_{210}(\mathbf{r}_1)\varphi_{310}(\mathbf{r}_1)d\mathbf{r}_1 = 0$ being sufficient for the orthogonality of the singlet wave functions $\psi_{22}(\mathbf{r}_1, \mathbf{r}_2)$ and $\psi_{33}(\mathbf{r}_1, \mathbf{r}_2)$ given by Eq.(8). The condition leads to the expression for $b$:

$$b(z_0) = \frac{\beta\left(2\exp(z_0(\alpha+\beta))\left(20 + 3z_0^2(\alpha+\beta)^2\right) - 20 - z_0(\alpha+\beta)(8 + z_0(\alpha+\beta))\right)}{(\alpha+\beta)\left(2\exp(z_0(\alpha+\beta))\left(4 + z_0^2(\alpha+\beta)^2\right) - 4 - z_0(\alpha+\beta)\right)}$$

$$\to \begin{cases} \dfrac{5\tilde{\beta}}{\tilde{\alpha}+\tilde{\beta}} & \text{for } z_0 = 0, \text{ thus } b = 2 \text{ at } \tilde{\alpha} = \dfrac{3}{2}\tilde{\beta} \text{ (surface, see Table 1)} \\ \dfrac{3\tilde{\beta}}{\tilde{\alpha}+\tilde{\beta}} & \text{for } z_0 >> a_B^*\sqrt{\dfrac{20}{3\left(\tilde{\alpha}+\tilde{\beta}\right)^2}}, \text{ thus } b \to 1 \text{ at } z_0 \to \infty, \tilde{\alpha} = 2\tilde{\beta} \text{ (bulk)} \end{cases} \tag{17}$$

It is seen from Eq.(17) that parameter $b(0) = 2$ at the surface ($z_0 = 0$), where $\tilde{\alpha} \sim 1/2, \tilde{\beta} \sim 1/3$ (see **Table 1**), while $b(z_0) \to 1$ in the bulk ($z_0 \to \infty$), where $\tilde{\alpha} = 1, \tilde{\beta} = 1/2$ (see e.g. Ref. [30]).

Thus $\varphi_{310}(\mathbf{r}) \to \exp(-\beta|\mathbf{r}-\mathbf{r}_0|)(1 - \beta|\mathbf{r}-\mathbf{r}_0|)\left(1 + \dfrac{z - z_0}{z_0}\right) \to \exp(-\beta|\mathbf{r}-\mathbf{r}_0|)(1 - \beta|\mathbf{r}-\mathbf{r}_0|)$ at $z_0 \to \infty$ and $|z - z_0| \sim \beta^{-1}$, so that the excited function tends to the 2s wave function in the bulk of material as it can be expected, but not into 2p wave function as appeared in [24] (see **Fig. 7**). We would like to underline, that the function (16) is different from the one used in Ref.[24] everywhere, due to the presence of factor $\left(b(z_0) - \beta|\mathbf{r}-\mathbf{r}_0|\right)$.

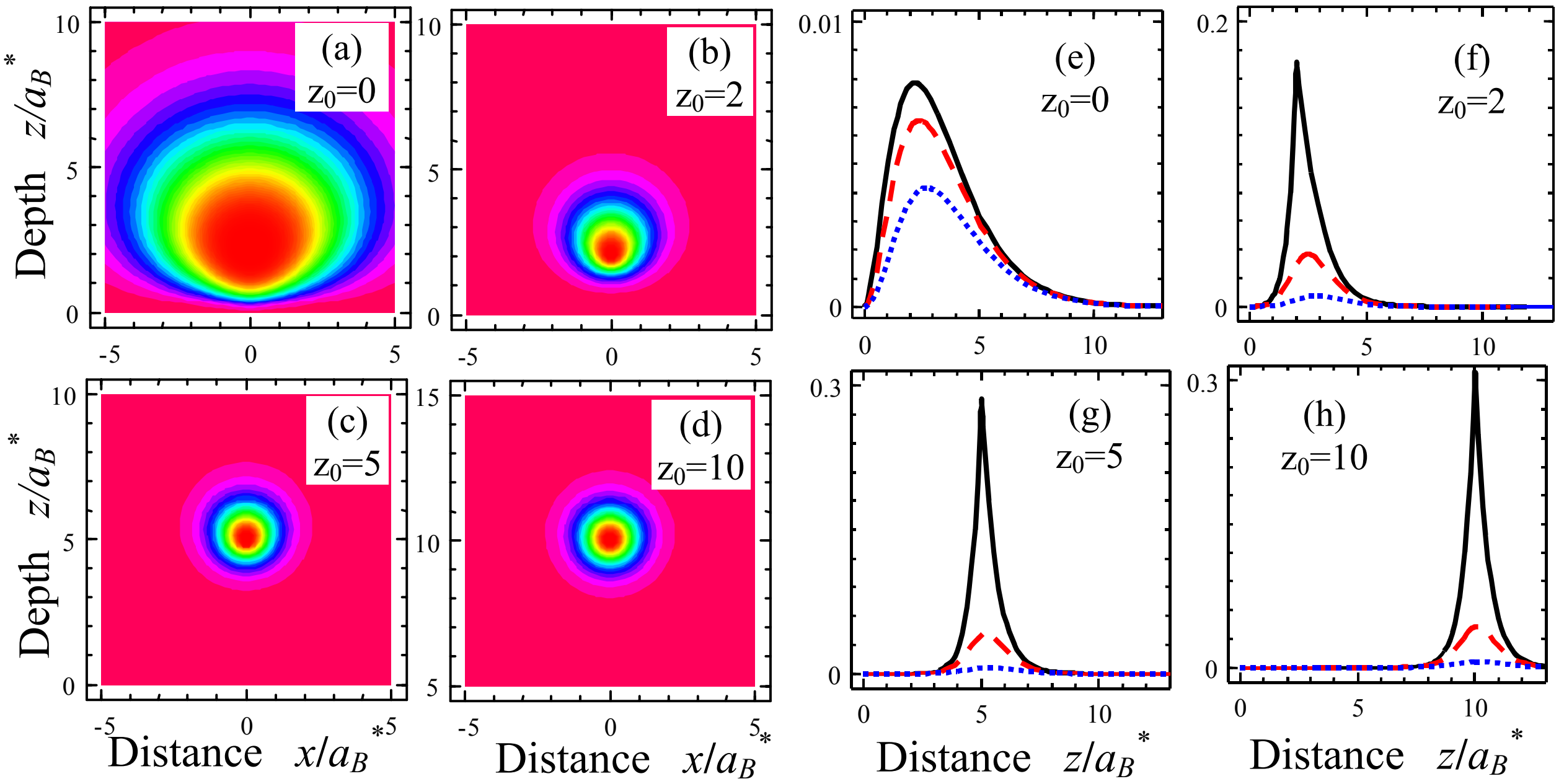

**Fig. 5.** Wave-function density $|\varphi_{210}(\mathbf{r})|^2$ contour maps (a-d) and its $z$-dependence (e-h) calculated at different distances $\tilde{z}_0 = 0,\ 2,\ 5,\ 10$ from the surface ($\tilde{z}_0 = z_0 / a_B^*$). For (e-h) $y = x$ and $x = 0$ (solid curves), $x = a_B^*$ (dashed curves), $x = 2a_B^*$ (dotted curves).

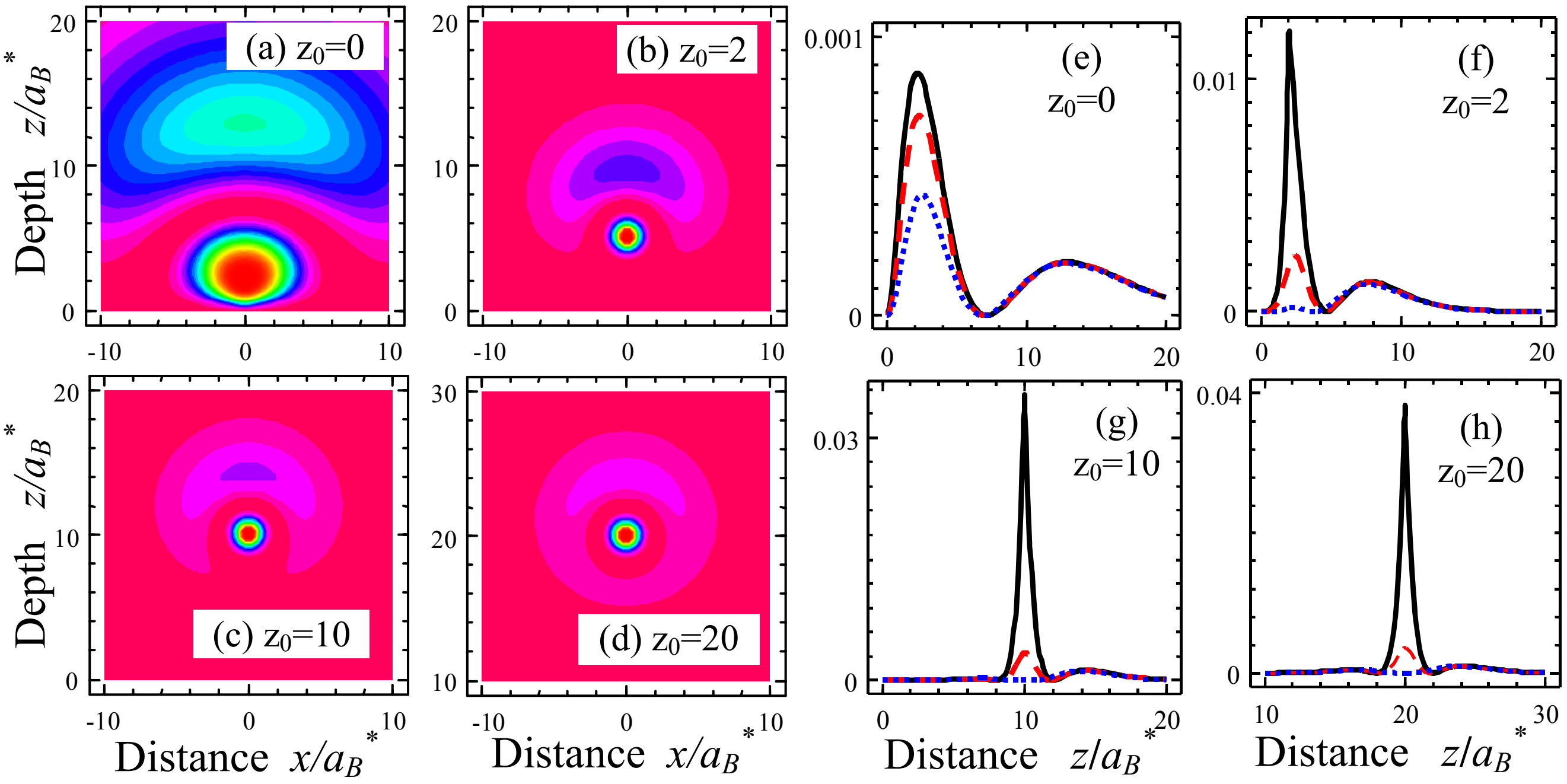

**Fig. 6.** Wave-function density $|\varphi_{310}(\mathbf{r})|^2$ contour maps (a-d) and its $z$-dependence (e-h) calculated at different distances $\tilde{z}_0 = 0,\ 2,\ 10,\ 20$ from the surface ($\tilde{z}_0 = z_0 / a_B^*$). For (e-h) $y = x$ and $x = 0$ (solid curves), $x = a_B^*$ (dashed curves), $x = 2a_B^*$ (dotted curves).

It is seen from the **Fig. 5, 6** that the one-particle wave functions (15) and (16) transform to the spherically-symmetric wave functions of 1s and 2s states at distances $z_0 > 5a_B^*$ and $z_0 > 10a_B^*$ correspondingly. Note, that 1s and 2s states are the lowest states of He atom in the bulk, see e.g. Ref. [25, 30]. For chosen material parameters the mixture of s-p states is observed at distances $0.5a_B^* < z_0 < 10a_B^*$, at that the p-type contribution being dominant at distances $0 < z_0 < 2a_B^*$. The transition to the bulk s-type wave functions can be expected for the distances more than $10a_B^*$.

Eqs. (11) are valid for the singlet and triplet energy levels at arbitrary $z_0$. Corresponding overlap integral $S_{23}$, matrix elements $A_{mn}$, exchange integral $J_{23}$ and Coulomb integrals $C_{mn}$ are calculated on functions (15)-(16) in **Appendix C**. For considered case $S_{23} = 0$, so that the lowest energy levels difference $\tilde{E}_{23}^S - \tilde{E}_{23}^T = 2J_{23}$.

Dependences of energy levels $E_{ij}$ and variational parameters *b*, $\tilde{\alpha}$ and $\tilde{\beta}$ on the distance $z_0$ from the surface are shown in **Figs. 7a,b** and **7c** correspondingly**.** The highest surface singlet state 33 transforms into the excited singlet state 2s2s in the bulk of material. Surface singlet state 22 transforms into the lowest singlet state 1s1s in the bulk of material. Surface singlet state 23 transforms into the excited triplet state 1s2s in the bulk of material. The ground surface triplet state 23 transforms into the excited singlet state 1s2s in the bulk of material. The energy values linearly scales with the ratio $\mu/m_e$, since $E_{ij} \sim \mu$ [compare **Fig. 7a** for $\mu = m_e$ with **Fig. 7b** for $\mu = 0.1m_e$]. The distance between the lowest sub-surface states $\tilde{E}_{23}^S$ and $\tilde{E}_{23}^T$ are higher than the thermal activation energy (0.025 $k_BT$) for effective mass $\mu \sim m_e$, but it becomes much lower than the thermal energy for $\mu \leq 0.1m_e$, thus both states should be occupied at room temperature for light effective mass.

It is seen from the **Figs. 7c**, that despite the function $b(\tilde{z}_0)$ transforms into its bulk value 1 at distances $z_0 > 10a_B^*$, the triplet state $\tilde{E}_{23}^T$ is the lowest one only in the immediate vicinity of the surface, i.e. at $z_0 < 2a_B^*$ (see vertical line position at $z_0 \approx 2a_B^*$ in **Fig.7a** and **Fig. 7b**). At higher distances $z_0 > 2a_B^*$ the singlet state $\tilde{E}_{22}$ becomes the lowest one and transforms into the lowest 1s1s state in the bulk.

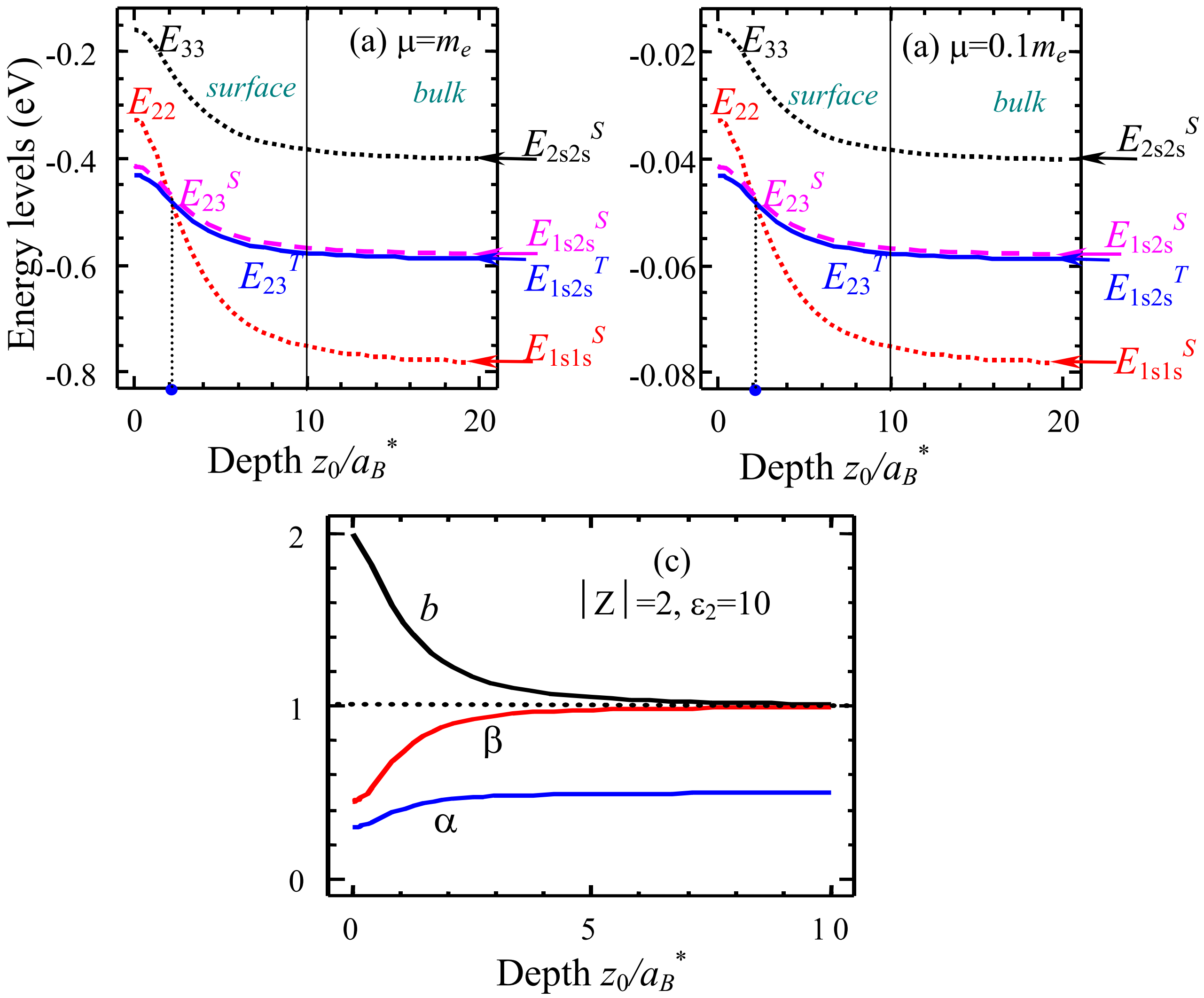

**Fig. 7**. Dependence of energy levels $E_{ij}$ (in eV) on the distance $z_0$ from the surface calculated for μ=$m_e$ (a) and μ=0.1$m_e$ (b) and $\varepsilon_1$=1, $\varepsilon_2$=10 |Z|=2. Right arrows indicate the transition to the bulk energy levels: 1s1s, 1s2s (singlet), 1s2s (triplet) and 2s2s correspondingly. (c) Dependence of the variational parameters $b$, $\tilde{\alpha}$ and $\tilde{\beta}$ on the distance $z_0$ from the surface.

Analyzing results of Eqs.(11) minimization (shown in **Figs. 7**), we obtained that the following Pade approximations for $b(\tilde{z}_0)$, $\alpha(\tilde{z}_0)$, $\beta(\tilde{z}_0)$ and $E_{ij}(\tilde{z}_0)$ dependences are valid:

$$b(\tilde{z}_0)=\frac{1}{1+\tilde{z}_0^2}+1, \qquad \tilde{\alpha}(\tilde{z}_0)=\frac{\tilde{\alpha}(0)-0.5}{1+\tilde{z}_0^2}+0.5, \qquad \tilde{\beta}(\tilde{z}_0)=\frac{\tilde{\beta}(0)-1}{1+\tilde{z}_0^2}+1, \tag{18}$$

$$E_{ij}^m(\tilde{z}_0)=\frac{E_{ij}^m(0)-E_{kl}^m(\infty)}{1+0.15\cdot\tilde{z}_0^2\left(\tilde{\alpha}+\tilde{\beta}\right)^2}+E_{kl}^m(\infty). \tag{19}$$

Here subscripts $ij=22,\ 23,\ 33$ and $kl=1s1s,\ 1s2s,\ 2s2s$, superscript $m=S,T$ indicate the singlet or triplet state.

The approximate expressions given by Eqs.(18)-(19) have the advantage of the analytical form of the energy levels and parameters of the wave functions dependence on the distance from the surface.

Interestingly, that the surface-perturbed wave functions ($2p_z$ and $3p_z$) transform into their bulk limits (1s and 2s) at distances about 10 $a_B^*$, while the energy level $\tilde{E}_{23}^T$ remains the lowest one at distances less than 2 $a_B^*$. Thus the wave function relaxation to the bulk shape is more slow than the energy difference $\tilde{E}_{23}^T - \tilde{E}_{22}$ relaxation, the latter was obtained after the wave functions integration along with Hamiltonian. However the energy levels per se tend to its bulk values at the distances ~10 $a_B^*$, i.e. at the same distance as wave functions relax [compare **Figs. 7** with **Figs. 5-6**].

Note, that effective Bohr radius $a_B^*$ strongly depends on the effective mass μ and dielectric permittivity $\varepsilon_2$: $a_B^* \sim \varepsilon_2/\mu$ [see **Fig. 8**]. It is seen from the **Fig. 8** that $a_B^*$~1–10 nm for light effective mass $\mu < 0.1\ m_e$ and $\varepsilon_2 > 5$. Thus predicted surface-induced magnetic triplet state could exist not only at the surface, but up to the distances of 2-20 nm under the surface.

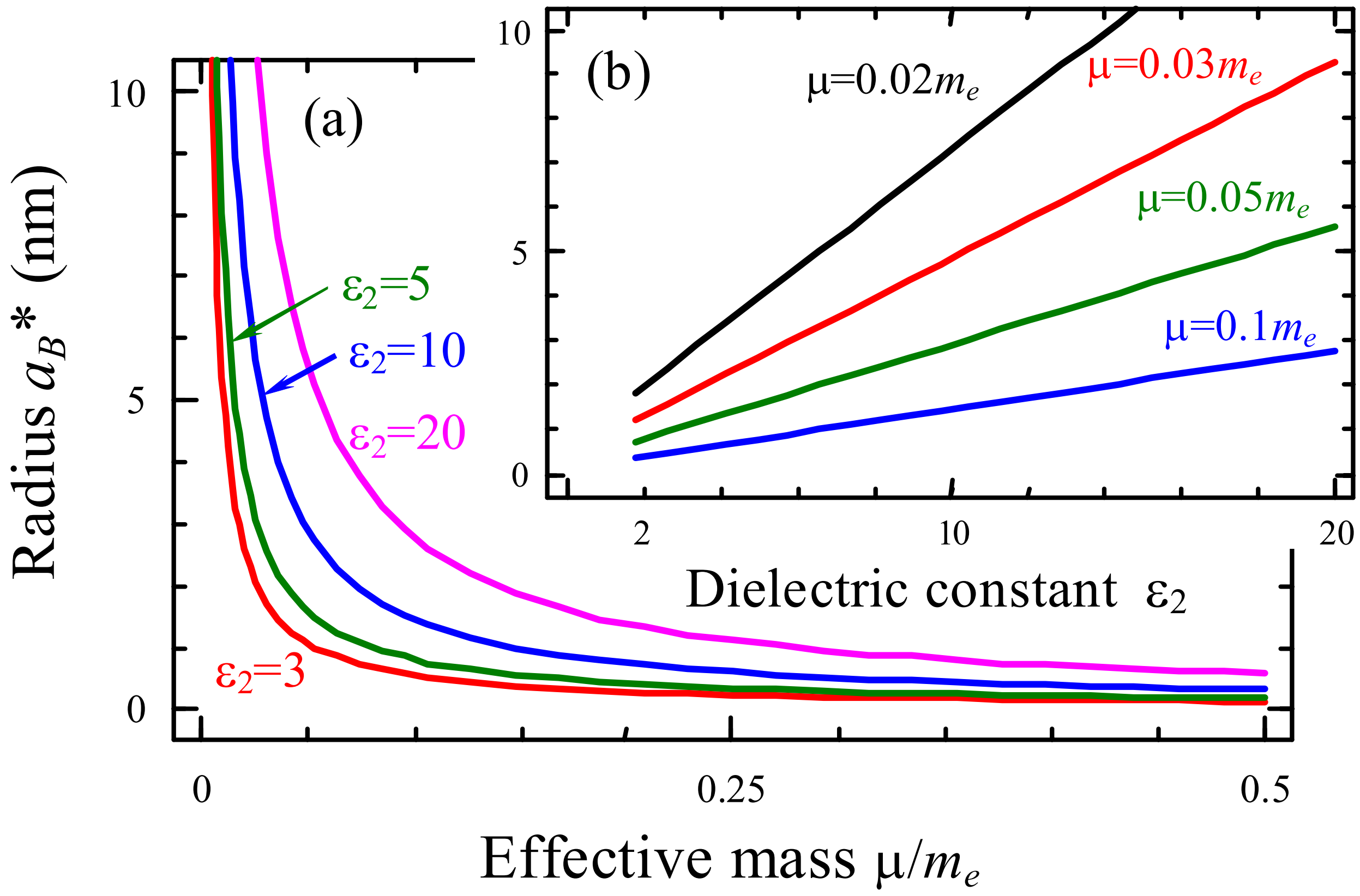

**Fig. 8.** Effective radius $a_B^*$ dependence on (a) effective mass calculated for $\varepsilon_2$=3, 5, 10, 20 and (b) dielectric permittivity for $\mu/m_e$ = 0.02, 0.03, 0.05, 0.1 and $\varepsilon_1$=1.

To summarize the section, the magnetic triplet state 23 ($2p_z3p_z$) remained the lowest one up to the distances 2-10 nm under the surface. At higher distances the nonmagnetic spherically-symmetric singlet state 1s1s becomes the lowest one (as anticipated for the He-like atoms in the bulk of material).

## 5. Pair of defects located at solid surface: spins interaction and energy levels

### *5.1 Pair interaction between the magnetic defects*

Antiferromagnetic or ferromagnetic spin state of the defect pair is determined by the ***sign of exchange integral*** in the effective Hamiltonian that describes interaction between the four electrons (or holes) spins of the neighboring surface defects.

Let us calculate the exchange integral between the two defects $a$ and $b$ located in the points $\mathbf{R}_a = (-R/2, 0, 0)$ and $\mathbf{R}_b = (+R/2, 0, 0)$ at the surface of solid (see **Fig. 1f** and **g**). Both defects are in the lowest triplet state $\tilde{E}_{23}^{T}$ as calculated in the Section 3. Hereinafter we used the condition $S_{23} = 0$ for the orthogonality of the considered 210- and 310-wave functions of the carriers localized at the one defect center. Under these condition the exchange contribution to the energy is determined as

$$J_{23} = \int_{z_2>0} \int_{z_1>0} \tilde{V}_{12}(\tilde{\mathbf{r}}_1, \tilde{\mathbf{r}}_2) \varphi_{210}(\tilde{\mathbf{r}}_1 - \tilde{\mathbf{R}}_a) \varphi_{310}(\tilde{\mathbf{r}}_2 - \tilde{\mathbf{R}}_b) \varphi_{310}(\tilde{\mathbf{r}}_2 - \tilde{\mathbf{R}}_a) \varphi_{210}(\tilde{\mathbf{r}}_1 - \tilde{\mathbf{R}}_b) d\tilde{\mathbf{r}}_1 d\tilde{\mathbf{r}}_2 \qquad (20)$$

The integral (20) is calculated analytically in the **Appendix D**.

Dependence of the exchange integral vs. the distance between the defects is shown in **Fig. 9**. It is seen from the figure that the pair exchange interaction of the identical surface defects (two doubly charged defects with four shared electrons or holes) reveals ferromagnetic spin state independently on the distance between the defects, but the exchange integral value is significant at distances $R$~2-10 $a_B^*$ and has pronounced maximum at distances $R$~5 $a_B^*$ (that is about 2 - 50 nm and 5 – 25 nm correspondingly in accordance with Fig.8 for $\mu < 0.1 m_e$ and $\varepsilon_2 > 5$).

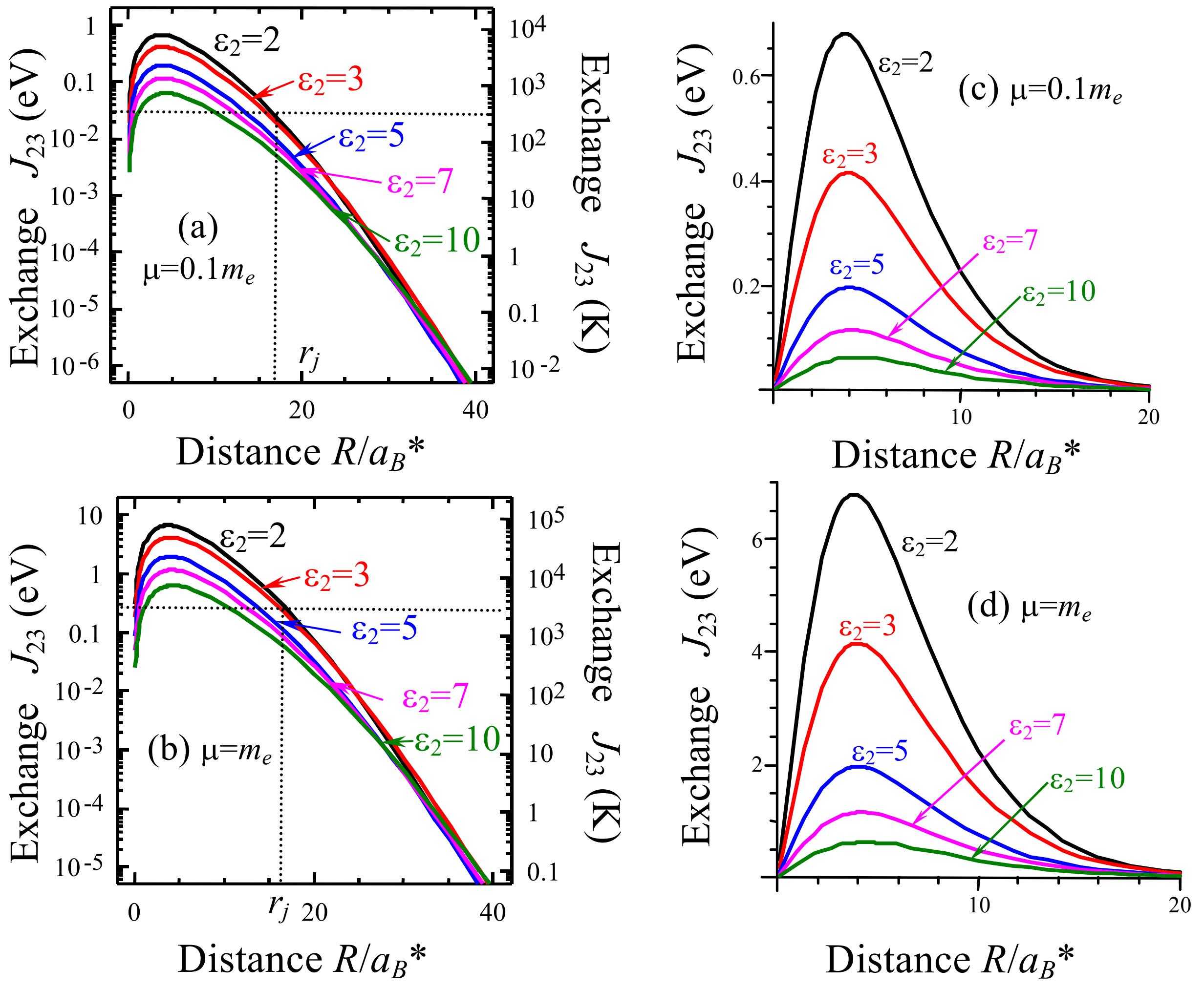

**Fig. 9.** Dependence of the exchange integral (left scale in eV, right scale in K) vs. the distance $R$ between the defects calculated for $\mu=0.1m_e$ (a), $\mu=m_e$ (b), $\varepsilon_1=1$ and different values of $\varepsilon_2$ specified near the curves. Graphs are plotted in log-linear scale, insets are plotted in linear scale.

Now let us discuss semi-quantitatively the possible mechanisms of the appearance of ferromagnetic long-range order between the defects in the vicinity of surface.

**(a)** The defect-induced ferromagnetism can have percolation nature [7], especially in thin films and nanoparticles, when the problem dimensionality reduces up to 2D by the spatial confinement. For continuous media approximation used in our paper the most appropriate is the well-known problem of spheres for a bulk three-dimensional (3D) case or circles for the surface two-dimensional (2D) case. Considering the problem of percolation in the system with random sites magnetic defects, placed in the center of a sphere or a circle, we use the conventional expressions $B_{3D}=\frac{4}{3}\pi N_{3D}r_j^3$ and $B_{2D}=\pi N_{2D}r_j^2$ for the number of overlapping spheres or circles equal to the average number of the interacting defects inside a sphere or a circle [26]. Here $r_j$, $N_{3D}$ and $N_{2D}$ are respectively the exchange radius and the concentration of the magnetic defects for 3D and 2D cases. Critical numbers of overlapping spheres or circles were calculated as $B_{3D}^c = (2.7 - 3.3) \approx 3$ and $B_{2D}^c = (3.8 - 4.2) \approx 4$ independently on their concentration and radius [26]. Allowing for temperature dependence of $r_j$ we will take its value at room temperature $T_r = 300\,\text{K}$ we are interested in, namely from the condition of the exchange integral equality to the thermal energy at room temperature $J_{23}(r_j) = k_B T_r$ (see vertical lines in **Figs. 9**). Then the percolation concentration of random sites at the surface and under the surface, $N_{2D}^c$ and $N_{3D}^c$ respectively, should be determined for a known radius $r_j$. Percolation critical concentrations $N_{3D}^c$ and $N_{2D}^c$ calculated for several values of effective mass μ and permittivity $\varepsilon_2$ are summarized in the **Table 2**.

**Table 2.** Percolation critical concentrations $N_{3D}^c$ and $N_{2D}^c$

| | Effective mass μ=0.1 $m_e$ | | Effective mass μ=1 $m_e$ | |
|---|---|---|---|---|
| Permittivity $\varepsilon_1$=1 | $\varepsilon_2$=3 | $\varepsilon_2$=10 | $\varepsilon_2$=3 | $\varepsilon_2$=10 |
| $a_B$*(nm) | 0.53 | 1.46 | 0.05 | 0.15 |
| $r_j$ (nm) | 8.0 | 15 | 1.2 | 2.6 |
| $N_{3D}^c$ ($10^{19}$ cm$^{-3}$) | 0.14 | 0.02 | 41.14 | 4.08 |
| $N_{3D}^c$ (%) | 0.017 | 0.002 | 5.18 | 0.51 |
| $N_{2D}^c$ ($10^{13}$ cm$^{-2}$) | 0.20 | 0.06 | 8.84 | 1.88 |
| $N_{2D}^c$ (%) | 0.50 | 0.14 | 22.11 | 4.71 |

It is seen from the table that for the small effective mass μ=0.1 $m_e$ and $\varepsilon_2 \geq 3$ corresponding critical percolation concentrations $N_{3D}^c$ and $N_{2D}^c$ are quite reasonable (less than 1%, that is less than $10^{18}$ cm$^{-3}$ (3D case) or $10^{12}$ cm$^{-2}$ (2D case)), while for the case μ=1 $m_e$ they may be very high (more than 20% for $\varepsilon_2 = 3$ and more that 5% for $\varepsilon_2 = 10$). Note, that in accordance with the results presented by Volnianska and Boguslawski [2] and obtained by Osorio-Guillen et al [17] from the first principles calculations, the concentration of Ca vacancies in CaO percolation threshold is 4,9% that is in qualitatively agreement with $N_{3D}^c = 5.18\%$ in **Table 2**. We hope that for the values $\varepsilon_2$ and μ characteristic for CaO the agreement could be better.

The percolation critical concentration of surface defects is always several times higher that the bulk one as anticipated from the percolation theory [26]. Fortunately the concentration of defects at surface may be much higher than far from it. Actually, let us estimate the increase of the vacancies concentration at the surface in comparison with a bulk of material using the results of the density functional calculations [1, 15, 16]. Within the framework of the activation theory, the probability of the vacancy formation is $W(z) = w_0 \exp\left(-E_a(z)/k_B T_f\right)$, where $E_a$ is the formation energy at distance z under the surface, $T_f$ is the material formation temperature. Using the difference between the vacancy formation energy on the surface and in the bulk ($\Delta E = E_a(\infty) - E_a(0)$), which is about 3 eV for GaN [16], 1.5 eV for $SrTiO_3$ [15] and 0.28 eV for MgO [1], one could estimate the ratio of the defects concentration on the surface to those in the bulk $\frac{W(0)}{W(\infty)} = \exp\left(\frac{\Delta E}{k_B T_f}\right)$ as 1.3 $10^{15}$ for GaN, 3.6 $10^{7}$ for $SrTiO_3$ and 26 for MgO at typical formation temperature $T_f = 1000$ K. Thus high concentration of the surface defects necessary for the room-temperature ferromagnetism appearance at $\mu \sim m_e$ can be achieved without any external stimuli.

However for the more rigorous calculations than listed in the table 2, the defect concentration gradient from the surface into the bulk should be taken into account for calculations of magnetization both on the surface and subsurface layers.

**(b)** To explore whether the surface defects (impurities or vacancies) can support ferromagnetism above room temperature, we followed Jin et al [16] and estimate the Curie temperature $T_C$ using a classical Heisenberg Hamiltonian [36]. In this approach the magnetic energy difference $\Delta E = \tilde{E}^T_{AFM} - \tilde{E}^T_{FM} = 2J_{23}$ is equal to the mean-field value $3k_B T_C/2$, so that $T_C = 4J_{23}/3k_B$. It is seen, that $T_C$ can be larger than the room temperature $T_r$ or smaller than $T_r$, which depends on the distance between defects and so on their concentration (as follows from **Fig. 9**). Since exchange integral $J_{23}$ is dependent on the defects separation (and so on their concentration) one could find the critical value of concentration or separation. In particular, the minimal concentration of surface defects for ferromagnetism appearance is $N_{2D}^{\min}(\%) = a^2/\pi r_j^2$ ($a$ is the lattice constant).

Calculated mean field phase diagrams in the coordinates "surface defects separation – dielectric permittivity $\varepsilon_2$" and "surface defects concentration – dielectric permittivity $\varepsilon_2$" are shown in the **Figs. 10a** and **b** respectively for the same parameters as used in the **Fig. 9**. It is seen that for the high separation the exchange integral is lower than the corresponding mean field (the region $T_C < T_r$ in the **Fig. 10a**), the equilibrium curve shifts to the higher values of defects separation with permittivity increase, while effective mass increase leads to the decrease of critical separation. The situation with defects concentration (the region $T_C > T_r$ in the **Fig. 10b**) is vice versa, since the ferromagnetic ordering could arise with the increase of concentration.

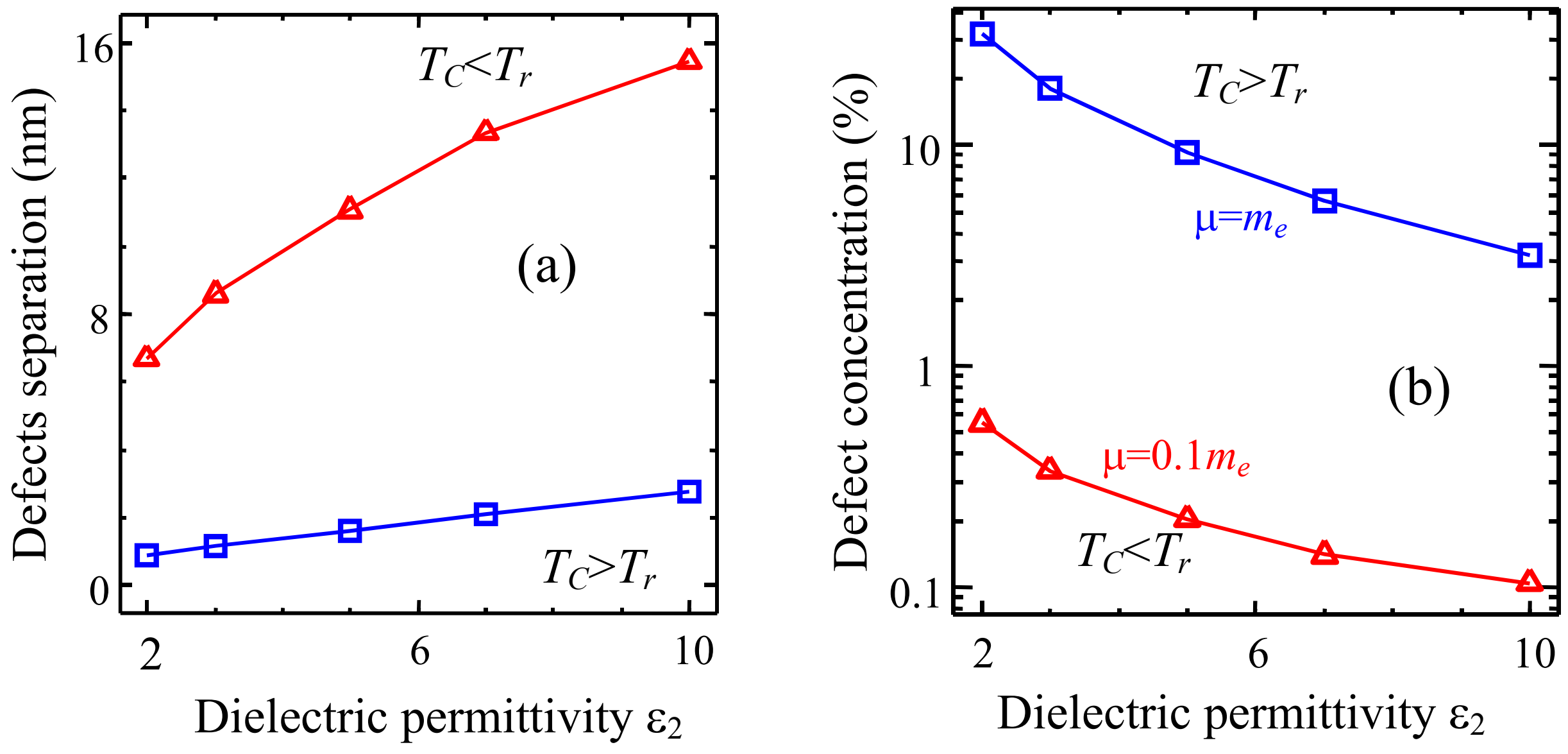

**Fig. 10.** Mean field phase diagrams in the coordinates "surface defects separation – dielectric permittivity $\varepsilon_2$" (a) and "surface defects concentration – dielectric permittivity $\varepsilon_2$" (b) for two values of effective mass (μ=0.1 $m_e$ and μ=1 $m_e$) specified near the curves.

**(c)** In semiconductor materials (like ZnO) defect-induced ferromagnetic long-range order can originate from the magnetic defects with spins $S_i$ and $S_j$ indirect *Ruderman-Kittel-Kasuya-Yosida* interaction via charge carriers (RKKY) [37, 38, 39]. It can be written in the conventional form: $J_{ss}(\vec{r}_{ij}) = J_0 \dfrac{\sin x - x\cos x}{x^4}$, where $x = 2k_F r_{ij}$, $k_F$ is the wave vector on Fermi level and $J_0$ is the RKKY interaction amplitude (see [40, 41] for details). The value of $k_F$~ (0.15–1.3)×10$^7$cm$^{-1}$ has been calculated on the basis of the experimental data for ZnO (see [42]). Calculated in [42] $J_0$ value is about 100 K, which allows estimating that the ferromagnetic phase transition temperature in semiconductors like ZnO could be a few hundred K. Unfortunately, technology driven scattering of donors and acceptors concentration and so $k_F$ makes it hard to be sure if RKKY interaction could lead to room temperature ferromagnetism in ZnO with defects.

**(d)** Surface-enhanced Dzyaloshinskii-Moriya interaction can influence the magnetization order. Indeed, the symmetry lowering near the surface could strongly increase the symmetry related Dzyaloshinskii-Moriya vector value and change its direction [32], so one can expect weak ferromagnetism even for the antiferromagnetic case $J_{23} < 0$. The latter case could be essential at distances $z \ge a_B^*$ under the surface, where the symmetry is lower than in the bulk and where we did not performed the analytical calculations of the exchange integral.

Therefore, one can expect that the interaction between the considered defects should exhibit the ferromagnetic long-range order at the surface of solid and in its vicinity up to the distances $z \approx 10a_B^*$. One of the important sequences of the surface-induced ferromagnetic long-range order is the inevitable appearance of the piezomagnetic and linear magnetoelectric effects in the vicinity of surface as predicted earlier for nanosystems [43].

### *5.2. Ground state of a pair single-charged defects with two shared electrons*

Let us consider the pair of the identical surface impurities (*a* and *b*) with two electrons [as shown in **Fig. 1e**]. Hereinafter $\mathbf{r}_i = (x_i, y_i, z_i)$ are the coordinates of the electrons 1 and 2. Their images coordinates are $\mathbf{r}_i^/ = (x_i, y_i, -z_i)$. The surface defects *a* and *b* are located in the points $\mathbf{R}_a = (-R/2, 0, 0)$ and $\mathbf{R}_b = (+R/2, 0, 0)$.

In the effective mass approximation the Schrödinger equation for the two-fermions the wave function has the form:

$$\left(-\frac{\hbar^2}{2\mu}(\Delta_1 + \Delta_2) + V(\mathbf{r}_1, \mathbf{r}_2, \mathbf{R}_a, \mathbf{R}_b)\right)\varphi(\mathbf{r}_1, \mathbf{r}_2) = E\varphi(\mathbf{r}_1, \mathbf{r}_2) \tag{21}$$

The boundary condition $\varphi(z_1 = 0) = \varphi(z_2 = 0) = 0$ is again used, as argued in the Section 3.

There are several contributions into the electrostatic potential $V$ of the carriers interaction with neighboring defects and polarized half-space (image charges), namely:

$$V(\mathbf{r}_1, \mathbf{r}_2, \mathbf{R}_a, \mathbf{R}_b) = V_{dd} + V_{dc} + V_{\mathrm{dim}} + V_{cim} + V_{12}(\mathbf{r}_1, \mathbf{r}_2) \tag{22}$$

1) $V_{dd}$ is the interaction energy of the two surface defects 1 and 2:

$$V_{dd} = \frac{+Z^2 e^2}{2\pi\varepsilon_0(\varepsilon_1 + \varepsilon_2)|R_a - R_b|}. \tag{23}$$

2) $V_{dc}$ is the carrier interaction with the surface defects, $V_{dim}$ is their images interaction with the defects, which sum has the form

$$V_{dc} + V_{\mathrm{dim}} = -\frac{|Z|e^2}{2\pi\varepsilon_0(\varepsilon_1 + \varepsilon_2)} \sum_{\substack{i=1,2\\ j=a,b}} \frac{1}{|R_j - \mathbf{r}_i|}. \tag{24}$$

In Eq.(24) we used that $|R_j - \mathbf{r}_i| = |R_j - \mathbf{r}_i^/|$ for the considered surface defects.

4) $V_{cim}$ is the interaction of the carriers with their images:

$$V_{cim} = \frac{e^2}{4\pi\varepsilon_0\varepsilon_2} \frac{\varepsilon_2 - \varepsilon_1}{\varepsilon_2 + \varepsilon_1}\left(\frac{1}{2z_1} + \frac{1}{2z_2}\right). \tag{25}$$

3) $V_{12}$ is the electron-electron interaction energy and the interaction of electrons with their images:

$$V_{12}(\mathbf{r}_1,\mathbf{r}_2)=+\frac{e^2}{4\pi\varepsilon_0\varepsilon_2}\left(\frac{1}{|\mathbf{r}_1-\mathbf{r}_2|}+\frac{\varepsilon_2-\varepsilon_1}{\varepsilon_2+\varepsilon_1}\frac{1}{|\mathbf{r}_1-\mathbf{r}_2^{/}|}\right), \tag{26}$$

In Eq.(26) we used that $|\mathbf{r}_1-\mathbf{r}_2^{/}|=|\mathbf{r}_2-\mathbf{r}_1^{/}|$.

We are looking for the ground state of the system by direct variational method. The one-electron wave function localized near the each defect site was chosen in the form [44]:

$$\varphi_j(\mathbf{r}_i)\equiv\varphi(\mathbf{r}_i-\mathbf{R}_j)=\sqrt{\frac{2\alpha^5}{\pi}}z_i\exp\left(-\alpha\sqrt{(x_i-X_j)^2+y_i^2+z_i^2}\right), \tag{27}$$

Here $i$ = 1,2, $z_i>0$ and $X_a=-R/2$, $X_b=+R/2$. Parameter α will be obtained by direct variational method. Note the function (27) is normalized on the half-space z > 0.

Eigen values of the symmetric (singlet, Σ = 0) and antisymmetric (triplet, Σ = 1) coordinate functions $\varphi_{S,T}(\mathbf{r}_1,\mathbf{r}_2)=\frac{1}{\sqrt{2}}\left(\varphi(\mathbf{r}_1-\mathbf{R}_a)\varphi(\mathbf{r}_2-\mathbf{R}_b)\pm\varphi(\mathbf{r}_2-\mathbf{R}_a)\varphi(\mathbf{r}_1-\mathbf{R}_b)\right)$ was obtained in the first approximation of perturbation theory by the conventional way (see [25, 30, 45, 46, 47, 48]):

$$\begin{aligned}\widetilde{E}_{S,T}&=\frac{a_B^2\mu}{\hbar^2}\int\limits_{z>0}\varphi_{S,A}^*(\mathbf{r}_1,\mathbf{r}_2)\left(-\frac{\hbar^2}{2\mu}(\Delta_1+\Delta_2)+V\right)\varphi_{S,A}(\mathbf{r}_1,\mathbf{r}_2)d\mathbf{r}_1d\mathbf{r}_2,\\&=\frac{Z}{\widetilde{r}_{ab}}+\frac{2(A\pm S(A_{ab}+J_{ab})+C_{ab})+C_{12}\pm J_{12}}{1\pm S^2}\end{aligned}. \tag{28}$$

Details of calculations can be found in **Appendix E**.

Hereinafter we introduced the dimensionless coordinates $\widetilde{\mathbf{r}}=\mathbf{r}/a_B^*$, distance $\widetilde{R}=R/a_B^*$ between the defects, variational parameters $\widetilde{\alpha}=\alpha a_B^*$ and $\widetilde{\rho}=\widetilde{\alpha}\widetilde{R}$ renormalized on Bohr radius $a_B^*=2\pi\varepsilon_0(\varepsilon_1+\varepsilon_2)\hbar^2/Z\mu e^2$, dimensionless energies $\widetilde{E}=\frac{a_B^{*2}\mu}{\hbar^2}E$ and $\zeta=(\varepsilon_2-\varepsilon_1)/(\varepsilon_2+\varepsilon_1)$.

Introducing the designation $\varphi_{ji}\equiv\varphi(\mathbf{r}_i-\mathbf{R}_j)$, the overlap integral is

$$S=\int\limits_{z_i>0}(\varphi_{ai}\varphi_{bi})d\widetilde{\mathbf{r}}_i=\exp(-\rho)\left(1+\rho+\frac{\rho^2}{15}(6+\rho)\right), \tag{29a}$$

Hereinafter $\widetilde{\alpha}\widetilde{R}\equiv\rho$.

The matrix elements of the one-particle operators of kinetic energy and external potential are

$$A=\int\limits_{z_1>0}\varphi_{a1}\left(-\frac{1}{2}\widetilde{\Delta}_1-\frac{1}{\widetilde{r}_{a1}}\right)\varphi_{a1}d\widetilde{\mathbf{r}}_1=-\frac{\widetilde{\alpha}}{2}+\frac{\widetilde{\alpha}^2}{2}, \tag{29b}$$

$$A_{ab}=\int\limits_{z_1>0}\varphi_{a1}\left(-\frac{1}{2}\widetilde{\Delta}_1-\frac{1}{\widetilde{r}_{b1}}\right)\varphi_{b1}d\widetilde{\mathbf{r}}_1=-\frac{\widetilde{\alpha}}{2}\left((1-\widetilde{\alpha})(1+\rho)+\frac{\rho^2}{15}(5-4\widetilde{\alpha}+\widetilde{\alpha}\rho)\right)\exp(-\rho). \tag{29c}$$

When deriving (29b-c) we used the identity $-\left(\widetilde{\Delta}_1/2+1/\widetilde{r}_{a1}\right)\varphi_{a1}=-\varphi_{a1}\left(\widetilde{\alpha}^2/2+(1-2\widetilde{\alpha})/\widetilde{r}_{a1}\right)$.

The one-particles Coulomb ($C_{ab}$) and exchange ($J_{ab}$) integrals are

$$C_{ab} = \int_{z_1>0} (\varphi_{a1})^2 \left( \frac{1}{2\tilde{z}_1} - \frac{1}{\tilde{r}_{b1}} \right) d\tilde{\mathbf{r}}_1 = \frac{\zeta}{Z(1+\zeta)} \frac{3\tilde{\alpha}}{8} + \frac{\tilde{\alpha}}{2\rho^3} \left( 3 - 2\rho^2 - \exp(-\rho)(1+\rho)(3+\rho(3+\rho)) \right), \quad (29d)$$

$$J_{ab} = \int_{z_1>0} \frac{\varphi_{a1}\varphi_{b1}}{\tilde{r}_{a1}} \left( \frac{1}{2\tilde{z}_1} - \frac{1}{\tilde{r}_{a1}} \right) d\tilde{\mathbf{r}}_1 = -\frac{\tilde{\alpha}}{6} \exp(-\rho)(3+\rho(3+\rho)) + \frac{3\tilde{\alpha}\rho^2}{64} \frac{\zeta(\rho K_1(\rho) + 4K_2(\rho))}{|Z|(1+\zeta)}, \quad (29e)$$

Where $K_{1,2}(x)$ are the modified Bessel functions of the second kind, which exponentially vanish at $x \to \infty$.

The two-particles Coulomb ($C_{12}$) and exchange ($J_{12}$) integrals have been calculated analytically as the functions of ρ:

$$C_{12} = \int_{z_2>0} \int_{z_1>0} \tilde{V}_{12}(\tilde{\mathbf{r}}_1, \tilde{\mathbf{r}}_2)(\varphi_{a1})^2 (\varphi_{b2})^2 d\tilde{\mathbf{r}}_1 d\tilde{\mathbf{r}}_2 = \int_{z_2>0} \int_{z_1>0} \left( \frac{1}{\tilde{r}_{12}} + \frac{\zeta}{\tilde{r}_{1'2}} \right) \frac{(\varphi_{a1})^2 (\varphi_{b2})^2}{|Z|(1+\zeta)} d\tilde{\mathbf{r}}_1 d\tilde{\mathbf{r}}_2 \approx$$

$$\approx \frac{\zeta}{|Z|(1+\zeta)} \frac{3\tilde{\alpha}}{4} + \frac{1}{|Z|} \frac{\tilde{\alpha}}{\rho} \left( \begin{aligned} & \frac{261}{2048\rho^5} - \frac{679}{2048\rho^3} + \frac{441}{1024\rho^2} + \frac{63}{256\rho} - \frac{125}{256} - \\ & - e^{-2\rho} \left( \frac{261}{1024\rho^5} + \frac{261}{512\rho^4} - \frac{157}{1024\rho^3} - \frac{121}{512\rho^2} + \frac{107}{128\rho} + \frac{13}{128} \right) + \\ & + e^{-4\rho} \left( \frac{261}{2048\rho^5} + \frac{261}{512\rho^4} + \frac{1409}{2048\rho^3} + \frac{361}{1024\rho^2} + \frac{59}{256\rho} + \frac{3}{256} \right) - \\ & - \mathrm{Ei}(-2\rho) + \mathrm{Ei}(-4\rho) \frac{1}{2} + \frac{\mathbf{C} + \ln(\rho)}{2} \end{aligned} \right) \quad (29f)$$

$$J_{12} = \int_{z_2>0} \int_{z_1>0} \tilde{V}_{12}(\tilde{\mathbf{r}}_1, \tilde{\mathbf{r}}_2)\varphi_{a1}\varphi_{b2}\varphi_{a2}\varphi_{b1} d\tilde{\mathbf{r}}_1 d\tilde{\mathbf{r}}_2 = \int_{z_2>0} \int_{z_1>0} \left( \frac{1}{\tilde{r}_{12}} + \frac{\zeta}{\tilde{r}_{1'2}} \right) \frac{\varphi_{a1}\varphi_{b2}\varphi_{a2}\varphi_{b1}}{|Z|(1+\zeta)} d\tilde{\mathbf{r}}_1 d\tilde{\mathbf{r}}_2 \approx$$

$$\approx \frac{1}{|Z|} \frac{\tilde{\alpha}}{\rho} \left( \frac{\mathbf{C} + \ln(\rho)}{2} S^2 - \exp(-2\rho)\rho \left( -\frac{93}{512} + \frac{163\rho}{256} + \frac{991\rho^2 + 401\rho^3}{2880} + \frac{\rho^4}{48} + \frac{\rho^5}{600} \right) \right. \quad (29g)$$

$$\left. + \mathrm{Ei}(-2\rho) \left( -1 + \frac{\rho^2}{5} + \frac{2\rho^4}{75} + \frac{\rho^6}{225} \right) + \exp(2\rho)\mathrm{Ei}(-4\rho) \frac{1}{2} \left( 1 - \rho + \frac{6}{15}\rho^2 - \frac{\rho^3}{15} \right)^2 \right)$$

Where **C**≈0.577216 is Euler's constant, $\mathrm{Ei}(x) = -\int_{-x}^{\infty} dt \frac{e^{-t}}{t}$ is the exponential integral function.

Accuracy of the series cut in Eqs.(29f) becomes surprisingly high (not less that several 1%) at distances $\rho > 2$. Accuracy of the series cut in Eqs.(29g) is not so high, it becomes satisfactory only when taking more than 3-4 terms (see **Appendix E** for mathematical details).

Energy levels (28) difference is

$$\tilde{E}_S - \tilde{E}_T = \frac{4S}{1 - S^4} (A_{ab} + J_{ab} - S\,A - S\,C_{ab}) + 2 \frac{J_{12} - S^2 C_{12}}{1 - S^4} \quad (30)$$

Energy levels (28) were minimized with respect to the variational parameter $\tilde{\alpha}$ analytically. From equations $\partial\tilde{E}_S(\tilde{\alpha},\rho)/\partial\tilde{\alpha}=0$ and $\partial\tilde{E}_T(\tilde{\alpha},\rho)/\partial\tilde{\alpha}=0$ we obtained the functions $\tilde{\alpha}_S(\rho)$ and $\tilde{\alpha}_T(\rho)$. Then we plotted the dependences of $\tilde{E}_S$ and $\tilde{E}_T$ on the distance $R$ between defects as shown in **Fig. 11**. The curves were calculated numerically using exact series for exchange and Coulomb integrals up to forth terms in Eqs.(29f) and (29g). Note that for the defect charge $|Z|>2$ minima disappear indicating the charged molecule instability.

It is seen from the figure that singlet state is the lowest one for the defect molecule. However, the hydrogen-like molecule in the bulk has no minimum for the triplet state at all, while we predicted the deep negative minimum appearance at the surface. The energy difference $\tilde{E}_S-\tilde{E}_T$ between the ground singlet state and metastable triplet state is relatively low for the case $\mu=0.1m_e$, $\varepsilon_2=10$, $|Z|=1$ and $\varepsilon_1=1$ (**Fig. 11a**). The difference strongly increases with the effective mass increase (compare energy values in **Fig. 11a** with **Fig. 11b** and **Fig. 11c** with **Fig. 11d**) or permittivity $\varepsilon_2$ decrease (compare energy values in **Fig. 11a** with **Fig. 11c** and **Fig. 11b** with **Fig. 11d**).

Thus we obtained that the nonmagnetic singlet state is the lowest one for a molecule with two electrons formed by a pair of identical surface impurities (like surface hydrogen molecule), while its next state with deep enough negative minimum is the magnetic triplet. The appearance of the metastable magnetic triplet state for such molecule at the surface indicates the possibility of metastable orto-states of the hydrogen-like molecules, while they are absent in the bulk of material. Coexistence of para- and orto-states for the hydrogen-like molecules at the surface could be revealed experimentally by the spectroscopic methods of surface investigation, namely two series of spectral terms could be observed similarly to the He atom four fundamental series.

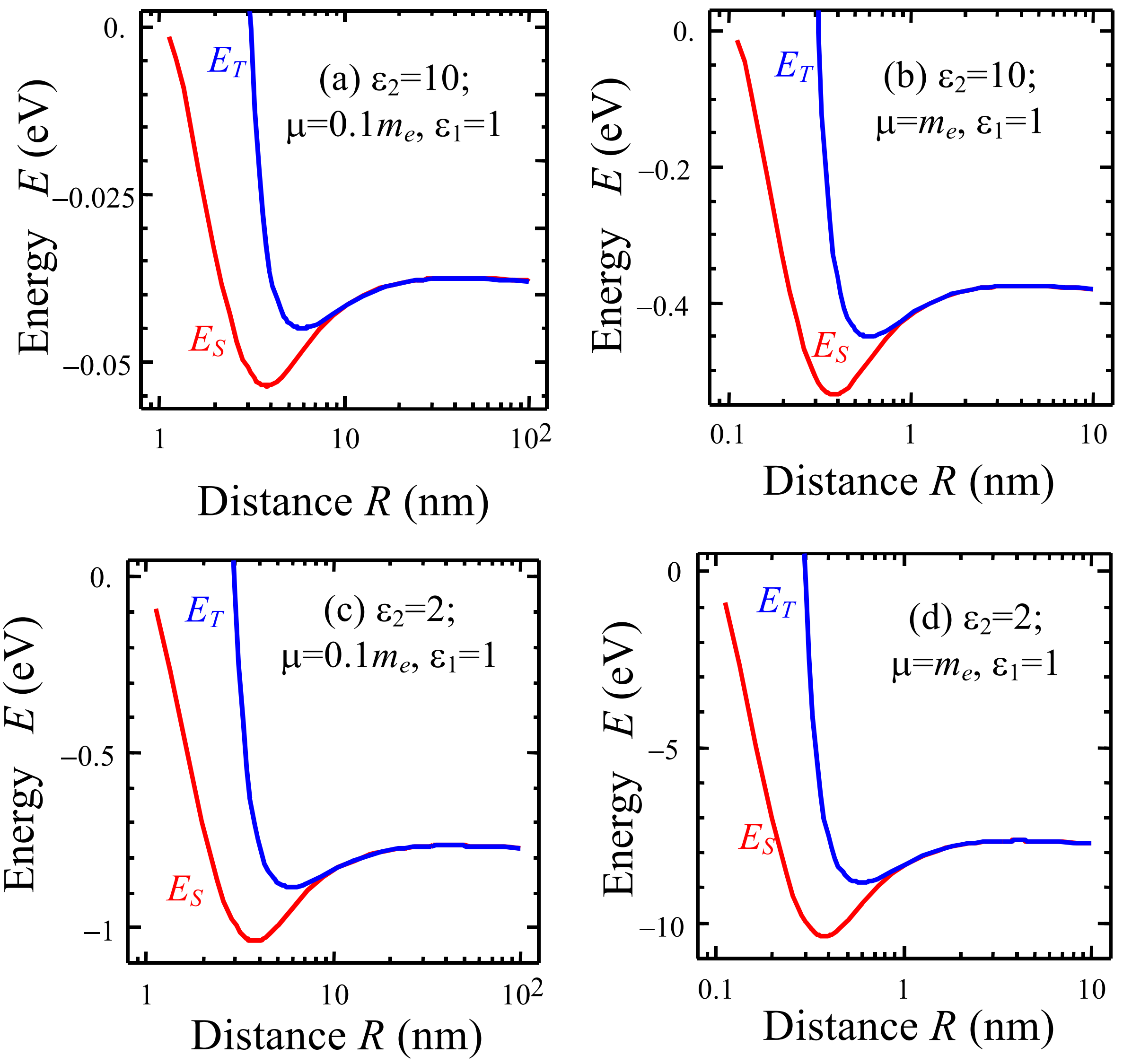

**Fig. 11.** The dependences of $\widetilde{E}_S$ and $\widetilde{E}_T$ vs the distance $R$. In the absolute minimum variational parameters $\widetilde{\alpha}_S(\rho_{\min}) = 0.488835$, $\widetilde{\alpha}_T(\rho_{\min}) = 0.394358$ for $\varepsilon_2=2$ (c,d) and $\widetilde{\alpha}_S(\rho_{\min}) = 0.407118$, $\widetilde{\alpha}_T(\rho_{\min}) = 0.322618$ for $\varepsilon_2=10$ (a,b). For all plots $\varepsilon_1=1$ and $|Z|=1$; $\mu=0.1m_e$ for plots (a,c) and $\mu=m_e$ for plots (b,d).

**6. Summary**

To summarize our analytical results, obtained by quantum-mechanical calculations performed in the continuous media approach, effective mass approximation and direct variational method, we predicted the surface-induced magnetism of impurities, anionic and cationic vacancies and established that:

1) The magnetic triplet $2p_z3p_z$ is the ground state of impurity atoms with two electrons (like He, $Li^{1+}$, $Be^{2+}$, etc) located at the surface. The triplet state remained the lowest one under the surface up to the distances 2-10 nm, at higher distances the nonmagnetic spherically-symmetric singlet state 1s1s becomes the lowest one (as anticipated for the He-like atoms in the bulk of material). The triplet state $2p_z3p_z$ should become the magnetic one ($\Sigma_z = \pm1$) allowing for the Hund's rule that orient two fermions (electrons or holes) spins in the same direction.

2) Similarly to impurity atoms, the magnetic triplet $2p_z3p_z$ is the ground state of neutral vacancies, both anionic and cationic located at the surface and under the surface up to the distances 2-10 nm, while it is not the ground state in the bulk of material.

3) The exchange interaction between the nearest surface defects in the triplet state $2p_z3p_z\times2p_z3p_z$ belong to the ferromagnetic type: exchange integral $J$ between the four electrons or holes of the two identical vacancies or impurity atoms is positive at any distance $R$ between them, but $J$ value is significant at distances $R$~2-10 effective Bohr radii (~2 - 50 nm), with pronounced maximum at $R$~5 effective Bohr radii (~5 – 25 nm).

4) Located at the surface hydrogen-like molecules (two defects with two shared electrons) have the singlet ground state (para-state). The next level is the magnetic triplet state (orto-state), which has the deep minimum with negative value of energy at the distance $R \sim 0.5 - 5$ nm between the hydrogen-like atoms. In the bulk of material the triplet state has no minimum at all as well known for $H_2$-molecule. The energy gap between these lowest states can be small enough (~0.01 eV) for light effective mass $\mu\sim0.1m_e$ and dielectric permittivity $\varepsilon_2\sim10$, so that both orto- and para-states can be occupied at room temperatures. Predicted orto- and para-states of the hydrogen-like molecules at the solid surfaces could be revealed from the modern spectroscopy methods, similarly to the He-like atom case.

5) Different mechanisms (percolation threshold, RKKY interaction and mean field) of the surface defect-induced ferromagnetic long-range order appearance are discussed and estimated. The surface long-range order could be revealed from the spin-polarized scanning tunnelling microscopy, similarly as was done by Bode et al [31]. One of the important sequences of the surface-induced ferromagnetic long-range order is the inevitable appearance of the piezomagnetic and linear magnetoelectric effects in the vicinity of surface as predicted earlier for nanosystems [43].

6) Surface-induced magnetism in confined geometries (e.g. thin films, nanorods, nanotubes and spherical nanoparticles) can be considered by the same approach as done for the flat surface, but it could be much more cumbersome, since the image charges series appear in the calculations [49]. On the other hand the surface curvature would strongly affect on the wave-function form in the vicinity of surface, via mechanical strains and surface tension, which should be also considered. Thus we hope that obtained results could describe the ferromagnetism in thin $TiO_2$, $HfO_2$, and $In_2O_3$ films related with the contribution of the oxygen vacancies semi-quantitatively and explain the ferromagnetic properties of spherical nanoparticles of nonmagnetic oxides $CeO_2$, $Al_2O_3$, ZnO as well as their possible superparamagnetic behavior at least qualitatively.

7) The calculations of two-electrons (or holes) defect wave functions localized at the surface and its vicinity showed that these functions are pure p-type at the surface, and the mixture of s-type and p-type in sub-surface layers, at that the p-type contribution being dominant up to distances ~5 nm at reasonable material parameters. The transition to the bulk s-type wave functions can be expected for the distances more than 10-20 nm.

**Acknowledgements**

Authors are very grateful to R.O. Kuzian for useful discussions and critical remarks.

**Appendix A. Effective mass approximation validity for p-states**

It is known that the condition $r_d >> a$ of the effective mass approximation validity is strict only for the well-localized carriers s-states, while for the p-states, which are zero at the defect size, the condition $r_d > 2a$ is enough as well as corresponding wave functions are not sensitive to the concrete short-range peculiarities of the defect potential. As a result the effective mass approximation typically describes p-states much better than s-states [26]. Actually the coordinate part of the localized carrier wave function expansion on the plane waves is $\psi(\mathbf{r}) = \frac{1}{V_0^{1/2}} \sum_{\mathbf{k}} B_n(\mathbf{k}) \exp(i\mathbf{kr})$, where $B_n(\mathbf{k})$ should be nonzero only in the small region near the center of the Brillion zone, e.g. at $k << 2\pi/a$. For the bulk s-states with $\psi(\mathbf{r}) = \sqrt{\frac{\alpha^3}{\pi}} \exp(-\alpha r)$ the Fourier image $B_n(\mathbf{k}) \approx \frac{8\alpha^{5/2}\sqrt{\pi/V_0}}{(\alpha^2 + k^2)^2}$ has maximum at $k = 0$, then rapidly vanishes at $k/\alpha > 1$ ($\alpha$ is the effective Bohr radius). So the strong inequality $\frac{1}{\alpha} >> a$ should be valid in the effective mass approximation for s-states. For the bulk p-states with $\psi(\mathbf{r}) = \sqrt{\frac{\alpha^5}{\pi}} z \exp(-\alpha r)$ the Fourier image $B_n(\mathbf{k}) \approx \sqrt{\pi/V_0} \frac{32\alpha^{7/2} k_z}{(\alpha^2 + k^2)^3}$ is zero at $k_z = 0$, has maximum at $k_z/\alpha \approx 0.5$ and rapidly vanishes at $k_z/\alpha > 1.5$, i.e. the characteristic values of parameter $\frac{1}{\alpha} >> \frac{a}{4\pi}$ are required in the effective mass approximation validity for p-states (see **Fig. 1A**).

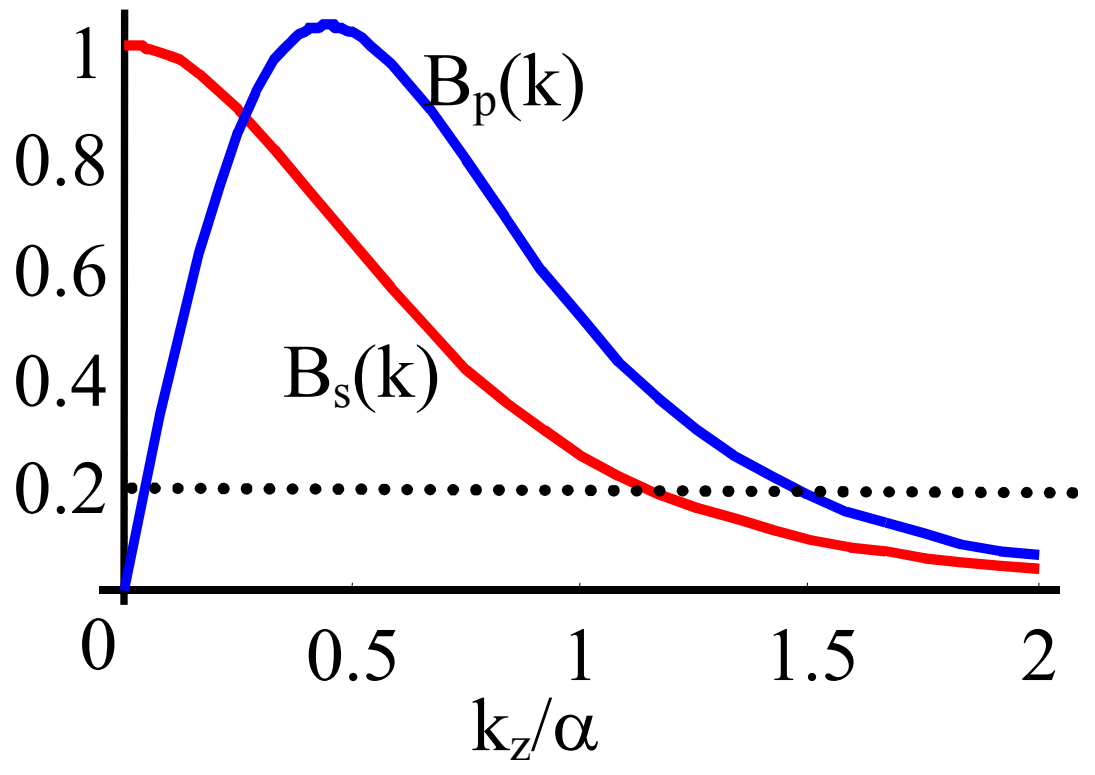

**Fig. 1A.** Fourier images $B_S(\mathbf{k})$ and $B_p(\mathbf{k})$ of the one-electron s- and p-functions.

From the first-principles Janotti and Walle [50] study the electronic structure of native point defects in the bulk of n-type ZnO samples. It was shown that oxygen vacancies are deep donors and have high formation energies. Zinc interstitials and antisites are shallow donors, but have high formation energies. Zinc vacancies are deep acceptors and have low formation energies. Using quantum molecular dynamics Boguslawski et al [51] show that substitutional C impurity cation is a relatively shallow donor in GaN and AlN, which can assume a metastable configuration in the neutral charge state. However we did not find any data about the electronic structure of the surface defects.

**The image-charge method**

The image-charge method leads to the potential created by a point charge $q$ located in the point $\mathbf{r}_i = (x_i, y_i, z_i)$ (z>0, ε = ε2) and its image $\mathbf{r}_i^{/} = (x_i, y_i, -z_i)$ (z>0, ε = ε1) in the form [52]:

$$\phi(\mathbf{r}) = \begin{cases} \frac{q}{4\pi\varepsilon_0\varepsilon_2}\left(\frac{1}{|\mathbf{r}-\mathbf{r}_i|} + \frac{\varepsilon_2-\varepsilon_1}{\varepsilon_1+\varepsilon_2}\frac{1}{|\mathbf{r}-\mathbf{r}_i^{/}|}\right), & z>0 \\ \frac{2q}{4\pi\varepsilon_0(\varepsilon_1+\varepsilon_2)}\frac{1}{|\mathbf{r}-\mathbf{r}_i|}, & z<0 \end{cases} \quad \text{(A.1)}$$

Interaction energy of the charge $q$ with another charge $Q$ located in the point $\mathbf{R} = (X, Y, Z)$ is $V = +Q\phi(\mathbf{r} = \mathbf{R})$. The interaction energy of the charge $q$ with polarized media is $V_{self} = \frac{q^2}{4\pi\varepsilon_0\varepsilon_2}\frac{\varepsilon_2-\varepsilon_1}{\varepsilon_1+\varepsilon_2}\frac{1}{2z_i}$.

$$\begin{aligned} V_e(\mathbf{r}_i) &= -\frac{e^2}{4\pi\varepsilon_0\varepsilon_2}\left(\frac{|Z|}{r_i} + \frac{\varepsilon_2-\varepsilon_1}{\varepsilon_1+\varepsilon_2}\frac{|Z|}{r_i^{/}}\right) + \frac{e^2}{4\pi\varepsilon_0\varepsilon_2^\infty}\frac{\varepsilon_2^\infty-\varepsilon_1^\infty}{\varepsilon_2^\infty+\varepsilon_1^\infty}\frac{1}{2z_i} \\ &\equiv -\frac{e^2}{2\pi\varepsilon_0}\left(\frac{|Z|}{(\varepsilon_1+\varepsilon_2)r_i} - \frac{\varepsilon_2^\infty-\varepsilon_1^\infty}{\varepsilon_2^\infty+\varepsilon_1^\infty}\frac{1}{4\varepsilon_2^\infty z_i}\right) \end{aligned} \quad \text{(A.2)}$$

$$\begin{aligned} V_{12}(\mathbf{r}_1,\mathbf{r}_2) &= \frac{e^2}{4\pi\varepsilon_0\varepsilon_2^\infty}\left(\frac{1}{|\mathbf{r}_1-\mathbf{r}_2|} + \frac{\varepsilon_2^\infty-\varepsilon_1^\infty}{\varepsilon_2^\infty+\varepsilon_1^\infty}\left(\frac{1}{2|\mathbf{r}_1-\mathbf{r}_2^{/}|} + \frac{1}{2|\mathbf{r}_2-\mathbf{r}_1^{/}|}\right)\right) \\ &= \frac{e^2}{4\pi\varepsilon_0\varepsilon_2^\infty}\left(\frac{1}{r_{12}} + \frac{\varepsilon_2^\infty-\varepsilon_1^\infty}{\varepsilon_2^\infty+\varepsilon_1^\infty}\frac{1}{r_{12^{/}}}\right) \end{aligned} \quad \text{(A.3)}$$

**Appendix B. Defect at the surface**

**B.1.** Since $\tilde{\alpha}$ and $\tilde{\beta}$ should be positive for the wave-functions to be square integrable, they could not satisfy one-particle Schrödinger equations, like $\left(\mathrm{sign}(\mu)\frac{\tilde{\Delta}_1}{2} + \frac{1}{\tilde{r}_1}\right)\varphi_{nlm} = E\varphi_{nlm}$, for negative effective mass μ, since e.g. $1 - 2\tilde{\alpha}\,\mathrm{sign}(\mu) > 0$ for any positive $\tilde{\alpha}$ and negative effective mass μ. This fact leads to the result that carriers with negative effective mass cannot create the stable localized states for chosen form of the wave function. Thus hereinafter we regard μ positive, which is true for the electrons with energies near the bottom of conductive band and holes near the top of valence band [27].

When calculating the integral $A_{mn}$:

$$A_{mn} = \int_{z_1>0} \varphi_{m10}(\tilde{\mathbf{r}}_1)\left(-\frac{\tilde{\Delta}_1}{2} + \tilde{V}_e(\tilde{\mathbf{r}}_1)\right)\varphi_{n10}(\tilde{\mathbf{r}}_1)d\tilde{\mathbf{r}}_1 = \int_{z_1>0} \varphi_{m10}(\tilde{\mathbf{r}}_1)\left(-\frac{\tilde{\Delta}_1}{2} - \frac{1}{\tilde{r}_1} + \frac{\zeta}{2|Z|(1+\zeta)}\frac{1}{\tilde{z}_1}\right)\varphi_{n10}(\tilde{\mathbf{r}}_1)d\tilde{\mathbf{r}}_1$$

we used the identities: $-\left(\mathrm{sign}(\mu)\frac{\tilde{\Delta}_1}{2} + \frac{1}{\tilde{r}_1}\right)\varphi_{210} = -\varphi_{210}\left(\mathrm{sign}(\mu)\frac{\tilde{\alpha}^2}{2} + \frac{1-2\tilde{\alpha}\,\mathrm{sign}(\mu)}{\tilde{r}_1}\right)$ and

$$-\left(\mathrm{sign}(\mu)\frac{\tilde{\Delta}_1}{2} + \frac{1}{\tilde{r}_1}\right)\varphi_{310} = \sqrt{\frac{4\tilde{\beta}^5}{3\pi}}z_i\ \exp\left(-\tilde{\beta}r_i\right)\left(\tilde{\beta} - 4\tilde{\beta}^2\mathrm{sign}(\mu) + \frac{\tilde{\beta}^3}{2}\tilde{r}_1 + \frac{-2+6\tilde{\beta}\,\mathrm{sign}(\mu)}{\tilde{r}_1}\right).$$

**B.2.** Calculations of the angular part in the Coulomb exchange and integrals

$$C_{mn} = \int_{z_2>0}\int_{z_1>0} \tilde{V}_{12}(\tilde{\mathbf{r}}_1,\tilde{\mathbf{r}}_2)\varphi^2_{m10}(\tilde{\mathbf{r}}_1)\varphi^2_{n10}(\tilde{\mathbf{r}}_2)d\tilde{\mathbf{r}}_1 d\tilde{\mathbf{r}}_2 = \frac{1}{|Z|(1+\zeta)}\int_{z_2>0}\int_{z_1>0}\left(\frac{1}{\tilde{r}_{12}} + \frac{\zeta}{\tilde{r}_{1'2}}\right)\varphi^2_{m10}(\tilde{\mathbf{r}}_1)\varphi^2_{n10}(\tilde{\mathbf{r}}_2)d\tilde{\mathbf{r}}_1 d\tilde{\mathbf{r}}_2$$

$$J_{mn} = \int_{z_2>0}\int_{z_1>0} \tilde{V}_{12}(\tilde{\mathbf{r}}_1,\tilde{\mathbf{r}}_2)\varphi_{m10}(\tilde{\mathbf{r}}_1)\varphi_{m10}(\tilde{\mathbf{r}}_2)\varphi_{n10}(\tilde{\mathbf{r}}_1)\varphi_{n10}(\tilde{\mathbf{r}}_2)d\tilde{\mathbf{r}}_1 d\tilde{\mathbf{r}}_2$$

1) Calculations of the angular part in the Coulomb exchange and integrals .

Using the expressions in spherical coordinates

$$\mathbf{r}_i = r_i(\sin\theta_i\cos\beta_i, \sin\theta_i\sin\beta_i, \cos\theta_i), \qquad \mathbf{r}_i^{/} = r_i(\sin\theta_i\cos\beta_i, \sin\theta_i\sin\beta_i, -\cos\theta_i), \tag{B.1a}$$

$$\cos(\vartheta_{12}) = \sin\theta_1\sin\theta_2\cos(\beta_1-\beta_2) + \cos\theta_1\cos\theta_2, \tag{B.1b}$$

$$\cos(\vartheta_{12}^{/}) = \sin\theta_1\sin\theta_2\cos(\beta_1-\beta_2) - \cos\theta_1\cos\theta_2, \tag{B.1c}$$

$$\tilde{r}_{12} = \sqrt{\tilde{r}_1^2 + \tilde{r}_2^2 - 2\tilde{r}_1\tilde{r}_2\cos(\vartheta_{12})}, \qquad \tilde{r}_{12}^{\ /} = \sqrt{\tilde{r}_1^2 + \tilde{r}_2^2 - 2\tilde{r}_1\tilde{r}_2\cos(\vartheta_{12}^{/})}. \tag{B.1d}$$

Expansions

$$\frac{1}{\tilde{r}_{12}} = \begin{cases} \dfrac{1}{\tilde{r}_1}\displaystyle\sum_{n=0}^{\infty}\left(\frac{\tilde{r}_2}{\tilde{r}_1}\right)^n P_n(\cos(\vartheta_{12})), & \tilde{r}_1 > \tilde{r}_2, \\ \dfrac{1}{\tilde{r}_2}\displaystyle\sum_{n=0}^{\infty}\left(\frac{\tilde{r}_1}{\tilde{r}_2}\right)^n P_n(\cos(\vartheta_{12})), & \tilde{r}_1 < \tilde{r}_2. \end{cases} \tag{B.2}$$

$$\begin{aligned} P_n(\cos(\vartheta_{12})) &= \frac{4\pi}{2n+1}\sum_{m=-n}^{n} Y^*_{nm}(\theta_1,\beta_1)Y_{nm}(\theta_2,\beta_2) \\ &= \sum_{m=-n}^{n}\frac{(n-m)!}{(n+m)!}P_n^m(\cos\theta_1)P_n^m(\cos\theta_2)e^{im(\beta_2-\beta_1)} \end{aligned} \tag{B.3a}$$

$$\begin{aligned} P_n(\cos(\vartheta_{12}^{/})) &= \frac{4\pi}{2n+1}\sum_{m=-n}^{n} Y^*_{nm}(\theta_1,\beta_1)Y_{nm}(\pi-\theta_2,\beta_2) \\ &= \sum_{m=-n}^{n}\frac{(n-m)!}{(n+m)!}P_n^m(\cos\theta_1)P_n^m(-\cos\theta_2)e^{im(\beta_2-\beta_1)} \end{aligned} \tag{B.3b}$$

2) Angular part integrals

$$I_n = \int_0^{2\pi} d\beta_1 \int_0^{2\pi} d\beta_2 \int_0^{\pi/2} \sin\theta_1 d\theta_1 \int_0^{\pi/2} \sin\theta_2 d\theta_2 \cos^2\theta_1 \cos^2\theta_2 P_n(\cos(\vartheta_{12}))$$

$$= 4\pi^2 \left( \int_0^{\pi/2} d\theta_1 \cos^2\theta_1 \sin\theta_1 P_n(\cos(\theta_1)) \right)^2 = 4\pi^2 \begin{cases} \frac{1}{9}, & \text{for } n=0, \quad \frac{1}{16}, \quad \text{for } n=1, \\ \frac{4}{225}, & \text{for } n=2, \quad \frac{1}{576}, \quad \text{for } n=3. \end{cases} \tag{B.4a}$$

$$I_n^{/} = \int_0^{2\pi} d\beta_1 \int_0^{2\pi} d\beta_2 \int_0^{\pi/2} \sin\theta_1 d\theta_1 \int_0^{\pi/2} \sin\theta_2 d\theta_2 \cos^2\theta_1 \cos^2\theta_2 P_n(\cos(\vartheta_{12}^{/}))$$

$$= (-1)^n 4\pi^2 \left( \int_0^{\pi/2} d\theta_1 \cos^2\theta_1 \sin\theta_1 P_n(\cos(\theta_1)) \right)^2 = (-1)^n I_n \tag{B.4b}$$

2) Calculations of the radial part in the Coulomb and exchange integrals

$$C_{22}^n(\tilde{\alpha}) = \left(\frac{2\tilde{\alpha}^3}{\pi}\right)^2 \int_0^\infty d\tilde{r}_1 \tilde{r}_1^2 (\tilde{\alpha}\tilde{r}_1)^2 \exp(-2\tilde{\alpha}\tilde{r}_1) \left( \int_0^{\tilde{r}_1} d\tilde{r}_2 \tilde{r}_2^2 \frac{\tilde{r}_2^n}{\tilde{r}_1^{n+1}} + \int_{\tilde{r}_1}^\infty d\tilde{r}_2 \tilde{r}_2^2 \frac{\tilde{r}_1^n}{\tilde{r}_2^{n+1}} \right) (\tilde{\alpha}\tilde{r}_2)^2 \exp(-2\tilde{\alpha}\tilde{r}_2)$$

$$= \frac{\tilde{\alpha}}{4\pi^2} \begin{cases} \frac{837}{256} & \text{for } n=0, \quad \frac{555}{256} \quad \text{for } n=1, \\ \frac{405}{256} & \text{for } n=2, \quad \frac{315}{256} \quad \text{for } n=3. \end{cases} \tag{B.5a}$$

$$C_{33}^n(\tilde{\beta}) = \left(\frac{4\tilde{\beta}^3}{3\pi}\right)^2 \int_0^\infty d\tilde{r}_1 \tilde{r}_1^2 (\tilde{\beta}\tilde{r}_1)^2 (2-\tilde{\beta}\tilde{r}_1)^2 \exp(-2\tilde{\beta}\tilde{r}_1)$$

$$\times \left( \int_0^{\tilde{r}_1} d\tilde{r}_2 \tilde{r}_2^2 \frac{\tilde{r}_2^n}{\tilde{r}_1^{n+1}} + \int_{\tilde{r}_1}^\infty d\tilde{r}_2 \tilde{r}_2^2 \frac{\tilde{r}_1^n}{\tilde{r}_2^{n+1}} \right) (\tilde{\beta}\tilde{r}_2)^2 (2-\tilde{\beta}\tilde{r}_2)^2 \exp(-2\tilde{\beta}\tilde{r}_2) \tag{B.5b}$$

$$= \frac{\tilde{\beta}}{4\pi^2} \begin{cases} \frac{1987}{1024} & \text{for } n=0, \quad \frac{3935}{3072} \quad \text{for } n=1, \\ \frac{995}{1024} & \text{for } n=2, \quad \frac{805}{1024} \quad \text{for } n=3. \end{cases}$$

$$C_{23}^n(\tilde{\beta}) = \left(\frac{4\tilde{\beta}^3}{3\pi}\right)\left(\frac{2\tilde{\alpha}^3}{\pi}\right)\int_0^\infty d\tilde{r}_1\tilde{r}_1^2(\tilde{\alpha}\tilde{r}_1)^2\exp(-2\tilde{\alpha}\tilde{r}_1)$$

$$\times\left(\int_0^{\tilde{r}_1} d\tilde{r}_2\tilde{r}_2^2\frac{\tilde{r}_2^n}{\tilde{r}_1^{n+1}} + \int_{\tilde{r}_1}^\infty d\tilde{r}_2\tilde{r}_2^2\frac{\tilde{r}_1^n}{\tilde{r}_2^{n+1}}\right)(\tilde{\beta}\tilde{r}_2)^2(2-\tilde{\beta}\tilde{r}_2)^2\exp(-2\tilde{\beta}\tilde{r}_2)$$

$$= \frac{\tilde{\beta}^4}{4\pi^2(\tilde{\alpha}+\tilde{\beta})^9}\begin{cases}\left(\begin{array}{l}2\tilde{\alpha}^6 + 18\tilde{\alpha}^5\tilde{\beta} + 72\tilde{\alpha}^4\tilde{\beta}^2 + 168\tilde{\alpha}^3\tilde{\beta}^3 + 140\tilde{\alpha}^2\tilde{\beta}^4 \\ + 252\tilde{\alpha}\tilde{\beta}^5 + 108\tilde{\beta}^6 + 27\dfrac{\tilde{\beta}^7}{\tilde{\alpha}} + 3\dfrac{\tilde{\beta}^8}{\tilde{\alpha}^2}\end{array}\right), & \text{for } n=0; \\ \dfrac{5}{3}\tilde{\beta}\left(\begin{array}{l}2\tilde{\alpha}^5 + 18\tilde{\alpha}^4\tilde{\beta} + 72\tilde{\alpha}^3\tilde{\beta}^2 - 56\tilde{\alpha}^2\tilde{\beta}^3 \\ + 180\tilde{\alpha}\tilde{\beta}^4 + 45\tilde{\beta}^5 + 5\dfrac{\tilde{\beta}^6}{\tilde{\alpha}}\end{array}\right), & \text{for } n=1, \\ 5\tilde{\beta}^2(3\tilde{\alpha}^4 + 27\tilde{\alpha}^3\tilde{\beta} - 52\tilde{\alpha}^2\tilde{\beta}^2 + 72\tilde{\alpha}\tilde{\beta}^3 + 8\tilde{\beta}^4), & \text{for } n=2, \\ 35\tilde{\alpha}\tilde{\beta}^3(5\tilde{\alpha}^2 - 11\tilde{\alpha}\tilde{\beta} + 12\tilde{\beta}^2), & \text{for } n=3.\end{cases} \tag{B.5c}$$

$$J_{23}^n(\tilde{\alpha},\tilde{\beta}) = \left(\frac{4\tilde{\beta}^3}{3\pi}\right)\left(\frac{2\tilde{\alpha}^3}{\pi}\right)\int_0^\infty d\tilde{r}_1\tilde{r}_1^2(\tilde{\alpha}\tilde{r}_1)\exp(-\tilde{\alpha}\tilde{r}_1)(\tilde{\beta}\tilde{r}_1)(2-\tilde{\beta}\tilde{r}_1)\exp(-\tilde{\beta}\tilde{r}_1)$$

$$\times\left(\int_0^{\tilde{r}_1} d\tilde{r}_2\tilde{r}_2^2\frac{\tilde{r}_2^n}{\tilde{r}_1^{n+1}} + \int_{\tilde{r}_1}^\infty d\tilde{r}_2\tilde{r}_2^2\frac{\tilde{r}_1^n}{\tilde{r}_2^{n+1}}\right)(\tilde{\alpha}\tilde{r}_2)\exp(-\tilde{\alpha}\tilde{r}_2)(\tilde{\beta}\tilde{r}_2)(2-\tilde{\beta}\tilde{r}_2)\exp(-\tilde{\beta}\tilde{r}_2) \tag{B.5d}$$

$$= \frac{16\tilde{\alpha}^2\tilde{\beta}^8}{4\pi^2(\tilde{\alpha}+\tilde{\beta})^{11}}\begin{cases}186\tilde{\alpha}^2 - 465\tilde{\alpha}\tilde{\beta} + 314\tilde{\beta}^2 & \text{for } n=0, \\ \dfrac{5}{3}(74\tilde{\alpha}^2 - 185\tilde{\alpha}\tilde{\beta} + 137\tilde{\beta}^2) & \text{for } n=1, \\ 5(18\tilde{\alpha}^2 - 45\tilde{\alpha}\tilde{\beta} + 35\tilde{\beta}^2) & \text{for } n=2, \\ 35(2\tilde{\alpha}^2 - 5\tilde{\alpha}\tilde{\beta} + 4\tilde{\beta}^2) & \text{for } n=3.\end{cases}$$

Coulomb and exchange integrals acquire the form:

$$C_{23}(\tilde{\alpha},\tilde{\beta}) = \sum_{n=0}^\infty \frac{1+(-1)^n\zeta}{|Z|(1+\zeta)} I_n C_{23}^n(\tilde{\alpha},\tilde{\beta}), \qquad J_{23}(\tilde{\alpha},\tilde{\beta}) = \sum_{n=0}^\infty \frac{1+(-1)^n\zeta}{|Z|(1+\zeta)} I_n J_{23}^n(\tilde{\alpha},\tilde{\beta}), \tag{B.6}$$

$$C_{22}(\tilde{\alpha}) = \sum_{n=0}^\infty \frac{1+(-1)^n\zeta}{|Z|(1+\zeta)} I_n C_{22}^n(\tilde{\alpha}), \qquad C_{33}(\tilde{\beta}) = \sum_{n=0}^\infty \frac{1+(-1)^n\zeta}{|Z|(1+\zeta)} I_n C_{33}^n(\tilde{\beta}). \tag{B.7}$$

Since the coefficient $I_n$ falls to zero very rapidly with n increase (as $\sim 1/n^7$) it is natural to leave only a few terms in these series (summing up to $n$=3 would provide up to four decimal points). Moreover, $I_n$ is zero for all even $n$ except $n$=0 and $n$=2, so for the case $\zeta \approx 1$ the series (B.6) and (B.7) reduces to two nontrivial terms.

$$C_{23}(\tilde{\alpha},\tilde{\beta}) \approx \sum_{n=0}^3 \frac{1+(-1)^n\zeta}{|Z|(1+\zeta)} I_n C_{23}^n(\tilde{\alpha},\tilde{\beta}), \tag{B.8a}$$

$$C_{23}\left(\tilde{\alpha},\tilde{\beta}\right)\Big|_{\zeta=1}=\frac{1}{2|Z|}\frac{2\tilde{\beta}^4}{45\left(\tilde{\alpha}+\tilde{\beta}\right)^9}\begin{pmatrix}10\tilde{\alpha}^6+90\tilde{\alpha}^5\tilde{\beta}+372\tilde{\alpha}^4\tilde{\beta}^2+948\tilde{\alpha}^3\tilde{\beta}^3+492\tilde{\alpha}^2\tilde{\beta}^4\\ +1548\tilde{\alpha}\tilde{\beta}^5+572\tilde{\beta}^6+135\dfrac{\tilde{\beta}^7}{\tilde{\alpha}}+15\dfrac{\tilde{\beta}^8}{\tilde{\alpha}^2}\end{pmatrix} \tag{B.8b}$$

$$J_{23}\left(\tilde{\alpha},\tilde{\beta}\right)\approx\frac{\tilde{\alpha}^2\tilde{\beta}^8\left(19\left(4562\tilde{\alpha}^2-11405\tilde{\alpha}\tilde{\beta}+7980\tilde{\beta}^2\right)+\zeta\left(41578\tilde{\alpha}^2-103945\tilde{\alpha}\tilde{\beta}+67640\tilde{\beta}^2\right)\right)}{180|Z|(1+\zeta)\left(\tilde{\alpha}+\tilde{\beta}\right)^{11}}, \tag{B.9}$$

$$J_{23}\left(\tilde{\alpha},\tilde{\beta}\right)\Big|_{\zeta=1}=\frac{1}{2|Z|}\frac{32\tilde{\alpha}^2\tilde{\beta}^8\left(334\tilde{\alpha}^2-835\tilde{\alpha}\tilde{\beta}+570\tilde{\beta}^2\right)}{15\left(\tilde{\alpha}+\tilde{\beta}\right)^{11}}, \tag{B.10}$$

$$C_{22}\left(\tilde{\alpha}\right)\approx\frac{\tilde{\alpha}}{|Z|(1+\zeta)}\left(\frac{43339+20789\zeta}{81920}\right), \qquad C_{22}\left(\tilde{\alpha}\right)\Big|_{\zeta=1}=\frac{\tilde{\alpha}}{2|Z|}\frac{501}{640}, \tag{B.11}$$

$$C_{33}\left(\tilde{\beta}\right)\approx\frac{\tilde{\beta}}{|Z|(1+\zeta)}\left(\frac{926909+446659\zeta}{2949120}\right). \qquad C_{33}\left(\tilde{\beta}\right)\Big|_{\zeta=1}=\frac{\tilde{\beta}}{2|Z|}\frac{3577}{7680} \tag{B.12}$$

**Appendix C. Defect under the surface**

Schrödinger equation for the two-fermions wave-function coordinate part has the form of Eq. (2) and the same the boundary condition $\varphi(z_1=0)=\varphi(z_2=0)=0$ is used. The electrostatic potential $V(\mathbf{r}_1,\mathbf{r}_2)=V_e(\mathbf{r}_1)+V_e(\mathbf{r}_2)+V_{12}(\mathbf{r}_1,\mathbf{r}_2)$ of the carriers interaction with defect and polarized half-space (image charges) is given by Eq.(3), where $V_e$ is the sum of carrier (electron or hole) interaction with the defect, defect image and the carrier interaction with its own image $V_e(\mathbf{r}_i)=-\frac{e^2}{4\pi\varepsilon_0\varepsilon_2}\left(\frac{|Z|}{|\mathbf{r}_i-\mathbf{r}_0|}+\frac{\varepsilon_2-\varepsilon_1}{\varepsilon_1+\varepsilon_2}\frac{|Z|}{|\mathbf{r}_i-\mathbf{r}_0^{/}|}\right)+\frac{e^2}{4\pi\varepsilon_0\varepsilon_2}\frac{\varepsilon_2-\varepsilon_1}{\varepsilon_2+\varepsilon_1}\frac{1}{2z_i}$.

As argued in the section 4.1, we introduced the dimensionless coordinates $\tilde{\mathbf{r}}$, $\tilde{\mathbf{r}}_0$ normalized on the effective Bohr radius $a_B^*=(\varepsilon_1+\varepsilon_2)\frac{2\pi\varepsilon_0\hbar^2}{|Z\mu|e^2}$. Using dimensionless Hamiltonian $\tilde{H}=\frac{a_B^{*2}|\mu|}{\hbar^2}H$ and energy $\tilde{E}=\frac{a_B^{*2}|\mu|}{\hbar^2}E$ we go to Eqs.(9), where the electrostatic energy is $\tilde{V}_e(\tilde{\mathbf{r}}_i)=-\frac{\varepsilon_1+\varepsilon_2}{2\varepsilon_2}\left(\frac{1}{|\tilde{\mathbf{r}}_i-\tilde{\mathbf{r}}_0|}+\frac{\varepsilon_2-\varepsilon_1}{\varepsilon_1+\varepsilon_2}\frac{1}{|\tilde{\mathbf{r}}_i+\tilde{\mathbf{r}}_0|}\right)+\zeta\frac{\varepsilon_1+\varepsilon_2}{2\varepsilon_2|Z|}\frac{1}{2\tilde{z}_i}$.

Eqs. (11) are valid for the singlet and triplet energy levels at arbitrary $z_0$, namely

$$\tilde{E}_{22}=2A_{22}+C_{22}, \qquad \tilde{E}_{33}=2A_{33}+C_{33}, \tag{C.1a}$$

$$\tilde{E}_{23}^S=\frac{A_{22}+A_{33}+\left(A_{23}+A_{32}\right)S_{23}+C_{23}+J_{23}}{1+S_{23}^2}, \tag{C.1b}$$

$$\widetilde{E}_{23}^{T}=\frac{A_{22}+A_{33}-\left(A_{32}+A_{23}\right)S_{23}+C_{23}-J_{23}}{1-S_{23}^{2}}. \tag{C.1c}$$

Calculations performed for μ>0 on functions (15)-(16) give the matrix elements

$$A_{mn}=\int_{z_1>0}\varphi_{m10}\left(\widetilde{\mathbf{r}}_1\right)\left(-\frac{\widetilde{\Delta}_1}{2}+\widetilde{V}_e\left(\widetilde{\mathbf{r}}_1\right)\right)\varphi_{n10}\left(\widetilde{\mathbf{r}}_1\right)d\widetilde{\mathbf{r}}_1 \tag{C.2}$$

Exchange integral:

$$J_{23}=\frac{1}{|Z|(1+\zeta)}\int_{z_2>0}\int_{z_1>0}\left(\frac{1}{\widetilde{r}_{12}}+\frac{\zeta}{\widetilde{r}_{1'2}}\right)\varphi_{210}\left(\widetilde{\mathbf{r}}_1\right)\varphi_{210}\left(\widetilde{\mathbf{r}}_2\right)\varphi_{310}\left(\widetilde{\mathbf{r}}_1\right)\varphi_{310}\left(\widetilde{\mathbf{r}}_2\right)d\widetilde{\mathbf{r}}_1 d\widetilde{\mathbf{r}}_2=\sum_{n=0}^{\infty}\frac{1+(-1)^n\zeta}{|Z|(1+\zeta)}J_{23}^{n}\left(\widetilde{\alpha},\widetilde{\beta}\right) \tag{C.3a}$$

$$\frac{J_{23}^{n}\left(\widetilde{\alpha},\widetilde{\beta}\right)}{4\pi^2A^2(\widetilde{\alpha})B^2\left(\widetilde{\beta}\right)}=\int_0^{\infty}d\widetilde{r}_1\widetilde{r}_1^{\,2}\left(\widetilde{\alpha}\widetilde{r}_1\right)\exp\left(-\widetilde{\alpha}\widetilde{r}_1\right)\left(\widetilde{\beta}\widetilde{r}_1\right)\left(b-\widetilde{\beta}\widetilde{r}_1\right)\exp\left(-\widetilde{\beta}\widetilde{r}_1\right)\int_{\max\left[\frac{-z_0}{\widetilde{r}_1},-1\right]}^{1}dx_1x_1^2P_n\left(x_1\right)\times$$
$$\times\left(\int_0^{\widetilde{r}_1}d\widetilde{r}_2\widetilde{r}_2^{\,2}\frac{\widetilde{r}_2^{\,n}}{\widetilde{r}_1^{\,n+1}}+\int_{\widetilde{r}_1}^{\infty}d\widetilde{r}_2\widetilde{r}_2^{\,2}\frac{\widetilde{r}_1^{\,n}}{\widetilde{r}_2^{\,n+1}}\right)\left(\widetilde{\alpha}\widetilde{r}_2\right)\exp\left(-\widetilde{\alpha}\widetilde{r}_2\right)\left(\widetilde{\beta}\widetilde{r}_2\right)\left(b-\widetilde{\beta}\widetilde{r}_2\right)\exp\left(-\widetilde{\beta}\widetilde{r}_2\right)\int_{\max\left[\frac{-z_0}{\widetilde{r}_2},-1\right]}^{1}dx_2x_2^2P_n\left(x_2\right) \tag{C.3b}$$

$P_n(x)$ is Legendre polynomial.

Coulomb integrals:

$$C_{mn}=\int_{z_2>0}\int_{z_1>0}\widetilde{V}_{12}\left(\widetilde{\mathbf{r}}_1,\widetilde{\mathbf{r}}_2\right)\varphi_{m10}^2\left(\widetilde{\mathbf{r}}_1\right)\varphi_{n10}^2\left(\widetilde{\mathbf{r}}_2\right)d\widetilde{\mathbf{r}}_1 d\widetilde{\mathbf{r}}_2$$
$$=\frac{1}{|Z|(1+\zeta)}\int_{z_2>0}\int_{z_1>0}\left(\frac{1}{\widetilde{r}_{12}}+\frac{\zeta}{\widetilde{r}_{1'2}}\right)\varphi_{m10}^2\left(\widetilde{\mathbf{r}}_1\right)\varphi_{n10}^2\left(\widetilde{\mathbf{r}}_2\right)d\widetilde{\mathbf{r}}_1 d\widetilde{\mathbf{r}}_2=\sum_{k=0}^{\infty}\frac{1+(-1)^k\zeta}{|Z|(1+\zeta)}C_{mn}^{k} \tag{C.4}$$

The series terms are

$$C_{22}^{k}\left(\widetilde{\alpha},\widetilde{z}_0\right)=4\pi^2A^4\left(\widetilde{\alpha},\widetilde{z}_0\right)\int_0^{\infty}d\widetilde{r}_1\widetilde{r}_1^{\,2}\left(\widetilde{\alpha}\widetilde{r}_1\right)^2\exp\left(-2\widetilde{\alpha}\widetilde{r}_1\right)\int_{\max\left[\frac{-z_0}{\widetilde{r}_1},-1\right]}^{1}dx_1x_1^2P_k\left(x_1\right)$$
$$\times\left(\int_0^{\widetilde{r}_1}d\widetilde{r}_2\widetilde{r}_2^{\,2}\frac{\widetilde{r}_2^{\,k}}{\widetilde{r}_1^{\,k+1}}+\int_{\widetilde{r}_1}^{\infty}d\widetilde{r}_2\widetilde{r}_2^{\,2}\frac{\widetilde{r}_1^{\,k}}{\widetilde{r}_2^{\,k+1}}\right)\left(\widetilde{\alpha}\widetilde{r}_2\right)^2\exp\left(-2\widetilde{\alpha}\widetilde{r}_2\right)\int_{\max\left[\frac{-z_0}{\widetilde{r}_2},-1\right]}^{1}dx_2x_2^2P_k\left(x_2\right) \tag{C.5a}$$

$$C_{33}^{k}\left(\widetilde{\beta},\widetilde{z}_0\right)=4\pi^2B^4\left(\widetilde{\beta},\widetilde{z}_0\right)\int_0^{\infty}d\widetilde{r}_1\widetilde{r}_1^{\,2}\left(\widetilde{\beta}\widetilde{r}_1\right)^2\left(b-\widetilde{\beta}\widetilde{r}_1\right)^2\exp\left(-2\widetilde{\beta}\widetilde{r}_1\right)\int_{\max\left[\frac{-z_0}{\widetilde{r}_1},-1\right]}^{1}dx_1x_1^2P_k\left(x_1\right)$$
$$\times\left(\int_0^{\widetilde{r}_1}d\widetilde{r}_2\widetilde{r}_2^{\,2}\frac{\widetilde{r}_2^{\,k}}{\widetilde{r}_1^{\,k+1}}+\int_{\widetilde{r}_1}^{\infty}d\widetilde{r}_2\widetilde{r}_2^{\,2}\frac{\widetilde{r}_1^{\,k}}{\widetilde{r}_2^{\,k+1}}\right)\left(\widetilde{\beta}\widetilde{r}_2\right)^2\left(b-\widetilde{\beta}\widetilde{r}_2\right)^2\exp\left(-2\widetilde{\beta}\widetilde{r}_2\right)\int_{\max\left[\frac{-z_0}{\widetilde{r}_2},-1\right]}^{1}dx_2x_2^2P_k\left(x_2\right) \tag{C.5b}$$

$$C_{23}^{k}\left(\tilde{\alpha},\tilde{\beta},\tilde{z}_0\right)=4\pi^2 A^2\left(\tilde{\alpha},\tilde{z}_0\right)B^2\left(\tilde{\beta},\tilde{z}_0\right)\int_0^\infty d\tilde{r}_1\tilde{r}_1^{\,2}\left(\tilde{\alpha}\tilde{r}_1\right)^2\exp\left(-2\tilde{\alpha}\tilde{r}_1\right)\int_{\max\left[\frac{-z_0}{\tilde{r}_1},-1\right]}^{1}dx_1x_1^2P_k\left(x_1\right)$$

$$\times\left(\int_0^{\tilde{r}_1}d\tilde{r}_2\tilde{r}_2^{\,2}\frac{\tilde{r}_2^{\,k}}{\tilde{r}_1^{\,k+1}}+\int_{\tilde{r}_1}^{\infty}d\tilde{r}_2\tilde{r}_2^{\,2}\frac{\tilde{r}_1^{\,k}}{\tilde{r}_2^{\,k+1}}\right)\left(\tilde{\beta}\tilde{r}_2\right)^2\left(b-\tilde{\beta}\tilde{r}_2\right)^2\exp\left(-2\tilde{\beta}\tilde{r}_2\right)\int_{\max\left[\frac{-z_0}{\tilde{r}_2},-1\right]}^{1}dx_2x_2^2P_k\left(x_2\right) \tag{C.5c}$$

The overlap integral

$$S_{23}=\int_{z_1>0}\varphi_{210}\left(\tilde{\mathbf{r}}_1\right)\varphi_{310}\left(\tilde{\mathbf{r}}_1\right)d\tilde{\mathbf{r}}_1=2\pi\frac{A\left(\tilde{\alpha},\tilde{z}_0\right)B\left(\tilde{\beta},\tilde{z}_0\right)}{\left(\tilde{\alpha}+\tilde{\beta}\right)^6}\times$$
$$\times\left(\begin{array}{l}2\exp\left(-\tilde{z}_0\left(\tilde{\alpha}+\tilde{\beta}\right)\right)\left(20\tilde{\beta}+\left(\tilde{\alpha}+\tilde{\beta}\right)\left(\tilde{z}_0\tilde{\beta}\left(8+\tilde{z}_0\left(\tilde{\alpha}+\tilde{\beta}\right)\right)-b\left(4+\tilde{z}_0\left(\tilde{\alpha}+\tilde{\beta}\right)\right)\right)\right)\\+4\left(b\left(\tilde{\alpha}+\tilde{\beta}\right)\left(4+\tilde{z}_0^2\left(\tilde{\alpha}+\tilde{\beta}\right)^2\right)-\tilde{\beta}\left(20+3\tilde{z}_0^2\left(\tilde{\alpha}+\tilde{\beta}\right)^2\right)\right)\end{array}\right) \tag{C.6}$$

Since the wave functions, which correspond to the different eigen energy values, should be orthogonal, the condition $S_{32}=0$ is necessary for the orthogonality of the singlet wave functions $\psi_{22}\left(\mathbf{r}_1,\mathbf{r}_2\right)$ and $\psi_{33}\left(\mathbf{r}_1,\mathbf{r}_2\right)$. The condition leads to the expression (17) for *b*. Also we should demand the orthogonality of the singlet $\psi_{23}^{S}\left(\mathbf{r}_1,\mathbf{r}_2\right)$ to the singlet $\psi_{22}\left(\mathbf{r}_1,\mathbf{r}_2\right)$ and $\psi_{33}\left(\mathbf{r}_1,\mathbf{r}_2\right)$. This again leads to the condition $S_{32}=0$. Triplet $\psi_{23}^{T}\left(\mathbf{r}_1,\mathbf{r}_2\right)$ is always orthogonal to all singlet wave functions at arbitrary $\tilde{\alpha}$ and $\tilde{\beta}$, both due to antisymmetric coordinate part and symmetric spin functions.

## Appendix D. Exchange interaction between the defects

Hereinafter only the interaction between the nearest defects will be calculated.

Wave functions of the lowest state may be obtained from the determinant

$$\Psi=\begin{vmatrix}\varphi_{210}\left(\tilde{\mathbf{r}}_1-\tilde{\mathbf{R}}_a\right)\chi_k(1) & \varphi_{210}\left(\tilde{\mathbf{r}}_2-\tilde{\mathbf{R}}_a\right)\chi_k(2) & \varphi_{210}\left(\tilde{\mathbf{r}}_3-\tilde{\mathbf{R}}_a\right)\chi_k(3) & \varphi_{210}\left(\tilde{\mathbf{r}}_4-\tilde{\mathbf{R}}_a\right)\chi_k(4)\\ \varphi_{310}\left(\tilde{\mathbf{r}}_1-\tilde{\mathbf{R}}_a\right)\chi_l(1) & \varphi_{310}\left(\tilde{\mathbf{r}}_2-\tilde{\mathbf{R}}_a\right)\chi_l(2) & \varphi_{310}\left(\tilde{\mathbf{r}}_3-\tilde{\mathbf{R}}_a\right)\chi_l(3) & \varphi_{310}\left(\tilde{\mathbf{r}}_4-\tilde{\mathbf{R}}_a\right)\chi_l(4)\\ \varphi_{210}\left(\tilde{\mathbf{r}}_1-\tilde{\mathbf{R}}_b\right)\chi_k(1) & \varphi_{210}\left(\tilde{\mathbf{r}}_2-\tilde{\mathbf{R}}_b\right)\chi_k(2) & \varphi_{210}\left(\tilde{\mathbf{r}}_3-\tilde{\mathbf{R}}_b\right)\chi_k(3) & \varphi_{210}\left(\tilde{\mathbf{r}}_4-\tilde{\mathbf{R}}_b\right)\chi_k(4)\\ \varphi_{310}\left(\tilde{\mathbf{r}}_1-\tilde{\mathbf{R}}_b\right)\chi_l(1) & \varphi_{310}\left(\tilde{\mathbf{r}}_2-\tilde{\mathbf{R}}_b\right)\chi_l(2) & \varphi_{310}\left(\tilde{\mathbf{r}}_3-\tilde{\mathbf{R}}_b\right)\chi_l(3) & \varphi_{310}\left(\tilde{\mathbf{r}}_4-\tilde{\mathbf{R}}_b\right)\chi_l(4)\end{vmatrix} \tag{D.1}$$

Indexes *k* and *l* of the spin functions $\chi_{k,l}(i)$ correspond to the up (↑) and down (↓) spin orientations, in particular

$$\Psi_{\uparrow\uparrow}\left(\tilde{\mathbf{r}}_1,\tilde{\mathbf{r}}_2,\tilde{\mathbf{r}}_3,\tilde{\mathbf{r}}_4\right)=\chi_\uparrow(1)\chi_\uparrow(2)\chi_\uparrow(3)\chi_\uparrow(4)\Phi\left(\tilde{\mathbf{r}}_1,\tilde{\mathbf{r}}_2,\tilde{\mathbf{r}}_3,\tilde{\mathbf{r}}_4\right), \tag{D.2a}$$

$$\Psi_{\downarrow\downarrow}\left(\tilde{\mathbf{r}}_1,\tilde{\mathbf{r}}_2,\tilde{\mathbf{r}}_3,\tilde{\mathbf{r}}_4\right)=\chi_\downarrow(1)\chi_\downarrow(2)\chi_\downarrow(3)\chi_\downarrow(4)\Phi\left(\tilde{\mathbf{r}}_1,\tilde{\mathbf{r}}_2,\tilde{\mathbf{r}}_3,\tilde{\mathbf{r}}_4\right), \tag{D.2b}$$

$$\begin{aligned}\Psi_{\downarrow\uparrow}\left(\tilde{\mathbf{r}}_1,\tilde{\mathbf{r}}_2,\tilde{\mathbf{r}}_3,\tilde{\mathbf{r}}_4\right)&=\chi_\downarrow(1)\chi_\uparrow(2)\chi_\downarrow(3)\chi_\uparrow(4)\Phi\left(\tilde{\mathbf{r}}_1,\tilde{\mathbf{r}}_2,\tilde{\mathbf{r}}_3,\tilde{\mathbf{r}}_4\right)\\&=\chi_\uparrow(1)\chi_\uparrow(2)\chi_\downarrow(3)\chi_\downarrow(4)\Phi\left(\tilde{\mathbf{r}}_1,\tilde{\mathbf{r}}_2,\tilde{\mathbf{r}}_3,\tilde{\mathbf{r}}_4\right),\end{aligned} \tag{D.2c}$$

where

$$\Phi(\tilde{\mathbf{r}}_1,\tilde{\mathbf{r}}_2,\tilde{\mathbf{r}}_3,\tilde{\mathbf{r}}_4)=\begin{vmatrix}\varphi_{210}(\tilde{\mathbf{r}}_1-\tilde{\mathbf{R}}_a) & \varphi_{210}(\tilde{\mathbf{r}}_2-\tilde{\mathbf{R}}_a) & \varphi_{210}(\tilde{\mathbf{r}}_3-\tilde{\mathbf{R}}_a) & \varphi_{210}(\tilde{\mathbf{r}}_4-\tilde{\mathbf{R}}_a)\\ \varphi_{310}(\tilde{\mathbf{r}}_1-\tilde{\mathbf{R}}_a) & \varphi_{310}(\tilde{\mathbf{r}}_2-\tilde{\mathbf{R}}_a) & \varphi_{310}(\tilde{\mathbf{r}}_3-\tilde{\mathbf{R}}_a) & \varphi_{310}(\tilde{\mathbf{r}}_4-\tilde{\mathbf{R}}_a)\\ \varphi_{210}(\tilde{\mathbf{r}}_1-\tilde{\mathbf{R}}_b) & \varphi_{210}(\tilde{\mathbf{r}}_2-\tilde{\mathbf{R}}_b) & \varphi_{210}(\tilde{\mathbf{r}}_3-\tilde{\mathbf{R}}_b) & \varphi_{210}(\tilde{\mathbf{r}}_4-\tilde{\mathbf{R}}_b)\\ \varphi_{310}(\tilde{\mathbf{r}}_1-\tilde{\mathbf{R}}_b) & \varphi_{310}(\tilde{\mathbf{r}}_2-\tilde{\mathbf{R}}_b) & \varphi_{310}(\tilde{\mathbf{r}}_3-\tilde{\mathbf{R}}_b) & \varphi_{310}(\tilde{\mathbf{r}}_4-\tilde{\mathbf{R}}_b)\end{vmatrix} \quad \text{(D.3)}$$

The exchange contribution to the energy is given by expression $E_A=-\frac{1}{2}\left(\sum_{k=\uparrow,l=\uparrow}J_{kl}+\sum_{k=\downarrow,l=\downarrow}J_{kl}\right)$ [30].

Hereinafter we neglect the overlap integrals of the electron wave functions localized at the nearest defect sites and used the condition $S_{23}=0$ for the functions on one center. Under these assumptions the exchange contribution to the energy is determined as

$$J_{23}=\int\limits_{z_2>0}\int\limits_{z_1>0}\tilde{V}_{12}(\tilde{\mathbf{r}}_1,\tilde{\mathbf{r}}_2)\varphi_{210}(\tilde{\mathbf{r}}_1-\tilde{\mathbf{R}}_a)\varphi_{310}(\tilde{\mathbf{r}}_2-\tilde{\mathbf{R}}_b)\varphi_{310}(\tilde{\mathbf{r}}_2-\tilde{\mathbf{R}}_a)\varphi_{210}(\tilde{\mathbf{r}}_1-\tilde{\mathbf{R}}_b)d\tilde{\mathbf{r}}_1 d\tilde{\mathbf{r}}_2 \quad \text{(D.4)}$$

Then we will use the representation

$$\varphi_{210}(\mathbf{r})=\sqrt{\frac{2\alpha^5}{\pi}}\phi(\mathbf{r},\alpha),\quad \varphi_{310}(\mathbf{r})=\sqrt{\frac{4\beta^5}{3\pi}}(2-\beta r)\phi(\mathbf{r},\beta),\quad \phi(\mathbf{r},\gamma)=z\exp(-\gamma r) \quad \text{(D.5)}$$

in order to calculate the auxiliary integral:

$$J_{23}(\alpha,\beta)=\frac{8\alpha^5\beta^5}{3\pi^2}\int\limits_{z_2>0}\int\limits_{z_1>0}\left(\frac{1}{\tilde{r}_{12}}+\frac{\zeta}{\tilde{r}_{1'2}}\right)\phi_{a1}(\alpha)\phi_{b2}(\beta)\phi_{a2}(\beta)\phi_{b1}(\alpha)(2-\beta r_{b1})(2-\beta r_{b2})d\tilde{\mathbf{r}}_1 d\tilde{\mathbf{r}}_2 \quad \text{(D.6)}$$

In order to calculate (D.7) we followed seminal papers [45, 46, 47, 48]. We used the prolate ellipsoidal coordinates for both electrons:

$$\mu_i=\frac{\tilde{r}_{ai}+\tilde{r}_{bi}}{\tilde{R}},\quad \nu_i=\frac{\tilde{r}_{ai}-\tilde{r}_{bi}}{\tilde{R}},\quad \tan\gamma_i=\frac{z_i}{y_i},\quad i=1,2,$$

$$\tilde{z}_{ai}=\tilde{z}_{bi}=\tilde{z}_i=\frac{\tilde{R}}{2}\sqrt{(\mu_i^2-1)(1-\nu_i^2)}\sin\gamma_i,$$

$$\tilde{r}_{ai}=\left|\tilde{\mathbf{r}}_i-\tilde{\mathbf{R}}_a\right|=(\mu_i+\nu_i)\frac{\tilde{R}}{2},\ \tilde{r}_{bi}=\left|\tilde{\mathbf{r}}_i-\tilde{\mathbf{R}}_b\right|=(\mu_i-\nu_i)\frac{\tilde{R}}{2},$$

$$\phi_{ai}(\tilde{\alpha})=\tilde{z}_{ai}\exp(-\tilde{\alpha}\tilde{r}_{ai}),\quad \phi_{bi}(\tilde{\alpha})=\tilde{z}_{bi}\exp(-\tilde{\alpha}\tilde{r}_{bi}), \quad \text{(D.7)}$$

$$d\tilde{\mathbf{r}}_1 d\tilde{\mathbf{r}}_2=\frac{\tilde{R}^6}{64}(\mu_1^2-\nu_1^2)(\mu_2^2-\nu_2^2)d\mu_1 d\mu_2 d\nu_1 d\nu_2 d\gamma_1 d\gamma_2,$$

$$\int\limits_{z_1>0}W(\tilde{\mathbf{r}}_1,\tilde{\mathbf{r}}_2)d\tilde{\mathbf{r}}_1 d\tilde{\mathbf{r}}_2=\frac{\tilde{R}^6}{64}\int_1^\infty d\mu_1\int_1^\infty d\mu_2\int_{-1}^1 d\nu_1\int_{-1}^1 d\nu_2(\mu_1^2-\nu_1^2)(\mu_2^2-\nu_2^2)\int_0^\pi d\gamma_1\int_0^\pi d\gamma_2 W(\mu_i,\nu_i,\gamma_i)$$

Expressions for distances between the electrons:

$$\tilde{r}_{12}^{\,2}=\tilde{R}^2\begin{pmatrix}(\mu_1\nu_1-\mu_2\nu_2)^2+(\mu_1^2-1)(1-\nu_1^2)+(\mu_2^2-1)(1-\nu_2^2)\\ -2\sqrt{(\mu_1^2-1)(1-\nu_1^2)(\mu_2^2-1)(1-\nu_2^2)}\cos(\gamma_1-\gamma_2)\end{pmatrix}$$

$$=\mu_1^2+\mu_2^2+\nu_1^2+\nu_2^2-2-2\mu_1\nu_1\mu_2\nu_2-2\sqrt{(\mu_1^2-1)(1-\nu_1^2)(\mu_2^2-1)(1-\nu_2^2)}\cos(\gamma_1-\gamma_2),\qquad\text{(D.8)}$$

$$\tilde{r}_{1'2}^{\,2}=\tilde{r}_{12'}^{\,2}=\tilde{R}^2\begin{pmatrix}(\mu_1\nu_1-\mu_2\nu_2)^2+(\mu_1^2-1)(1-\nu_1^2)+(\mu_2^2-1)(1-\nu_2^2)\\ -2\sqrt{(\mu_1^2-1)(1-\nu_1^2)(\mu_2^2-1)(1-\nu_2^2)}\cos(\gamma_1+\gamma_2)\end{pmatrix}$$

Neumann's expansion for $\frac{1}{\tilde{r}_{12}}$ and $\frac{1}{\tilde{r}_{1'2}}$ factorization in the ellipsoidal coordinates is:

$$\frac{1}{\tilde{r}_{12}}=\frac{1}{\tilde{R}}\sum_{n=0}^{\infty}\sum_{m=0}^{\infty}D_n^mP_n^m\begin{pmatrix}\mu_1\\ \mu_2\end{pmatrix}Q_n^m\begin{pmatrix}\mu_2\\ \mu_1\end{pmatrix}P_n^m(\nu_2)P_n^m(\nu_1)\cos(m(\gamma_1-\gamma_2))\qquad\text{(D.9a)}$$

$$\frac{1}{\tilde{r}_{1'2}}=\frac{1}{\tilde{R}}\sum_{n=0}^{\infty}\sum_{m=0}^{\infty}D_n^mP_n^m\begin{pmatrix}\mu_1\\ \mu_2\end{pmatrix}Q_n^m\begin{pmatrix}\mu_2\\ \mu_1\end{pmatrix}P_n^m(\nu_2)P_n^m(\nu_1)\cos(m(\gamma_1+\gamma_2))\qquad\text{(D.9b)}$$

Where the upper variables $\mu_i$ should be used when $\mu_2>\mu_1$ and the lower when $\mu_1>\mu_2$ [45, 46, 47, 48]. Coefficients $D_n^0=2n+1$ and $D_n^m=(-1)^m2(2n+1)\left(\frac{(n-m)!}{(n+m)!}\right)^2$ for $m>0$.

Functions $P_n^m(t)$ and $Q_n^m(t)$ are associated Legendre functions $\{m, n\}$ of the first and second kind respectively, by definition:

$$P_n^m(t)=\begin{cases}(-1)^m(1-t^2)^{m/2}\dfrac{d^m}{dt^m}P_n(t), & -1<t<1\\ (t^2-1)^{m/2}\dfrac{d^m}{dt^m}P_n(t), & t>1\end{cases}\quad\text{and}\quad Q_n^m(t)=(t^2-1)^{m/2}\frac{d^m}{dt^m}Q_n(t),\quad t>1.\qquad\text{(D.10)}$$

Here $P_n(t)=F(-n,n+1;1;(1-t)/2)$ are tabulated Legendre polynomials and $Q_n(t)=\frac{n!}{(2n+1)!!}t^{-n-1}F\left(\frac{n+2}{2},\frac{n+1}{2};\frac{2n+3}{2};\frac{1}{t^2}\right)$, $Q_0(t)=\frac{1}{2}\ln\left(\frac{t+1}{t-1}\right)$ are the spherical functions of the second kind.

Substituting Eqs.(D.8,91) into (D.6)we obtained that

$$J_{23}=\frac{\tilde{R}^{10}}{2^{10}}\frac{8\tilde{\alpha}^5\tilde{\beta}^5}{3\pi^2}\int_1^{\infty}d\mu_1\int_1^{\infty}d\mu_2\int_{-1}^{1}d\nu_1\int_{-1}^{1}d\nu_2(\mu_1^2-\nu_1^2)(\mu_2^2-\nu_2^2)(\mu_1^2-1)(1-\nu_1^2)(\mu_2^2-1)(1-\nu_2^2)\times$$
$$\times\exp\left(-\tilde{R}(\tilde{\alpha}\mu_1+\tilde{\beta}\mu_2)\right)\left(2-\tilde{\beta}(\mu_2+\nu_2)\frac{\tilde{R}}{2}\right)\left(2-\tilde{\beta}(\mu_2-\nu_2)\frac{\tilde{R}}{2}\right)\int_0^{\pi}d\gamma_1\int_0^{\pi}d\gamma_2\cdot\left(\frac{1}{\tilde{r}_{12}}+\frac{\zeta}{\tilde{r}_{1'2}}\right)\sin^2\gamma_1\sin^2\gamma_2$$

(D.11)

Performing integration over the angles $\gamma_i$:

$$B_m = \int_0^\pi d\gamma_1 \int_0^\pi d\gamma_2 \sin^2\gamma_1 \sin^2\gamma_2 \cos(m(\gamma_1-\gamma_2)) = \begin{cases} \frac{\pi^2}{4} \quad (m=0); \quad \frac{16}{9} \quad (m=1); \quad \frac{\pi^2}{16} \quad (m=2), \\ \frac{8\left(1-(-1)^m\right)}{m^2\left(m^2-4\right)^2}, \quad m \geq 3, \end{cases}$$

$$B_m^{/} = \int_0^\pi d\gamma_1 \int_0^\pi d\gamma_2 \sin^2\gamma_1 \sin^2\gamma_2 \cos(m(\gamma_1+\gamma_2)) = \begin{cases} \frac{\pi^2}{4} \quad (m=0); \quad -\frac{16}{9} \quad (m=1); \quad \frac{\pi^2}{16} \quad (m=2), \\ \frac{8(-1)^m\left(1-(-1)^m\right)}{m^2\left(m^2-4\right)^2}, \quad m \geq 3, \end{cases}$$

(D.12)

Note, that $B_m^{/} = (-1)^m B_m$ and coefficients vanish rapidly with $m$ increase, so the series could be cut at small $m$ with enough high accuracy.

$$J_{23} = \frac{\tilde{R}^{10}}{2^{10}} \frac{8\tilde{\alpha}^5\tilde{\beta}^5}{3\pi^2} \int_1^\infty d\mu_1 \int_1^\infty d\mu_2 \int_{-1}^1 d\nu_1 \int_{-1}^1 d\nu_2 \left(\mu_1^2-\nu_1^2\right)\left(\mu_2^2-\nu_2^2\right) \exp\left(-\tilde{R}\left(\tilde{\alpha}\mu_1+\tilde{\beta}\mu_2\right)\right) \times$$
$$\times \left(\mu_1^2-1\right)\left(1-\nu_1^2\right)\left(\mu_2^2-1\right)\left(1-\nu_2^2\right)\left(2-\tilde{\beta}(\mu_2+\nu_2)\frac{\tilde{R}}{2}\right)\left(2-\tilde{\beta}(\mu_2-\nu_2)\frac{\tilde{R}}{2}\right) \times \quad \text{(D.13)}$$
$$\times \sum_{n=0}^\infty \sum_{m=0}^\infty D_n^m P_n^m \begin{pmatrix} \mu_1 \\ \mu_2 \end{pmatrix} Q_n^m \begin{pmatrix} \mu_2 \\ \mu_1 \end{pmatrix} P_n^m(\nu_2) P_n^m(\nu_1) \left(1+(-1)^m \zeta\right) B_m$$

After elementary transformations forth fold integral (D.13) can be factorized as

$$J_{23} = \frac{\tilde{R}^{10}}{2^{10}} \frac{32\tilde{\alpha}^5\tilde{\beta}^5}{3\pi^2} \sum_{n=0}^\infty \sum_{m=0}^\infty D_n^m \left(1+(-1)^m\zeta\right) B_m \times$$
$$\times \begin{pmatrix} \int_1^\infty d\mu_1 \left(\mu_1^2-1\right) \exp\left(-\tilde{\alpha}\tilde{R}\mu_1\right) Q_n^m(\mu_1) \int_1^{\mu_1} d\mu_2 \left(\mu_2^2-1\right) \exp\left(-\tilde{\beta}\tilde{R}\mu_2\right) P_n^m(\mu_2) \\ + \int_1^\infty d\mu_2 \left(\mu_2^2-1\right) \exp\left(-\tilde{\beta}\tilde{R}\mu_2\right) Q_n^m(\mu_2) \int_1^{\mu_2} d\mu_1 \left(\mu_1^2-1\right) \exp\left(-\tilde{\alpha}\tilde{R}\mu_1\right) P_n^m(\mu_1) \end{pmatrix} \quad \text{(D.14)}$$
$$\times \int_{-1}^1 d\nu_1 \left(1-\nu_1^2\right)\left(\mu_1^2-\nu_1^2\right) P_n^m(\nu_1) \int_{-1}^1 d\nu_2 \left(1-\nu_2^2\right)\left(\mu_2^2-\nu_2^2\right) P_n^m(\nu_2) \left(\left(1-\frac{\tilde{\beta}\tilde{R}\mu_2}{4}\right)^2 - \frac{\tilde{\beta}^2\tilde{R}^2}{16}\nu_2^2\right)$$

Using the integrals

$$I_{nm}(\mu,\tilde{\beta}) = \frac{(n-m)!}{(n+m)!} \int_{-1}^1 d\nu_2 \left(1-\nu_2^2\right)\left(\mu^2-\nu_2^2\right)\left(\left(1-\frac{\tilde{\beta}\tilde{R}\mu}{4}\right)^2 - \frac{\tilde{\beta}^2\tilde{R}^2}{16}\nu_2^2\right) P_n^m(\nu_2) =$$
$$= \frac{(n-m)!}{(n+m)!} \int_{-1}^1 d\nu_2 \left(1-\nu_2^2\right)\left(\mu^2\left(1-\frac{\tilde{\beta}\tilde{R}\mu}{4}\right)^2 - \nu_2^2\left(\left(1-\frac{\tilde{\beta}\tilde{R}\mu}{4}\right)^2 + \frac{\tilde{\beta}^2\tilde{R}^2}{16}\mu^2\right) + \frac{\tilde{\beta}^2\tilde{R}^2}{16}\nu_2^4\right) P_n^m(\nu_2) =$$
$$= I_{nm0}\mu^2\left(1-\frac{\tilde{\beta}\tilde{R}\mu}{4}\right)^2 - I_{nm2}\left(\left(1-\frac{\tilde{\beta}\tilde{R}\mu}{4}\right)^2 + \frac{\tilde{\beta}^2\tilde{R}^2}{16}\mu^2\right) + I_{nm4}\frac{\tilde{\beta}^2\tilde{R}^2}{16}$$

(D.15)

$$I_{nmp} = \frac{(n-m)!}{(n+m)!}\int_{-1}^{1} d\nu \left(1-\nu^2\right)\nu^p P_n^m(\nu)$$

$$J_{23} = \frac{\tilde{R}^{10}}{2^5}\frac{\tilde{\alpha}^5\tilde{\beta}^5}{3\pi^2}\sum_{n=0}^{\infty}\sum_{m=0}^{\infty} D_n^m\left(1+(-1)^m\zeta\right)B_m \times$$
$$\times\left(\begin{array}{l}\int_1^{\infty} d\mu_1\left(\mu_1^2-1\right)\exp\left(-\tilde{\alpha}\tilde{R}\mu_1\right)Q_n^m(\mu_1)I_{nm}(\mu_1,0)\int_1^{\mu_1} d\mu_2\left(\mu_2^2-1\right)\exp\left(-\tilde{\beta}\tilde{R}\mu_2\right)P_n^m(\mu_2)I_{nm}\left(\mu_2,\tilde{\beta}\right)+ \\ +\int_1^{\infty} d\mu_2\left(\mu_2^2-1\right)\exp\left(-\tilde{\beta}\tilde{R}\mu_2\right)Q_n^m(\mu_2)I_{nm}\left(\mu_2,\tilde{\beta}\right)\int_1^{\mu_2} d\mu_1\left(\mu_1^2-1\right)\exp\left(-\tilde{\alpha}\tilde{R}\mu_1\right)P_n^m(\mu_1)I_{nm}(\mu_1,0)\end{array}\right)$$

Using these integrals we integrated further:

$$L_{nm}\left(\mu,\tilde{\alpha},\tilde{\beta}\right)=\int_1^{\mu} dx\left(x^2-1\right)\exp\left(-\tilde{\alpha}\tilde{R}x\right)P_n^m(x)I_{nm}\left(x,\tilde{\beta}\right) \qquad \text{(D.16)}$$

After elementary transformations we obtained the quadrature:

$$J_{23} = \frac{\tilde{R}^{10}}{2^5}\frac{\tilde{\alpha}^5\tilde{\beta}^5}{3\pi^2}\sum_{n=0}^{\infty}\sum_{m=0}^{n}(-1)^m p_m(2n+1)\left(1+(-1)^m\zeta\right)B_m \times$$
$$\times\int_1^{\infty} d\mu\left(\mu^2-1\right)Q_n^m(\mu)\left(\exp\left(-\tilde{\alpha}\tilde{R}\mu\right)I_{nm}(\mu,0)L_{nm}\left(\mu,\tilde{\beta},\tilde{\beta}\right)+\exp\left(-\tilde{\beta}\tilde{R}\mu\right)I_{nm}\left(\mu,\tilde{\beta}\right)L_{nm}(\mu,\tilde{\alpha},0)\right) \qquad \text{(D.17)}$$

Coefficients $p_0 = 1$ and $p_m = 2$ for $m$ >0.

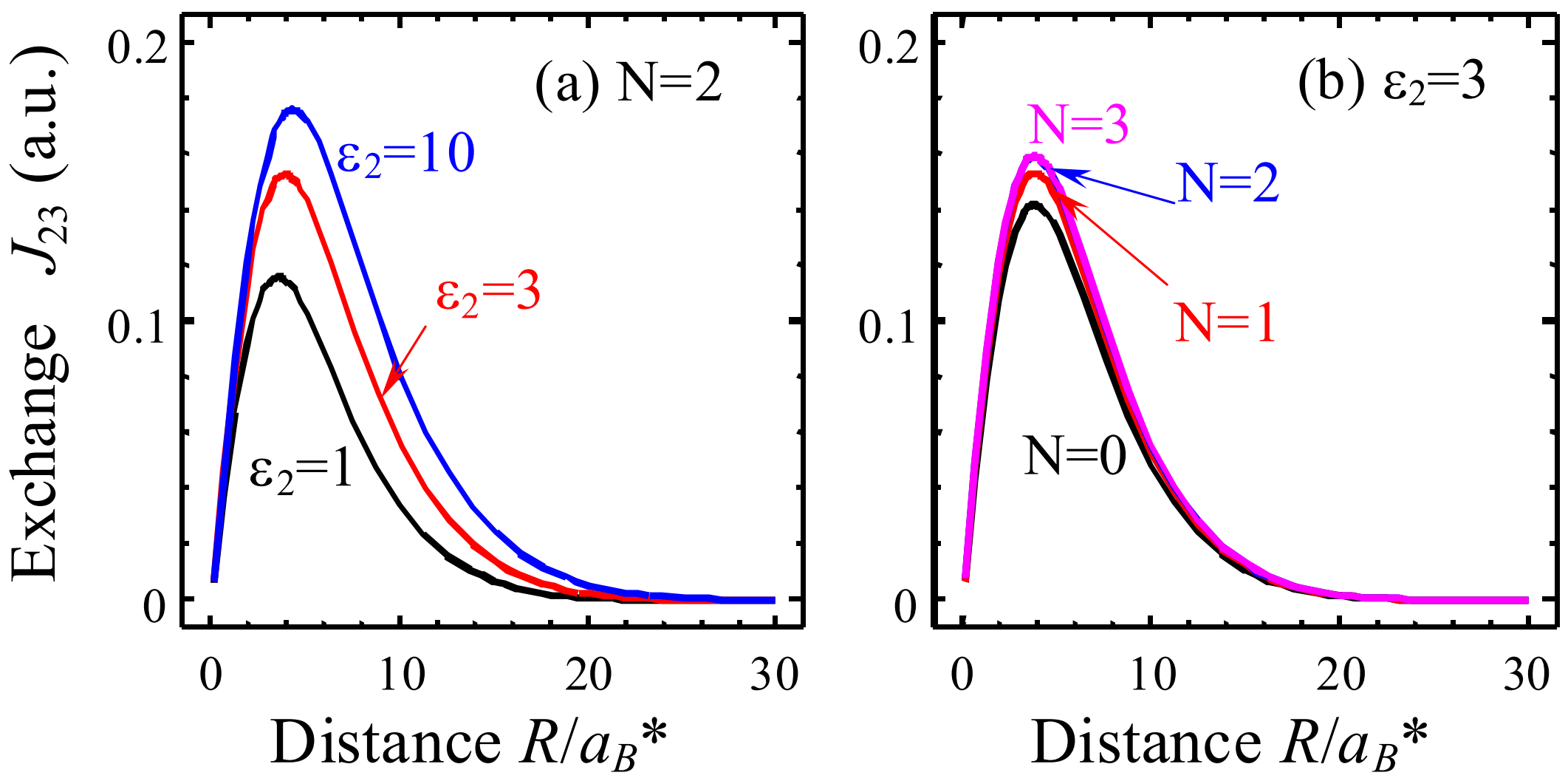

**Fig. D1.** Dependence of the exchange integral (in $a_B^{*2}|\mu|/\hbar^2$ units) vs. the distance between the defects. (a) Exchange integral for different values of $\varepsilon_2$ at $\varepsilon_1$=1. (b) The results of calculations for different maximal quantity of terms in a series for exchange integral.

**Appendix E. Energy levels of the hydrogen-like molecule at the surface**

Two-electron wave function obeys the Paul principle:

$$\psi(\mathbf{r}_1,\mathbf{r}_2,s_1,s_2)=\begin{cases}\frac{1}{\sqrt{2}}\left(\varphi_a(\mathbf{r}_1)\varphi_b(\mathbf{r}_2)+\varphi_a(\mathbf{r}_2)\varphi_b(\mathbf{r}_1)\right)\chi_A(s_1,s_2), & \Sigma=0,\\ \frac{1}{\sqrt{2}}\left(\varphi_a(\mathbf{r}_1)\varphi_b(\mathbf{r}_2)-\varphi_a(\mathbf{r}_2)\varphi_b(\mathbf{r}_1)\right)\chi_S(s_1,s_2), & \Sigma=1.\end{cases} \quad \text{(E.1)}$$

$\chi_{A,S}(s_1,s_2)$ are spin functions (spinors), the coordinate part could be either symmetric or antisymmetric: $\varphi_{S,A}(\mathbf{r}_1,\mathbf{r}_2)=\frac{1}{\sqrt{2}}\left(\varphi_a(\mathbf{r}_1)\varphi_b(\mathbf{r}_2)\pm\varphi_a(\mathbf{r}_2)\varphi_b(\mathbf{r}_1)\right)$.

Introducing dimensionless coordinates renormalized on Bohr radius: $\tilde{\mathbf{r}}=\frac{\mathbf{r}}{a_B}$, $a_B=2\pi\varepsilon_0(\varepsilon_1+\varepsilon_2)\hbar^2/Z\mu e^2$ and dimensionless Hamiltonian/energy $\tilde{H}=\frac{a_B^2\mu}{\hbar^2}H$, $\tilde{E}=\frac{a_B^2\mu}{\hbar^2}E$, we rewrite the Schrödinger equation for coordinate part in the form:

$$\left(-\frac{1}{2}\left(\tilde{\Delta}_1+\tilde{\Delta}_2\right)+\frac{Z}{\tilde{r}_{ab}}-\frac{1}{\tilde{r}_{a1}}-\frac{1}{\tilde{r}_{a2}}-\frac{1}{\tilde{r}_{b1}}-\frac{1}{\tilde{r}_{b1}}+\tilde{V}_{12}(\tilde{\mathbf{r}}_1,\tilde{\mathbf{r}}_2)\right)\varphi_{S,A}(\tilde{\mathbf{r}}_1,\tilde{\mathbf{r}}_2)=\tilde{E}\varphi_{S,A}(\tilde{\mathbf{r}}_1,\tilde{\mathbf{r}}_2) \quad \text{(E.2a)}$$

$$\tilde{V}_{12}(\tilde{\mathbf{r}}_1,\tilde{\mathbf{r}}_2)=\frac{1}{Z(1+\zeta)}\left(\frac{1}{\tilde{r}_{12}}+\frac{\zeta}{2}\left(\frac{1}{\tilde{z}_1}+\frac{1}{\tilde{z}_2}+\frac{1}{\tilde{r}_{1'2}}+\frac{1}{\tilde{r}_{12'}}\right)\right) \quad \text{(E.2b)}$$

Here $\zeta=(\varepsilon_2-\varepsilon_1)/(\varepsilon_2+\varepsilon_1)$.

Let us consider the equations for the energy levels determination:

$$\int\limits_{z_2>0}\int\limits_{z_1>0}\varphi^*_{S,A}(\tilde{\mathbf{r}}_1,\tilde{\mathbf{r}}_2)\left(\tilde{H}-\tilde{E}_{S,A}\right)\varphi_{S,A}(\tilde{\mathbf{r}}_1,\tilde{\mathbf{r}}_2)d\tilde{\mathbf{r}}_1 d\tilde{\mathbf{r}}_2=0 \quad \text{(E.3)}$$

Introducing the designation $\varphi_{ji}\equiv\varphi(\mathbf{r}_i-\mathbf{R}_j)$, the evident form for symmetric wave function is

$$\frac{1}{2}\int\limits_{z_2>0}\int\limits_{z_1>0}\left(\begin{array}{c}(\varphi_{a1}\varphi_{b2}\pm\varphi_{a2}\varphi_{b1})^*\\ \left(-\frac{1}{2}\left(\tilde{\Delta}_1+\tilde{\Delta}_2\right)+\frac{Z}{\tilde{r}_{ab}}-\frac{1}{\tilde{r}_{a1}}-\frac{1}{\tilde{r}_{a2}}-\frac{1}{\tilde{r}_{b1}}-\frac{1}{\tilde{r}_{b1}}+\tilde{V}_{12}(\tilde{\mathbf{r}}_1,\tilde{\mathbf{r}}_2)-\tilde{E}_{S,A}\right)\\ (\varphi_{a1}\varphi_{b2}\pm\varphi_{a2}\varphi_{b1})\end{array}\right)d\tilde{\mathbf{r}}_1 d\tilde{\mathbf{r}}_2=0 \quad \text{(E.5)}$$

Here the upper signs correspond to “S” state, while lower signs correspond to “A” state.

Introducing the designation $\left(-\frac{1}{2}\tilde{\Delta}-\frac{1}{\tilde{r}}\right)\varphi(\tilde{\mathbf{r}})\equiv F(\tilde{\mathbf{r}})$, Eq.(E.5) can be rewritten as

$$\frac{1}{2}\int\limits_{z_2>0}\int\limits_{z_1>0}\begin{pmatrix} \varphi_{a1}\varphi_{b2}F_{a1}\varphi_{b2}+\varphi_{a1}\varphi_{b2}\varphi_{a1}F_{b2}\pm\varphi_{a1}\varphi_{b2}F_{a2}\varphi_{b1}\pm\varphi_{a1}\varphi_{b2}\varphi_{a2}F_{b1}\pm \\ \pm\varphi_{a2}\varphi_{b1}F_{a1}\varphi_{b2}\pm\varphi_{a2}\varphi_{b1}\varphi_{a1}F_{b2}+\varphi_{a2}\varphi_{b1}F_{a2}\varphi_{b1}+\varphi_{a2}\varphi_{b1}\varphi_{a2}F_{b1}+ \\ -\left(\varphi_{a1}\varphi_{b2}\pm\varphi_{a2}\varphi_{b1}\right)\left(\frac{1}{\tilde{r}_{a2}}+\frac{1}{\tilde{r}_{b1}}\right)\varphi_{a1}\varphi_{b2}\mp\left(\varphi_{a1}\varphi_{b2}\pm\varphi_{a2}\varphi_{b1}\right)\left(\frac{1}{\tilde{r}_{a1}}+\frac{1}{\tilde{r}_{b2}}\right)\varphi_{a2}\varphi_{b1} \\ +\tilde{V}_{12}(\tilde{\mathbf{r}}_1,\tilde{\mathbf{r}}_2)\varphi_{S,A}(\tilde{\mathbf{r}}_1,\tilde{\mathbf{r}}_2)^2+\left(\frac{Z}{\tilde{r}_{ab}}-\tilde{E}_{S,A}\right)\left(\left(\varphi_{a1}\varphi_{b2}\right)^2\pm2\varphi_{a1}\varphi_{b2}\varphi_{a2}\varphi_{b1}+\left(\varphi_{a2}\varphi_{b1}\right)^2\right)\end{pmatrix}d\tilde{\mathbf{r}}_1d\tilde{\mathbf{r}}_2=0$$

(E.6)

Using the norm $\int\limits_{z_i>0}\left(\varphi_{a,bi}\right)^2d\tilde{\mathbf{r}}_i=1$, it could be rewritten as

$$\begin{aligned}&\frac{1}{2}\int\limits_{z_1>0}\left(\varphi_{a1}F_{a1}+\varphi_{b1}F_{b1}\right)d\tilde{\mathbf{r}}_1+\frac{1}{2}\int\limits_{z_2>0}\left(\varphi_{b2}F_{b2}+\varphi_{a2}F_{a2}\right)d\tilde{\mathbf{r}}_2+\\&\pm\frac{1}{2}\int\limits_{z_1>0}\varphi_{a1}\varphi_{b1}d\tilde{\mathbf{r}}_1\int\limits_{z_2>0}\varphi_{b2}F_{a2}d\tilde{\mathbf{r}}_2\pm\frac{1}{2}\int\limits_{z_1>0}\varphi_{a1}\varphi_{b1}d\tilde{\mathbf{r}}_1\int\limits_{z_2>0}\varphi_{a2}F_{b2}d\tilde{\mathbf{r}}_2\pm\\&\pm\frac{1}{2}\int\limits_{z_1>0}\varphi_{a1}F_{b1}d\tilde{\mathbf{r}}_1\int\limits_{z_2>0}\varphi_{b2}\varphi_{a2}d\tilde{\mathbf{r}}_2\pm\frac{1}{2}\int\limits_{z_1>0}F_{a1}\varphi_{b1}d\tilde{\mathbf{r}}_1\int\limits_{z_2>0}\varphi_{a2}\varphi_{b2}d\tilde{\mathbf{r}}_2-\\&-\frac{1}{2}\int\limits_{z_2>0}\frac{(\varphi_{b2})^2}{\tilde{r}_{a2}}d\tilde{\mathbf{r}}_2-\frac{1}{2}\int\limits_{z_1>0}\frac{(\varphi_{a1})^2}{\tilde{r}_{b1}}d\tilde{\mathbf{r}}_1-\frac{1}{2}\int\limits_{z_2>0}\frac{(\varphi_{b1})^2}{\tilde{r}_{a1}}d\tilde{\mathbf{r}}_1-\frac{1}{2}\int\limits_{z_1>0}\frac{(\varphi_{a2})^2}{\tilde{r}_{b2}}d\tilde{\mathbf{r}}_2\mp\\&\mp\frac{1}{2}\int\limits_{z_1>0}\varphi_{a1}\varphi_{b1}d\tilde{\mathbf{r}}_1\int\limits_{z_2>0}\frac{\varphi_{b2}\varphi_{a2}}{\tilde{r}_{a2}}d\tilde{\mathbf{r}}_2\mp\frac{1}{2}\int\limits_{z_1>0}\frac{\varphi_{a1}\varphi_{b1}}{\tilde{r}_{b1}}d\tilde{\mathbf{r}}_1\int\limits_{z_2>0}\varphi_{b2}\varphi_{a2}d\tilde{\mathbf{r}}_2\mp\\&\mp\frac{1}{2}\int\limits_{z_1>0}\frac{\varphi_{a1}\varphi_{b1}}{\tilde{r}_{a1}}d\tilde{\mathbf{r}}_1\int\limits_{z_2>0}\varphi_{b2}\varphi_{a2}d\tilde{\mathbf{r}}_2\mp\frac{1}{2}\int\limits_{z_1>0}\varphi_{a1}\varphi_{b1}d\tilde{\mathbf{r}}_1\int\limits_{z_2>0}\frac{\varphi_{b2}\varphi_{a2}}{\tilde{r}_{b2}}d\tilde{\mathbf{r}}_2+\\&+\frac{1}{2}\int\limits_{z_2>0}\int\limits_{z_1>0}\tilde{V}_{12}(\tilde{\mathbf{r}}_1,\tilde{\mathbf{r}}_2)\varphi_{S,A}(\tilde{\mathbf{r}}_1,\tilde{\mathbf{r}}_2)^2d\tilde{\mathbf{r}}_1d\tilde{\mathbf{r}}_2+\left(\frac{Z}{\tilde{r}_{ab}}-\tilde{E}_{S,A}\right)\left(1\pm S^2\right)=0\end{aligned}\tag{E.7}$$

Here overlap integral is $S=\int\limits_{z_i>0}\left(\varphi_{ai}\varphi_{bi}\right)d\tilde{\mathbf{r}}_i$.

Next we could use the change of the integration variable $\tilde{\mathbf{r}}_2\to\tilde{\mathbf{r}}_1$

$$\begin{aligned}&\int\limits_{z_1>0}\left(\varphi_{a1}F_{a1}+\varphi_{b1}F_{b1}\right)d\tilde{\mathbf{r}}_1\pm S\int\limits_{z_1>0}\varphi_{a1}F_{b1}d\tilde{\mathbf{r}}_1\pm S\int\limits_{z_1>0}F_{a1}\varphi_{b1}d\tilde{\mathbf{r}}_1-\int\limits_{z_1>0}\frac{(\varphi_{a1})^2}{\tilde{r}_{b1}}d\tilde{\mathbf{r}}_1-\int\limits_{z_2>0}\frac{(\varphi_{b1})^2}{\tilde{r}_{a1}}d\tilde{\mathbf{r}}_1\mp\\&\mp S\int\limits_{z_1>0}\frac{\varphi_{a1}\varphi_{b1}}{\tilde{r}_{b1}}d\tilde{\mathbf{r}}_1\mp S\int\limits_{z_1>0}\frac{\varphi_{a1}\varphi_{b1}}{\tilde{r}_{a1}}d\tilde{\mathbf{r}}_1+\frac{1}{2}\int\limits_{z_2>0}\int\limits_{z_1>0}\tilde{V}_{12}(\tilde{\mathbf{r}}_1,\tilde{\mathbf{r}}_2)\varphi_{S,A}(\tilde{\mathbf{r}}_1,\tilde{\mathbf{r}}_2)^2d\tilde{\mathbf{r}}_1d\tilde{\mathbf{r}}_2+\left(\frac{Z}{\tilde{r}_{ab}}-\tilde{E}_{S,A}\right)\left(1\pm S^2\right)=0\end{aligned}$$

Next we recall that $R_a$ and $R_b$ have only $x$-components, different in signs ($x\to-x$), we derived that:

$$\begin{aligned}&2\int\limits_{z_1>0}\varphi_{a1}F_{a1}d\tilde{\mathbf{r}}_1\pm2S\int\limits_{z_1>0}\varphi_{a1}F_{b1}d\tilde{\mathbf{r}}_1-2\int\limits_{z_1>0}\frac{(\varphi_{a1})^2}{\tilde{r}_{b1}}d\tilde{\mathbf{r}}_1\mp2S\int\limits_{z_1>0}\frac{\varphi_{a1}\varphi_{b1}}{\tilde{r}_{a1}}d\tilde{\mathbf{r}}_1+\\&+\int\limits_{z_2>0}\int\limits_{z_1>0}\tilde{V}_{12}(\tilde{\mathbf{r}}_1,\tilde{\mathbf{r}}_2)\frac{\varphi_{S,A}(\tilde{\mathbf{r}}_1,\tilde{\mathbf{r}}_2)^2}{2}d\tilde{\mathbf{r}}_1d\tilde{\mathbf{r}}_2+\left(\frac{Z}{\tilde{r}_{ab}}-\tilde{E}_{S,A}\right)\left(1\pm S^2\right)=0\end{aligned}\tag{E.8}$$

Corresponding energy:

$$\tilde{E}_{S,A} = \frac{Z}{\tilde{r}_{ab}} + \frac{1}{1 \pm S^2} \begin{pmatrix} 2 \int\limits_{z_1>0} \varphi_{a1} F_{a1} d\tilde{\mathbf{r}}_1 \pm 2S \int\limits_{z_1>0} \varphi_{a1} F_{b1} d\tilde{\mathbf{r}}_1 - 2 \int\limits_{z_1>0} \frac{(\varphi_{a1})^2}{\tilde{r}_{b1}} d\tilde{\mathbf{r}}_1 \mp 2S \int\limits_{z_1>0} \frac{\varphi_{a1}\varphi_{b1}}{\tilde{r}_{a1}} d\tilde{\mathbf{r}}_1 + \\ + \frac{1}{2} \int\limits_{z_2>0} \int\limits_{z_1>0} \tilde{V}_{12}(\tilde{\mathbf{r}}_1, \tilde{\mathbf{r}}_2) \left( (\varphi_{a1})^2 (\varphi_{b2})^2 + (\varphi_{a2})^2 (\varphi_{b1})^2 \right) d\tilde{\mathbf{r}}_1 d\tilde{\mathbf{r}}_2 \\ \pm \int\limits_{z_2>0} \int\limits_{z_1>0} \tilde{V}_{12}(\tilde{\mathbf{r}}_1, \tilde{\mathbf{r}}_2) \varphi_{a1}\varphi_{b2}\varphi_{a2}\varphi_{b1} d\tilde{\mathbf{r}}_1 d\tilde{\mathbf{r}}_2 \end{pmatrix} \quad \text{(E.9)}$$

Next we introduced the prolate ellipsoidal coordinates

$$\mu = \frac{r_{a1} - r_{b1}}{R} = \frac{\tilde{r}_{a1} + \tilde{r}_{b1}}{\tilde{R}}, \quad \nu = \frac{r_{a1} - r_{b1}}{R} = \frac{\tilde{r}_{a1} - \tilde{r}_{b1}}{\tilde{R}}, \quad \tan\beta = \frac{z}{y},$$

$$\tilde{x} = \frac{\tilde{R}}{2}(1 + \mu\nu), \quad \tilde{y} = \frac{\tilde{R}}{2}\sqrt{(\mu^2 - 1)(1 - \nu^2)}\cos\beta, \quad \tilde{z} = \frac{\tilde{R}}{2}\sqrt{(\mu^2 - 1)(1 - \nu^2)}\sin\beta,$$

$$\tilde{r}_{a1} = \frac{\tilde{R}}{2}(\mu + \nu), \quad \tilde{r}_{b1} = \frac{\tilde{R}}{2}(\mu - \nu), \quad \tilde{z}_{a1} = \tilde{z}_{b1} = \tilde{z}, \quad \text{(E.10)}$$

$$\varphi_{a1} = \sqrt{\frac{2\tilde{\alpha}^5}{\pi}} \tilde{z}_{a1} \exp(-\tilde{\alpha}\tilde{r}_{a1}), \quad \varphi_{b1} = \sqrt{\frac{2\tilde{\alpha}^5}{\pi}} \tilde{z}_{b1} \exp(-\tilde{\alpha}\tilde{r}_{b1}),$$

$$\int\limits_{z_1>0} W(\tilde{\mathbf{r}}_1) d\tilde{\mathbf{r}}_1 = \frac{\tilde{R}^3}{8} \int_1^\infty d\mu \int_{-1}^1 d\nu (\mu^2 - \nu^2) \int_0^\pi d\beta \cdot W(\mu, \nu, \beta)$$

Using the coordinates, the overlap integral:

$$S = \int\limits_{z_i>0} (\varphi_{ai}\varphi_{bi}) d\tilde{\mathbf{r}}_i = \frac{2\tilde{\alpha}^5}{\pi} \frac{\tilde{R}^3}{8} \int_1^\infty d\mu \int_{-1}^1 d\nu (\mu^2 - \nu^2) \int_0^\pi d\beta \cdot \exp(-\tilde{\alpha}\tilde{R}\mu) \frac{\tilde{R}^2}{4} (\mu^2 - 1)(1 - \nu^2) \sin^2\beta$$
$$= \tilde{\alpha}^5 \frac{\tilde{R}^5}{32} \int_1^\infty d\mu (\mu^2 - 1) \exp(-\tilde{\alpha}\tilde{R}\mu) \int_{-1}^1 d\nu (\mu^2 - \nu^2)(1 - \nu^2) = \exp(-\tilde{\alpha}\tilde{R}) \left( 1 + \tilde{\alpha}\tilde{R} + \frac{(\tilde{\alpha}\tilde{R})^2}{15} (6 + \tilde{\alpha}\tilde{R}) \right) \quad \text{(E.11)}$$

Other integrals

$$A = \int\limits_{z_1>0} \varphi_{a1} F_{a1} d\tilde{\mathbf{r}}_1 = -\left( \frac{\tilde{\alpha}^2}{2} + (1 - 2\tilde{\alpha}) \int\limits_{z_1>0} \frac{(\varphi_{a1})^2 d\tilde{\mathbf{r}}_1}{\tilde{r}_{a1}} \right) = -\frac{\tilde{\alpha}^2}{2} -$$
$$- (1 - 2\tilde{\alpha}) \frac{2\tilde{\alpha}^5}{\pi} \frac{\tilde{R}^3}{8} \int_1^\infty d\mu \int_{-1}^1 d\nu (\mu^2 - \nu^2) \int_0^\pi d\beta \cdot \frac{\exp(-\tilde{\alpha}\tilde{R}(\mu + \nu))}{\tilde{R}(\mu + \nu)} \frac{\tilde{R}^2}{2} (\mu^2 - 1)(1 - \nu^2) \sin^2\beta =$$
$$= -\frac{\tilde{\alpha}^2}{2} - (1 - 2\tilde{\alpha}) \tilde{\alpha}^5 \frac{\tilde{R}^4}{16} \int_1^\infty d\mu (\mu^2 - 1) \exp(-\tilde{\alpha}\tilde{R}\mu) \int_{-1}^1 d\nu (\mu - \nu)(1 - \nu^2) \exp(-\tilde{\alpha}\tilde{R}\nu) =$$
$$= -\frac{\tilde{\alpha}^2}{2} - (1 - 2\tilde{\alpha}) \frac{\tilde{\alpha}}{2} = -\frac{\tilde{\alpha}}{2} + \frac{\tilde{\alpha}^2}{2} \quad \text{, (E.12a)}$$

$$A_{ab} = \int\limits_{z_1>0} \varphi_{a1} F_{b1} d\tilde{\mathbf{r}}_1 = -\left( \frac{\tilde{\alpha}^2}{2} S + (1 - 2\tilde{\alpha}) \int\limits_{z_1>0} \frac{(\varphi_{a1})^2 d\tilde{\mathbf{r}}_1}{\tilde{r}_{b1}} \right) = -\frac{\tilde{\alpha}^2 S}{2} - (1 - 2\tilde{\alpha}) J_{ba} =$$
$$= -\frac{\tilde{\alpha}}{2} \left( (1 - \tilde{\alpha})(1 + \tilde{\alpha}\tilde{R}) + \frac{(\tilde{\alpha}\tilde{R})^2}{15} (5 - 4\tilde{\alpha} + \tilde{\alpha}^2 \tilde{R}) \right) \exp(-\tilde{\alpha}\tilde{R}) \quad \text{(E.12b)}$$

Coulomb integrals

$$C_{ab}=\int\limits_{z_1>0}\frac{(\varphi_{a1})^2 d\tilde{\mathbf{r}}_1}{\tilde{r}_{b1}}=\frac{2\tilde{\alpha}^5}{\pi}\frac{\tilde{R}^3}{8}\int\limits_1^\infty d\mu\int\limits_{-1}^1 d\nu\left(\mu^2-\nu^2\right)\int\limits_0^\pi d\beta\cdot\frac{\exp\left(-\tilde{\alpha}\tilde{R}(\mu+\nu)\right)}{\tilde{R}(\mu-\nu)}\frac{\tilde{R}^2}{2}\left(\mu^2-1\right)\left(1-\nu^2\right)\sin^2\beta$$

$$=\tilde{\alpha}^5\frac{\tilde{R}^4}{16}\int\limits_1^\infty d\mu\left(\mu^2-1\right)\exp\left(-\tilde{\alpha}\tilde{R}\mu\right)\int\limits_{-1}^1 d\nu(\mu+\nu)\left(1-\nu^2\right)\exp\left(-\tilde{\alpha}\tilde{R}\nu\right)=$$

$$=\frac{-3+2\left(\tilde{\alpha}\tilde{R}\right)^2+\exp\left(-\tilde{\alpha}\tilde{R}\right)\left(1+\tilde{\alpha}\tilde{R}\right)\left(3+\tilde{\alpha}\tilde{R}\left(3+\tilde{\alpha}\tilde{R}\right)\right)}{2\tilde{R}^3\tilde{\alpha}^2}$$

(E.13a)

$$C_{12}=\int\limits_{z_2>0}\int\limits_{z_1>0}\tilde{V}_{12}(\tilde{\mathbf{r}}_1,\tilde{\mathbf{r}}_2)(\varphi_{a1})^2(\varphi_{b2})^2 d\tilde{\mathbf{r}}_1 d\tilde{\mathbf{r}}_2\equiv\frac{1}{Z(1+\zeta)}C_{12}^r+\frac{\zeta}{Z(1+\zeta)}C_{12}^z \quad \text{(E.13b)}$$

Where:

$$C_{12}^z=\frac{1}{2}\int\limits_{z_2>0}\int\limits_{z_1>0}\left(\frac{1}{\tilde{z}_1}+\frac{1}{\tilde{z}_2}\right)(\varphi_{a1})^2(\varphi_{b2})^2 d\tilde{\mathbf{r}}_1 d\tilde{\mathbf{r}}_2=\int\limits_{z_1>0}\frac{(\varphi_{a1})^2}{\tilde{z}_1}d\tilde{\mathbf{r}}_1$$

$$C_{12}^r=\int\limits_{z_2>0}\int\limits_{z_1>0}\left(\frac{1}{\tilde{r}_{12}}+\frac{\zeta}{\tilde{r}_{1'2}}\right)(\varphi_{a1})^2(\varphi_{b2})^2 d\tilde{\mathbf{r}}_1 d\tilde{\mathbf{r}}_2 \quad \text{(E.13c)}$$

$$C_{12}^z=\frac{2\tilde{\alpha}^5}{\pi}\frac{\tilde{R}^3}{8}\int\limits_1^\infty d\mu\int\limits_{-1}^1 d\nu\left(\mu^2-\nu^2\right)\int\limits_0^\pi\exp\left(-\tilde{\alpha}\tilde{R}(\mu+\nu)\right)\frac{\tilde{R}}{2}\sqrt{\left(\mu^2-1\right)\left(1-\nu^2\right)}\sin\beta\, d\beta=$$

$$=\frac{2\tilde{\alpha}^5}{\pi}\frac{\tilde{R}^3}{8}\int\limits_1^\infty d\mu\int\limits_{-1}^1 d\nu\left(\mu^2-\nu^2\right)\exp\left(-\tilde{\alpha}\tilde{R}(\mu+\nu)\right)\tilde{R}\sqrt{\left(\mu^2-1\right)\left(1-\nu^2\right)}= \quad \text{(E.13d)}$$

$$=\frac{\tilde{\alpha}^3\tilde{R}^2}{4}\int\limits_1^\infty d\mu\exp\left(-\tilde{\alpha}\tilde{R}\mu\right)\tilde{R}\sqrt{\mu^2-1}\left(\tilde{\alpha}\tilde{R}\left(\mu^2-1\right)\mathrm{I}_1\left(\tilde{\alpha}\tilde{R}\right)+3\,\mathrm{I}_2\left(\tilde{\alpha}\tilde{R}\right)\right)=\frac{3\tilde{\alpha}}{4}$$

Exchange integrals:

$$J_{ab}=\int\limits_{z_1>0}\frac{\varphi_{a1}\varphi_{b1}}{\tilde{r}_{a1}}d\tilde{\mathbf{r}}_1=\frac{2\tilde{\alpha}^5}{\pi}\frac{\tilde{R}^3}{8}\int\limits_1^\infty d\mu\int\limits_{-1}^1 d\nu\left(\mu^2-\nu^2\right)\int\limits_0^\pi d\beta\cdot\frac{\exp\left(-\tilde{\alpha}\tilde{R}\mu\right)}{\tilde{R}(\mu+\nu)}\frac{\tilde{R}^2}{2}\left(\mu^2-1\right)\left(1-\nu^2\right)\sin^2\beta$$

$$=\tilde{\alpha}^5\frac{\tilde{R}^4}{16}\int\limits_1^\infty d\mu\left(\mu^2-1\right)\exp\left(-\tilde{\alpha}\tilde{R}\mu\right)\int\limits_{-1}^1 d\nu(\mu-\nu)\left(1-\nu^2\right)=\frac{\tilde{\alpha}}{6}\exp\left(-\tilde{\alpha}\tilde{R}\right)\left(3+\tilde{\alpha}\tilde{R}\left(3+\tilde{\alpha}\tilde{R}\right)\right)$$

, (E.14a)

$$J_{12}=\int\limits_{z_2>0}\int\limits_{z_1>0}\tilde{V}_{12}(\tilde{\mathbf{r}}_1,\tilde{\mathbf{r}}_2)\varphi_{a1}\varphi_{b2}\varphi_{a2}\varphi_{b1}d\tilde{\mathbf{r}}_1 d\tilde{\mathbf{r}}_2=\frac{1}{Z(1+\zeta)}J_{12}^r+\frac{\zeta}{Z(1+\zeta)}J_{12}^z \quad \text{(E.14b)}$$

Where:

$$J_{12}^z=\frac{1}{2}\int\limits_{z_2>0}\int\limits_{z_1>0}\left(\frac{1}{\tilde{z}_1}+\frac{1}{\tilde{z}_2}\right)\varphi_{a1}\varphi_{b2}\varphi_{a2}\varphi_{b1}d\tilde{\mathbf{r}}_1 d\tilde{\mathbf{r}}_2=S\int\limits_{z_1>0}\frac{\varphi_{a1}\varphi_{b1}}{\tilde{z}_1}d\tilde{\mathbf{r}}_1$$

$$J_{12}^r=\int\limits_{z_2>0}\int\limits_{z_1>0}\left(\frac{1}{\tilde{r}_{12}}+\frac{\zeta}{\tilde{r}_{1'2}}\right)\varphi_{a1}\varphi_{b2}\varphi_{a2}\varphi_{b1}d\tilde{\mathbf{r}}_1 d\tilde{\mathbf{r}}_2 \quad \text{(E.14c)}$$

$$J_{12}^{z}=S\frac{2\tilde{\alpha}^{5}}{\pi}\frac{\tilde{R}^{3}}{8}\int_{1}^{\infty}d\mu\int_{-1}^{1}d\nu\left(\mu^{2}-\nu^{2}\right)\int_{0}^{\pi}\exp\left(-\tilde{\alpha}\tilde{R}\mu\right)\frac{\tilde{R}}{2}\sqrt{\left(\mu^{2}-1\right)\left(1-\nu^{2}\right)}\sin\beta\,d\beta=$$

$$=S\frac{2\tilde{\alpha}^{5}}{\pi}\frac{\tilde{R}^{3}}{8}\int_{1}^{\infty}d\mu\int_{-1}^{1}d\nu\left(\mu^{2}-\nu^{2}\right)\exp\left(-\tilde{\alpha}\tilde{R}\mu\right)\tilde{R}\sqrt{\left(\mu^{2}-1\right)\left(1-\nu^{2}\right)}= \quad \text{(E.14c)}$$

$$=S\frac{\tilde{\alpha}^{5}\tilde{R}^{4}}{32}\int_{1}^{\infty}d\mu\exp\left(-\tilde{\alpha}\tilde{R}\mu\right)\sqrt{\mu^{2}-1}\left(4\mu^{2}-1\right)=S\frac{3\tilde{\alpha}^{3}\tilde{R}^{2}}{32}\left(\tilde{\alpha}\tilde{R}\,K_{1}\left(\tilde{\alpha}\tilde{R}\right)+4K_{2}\left(\tilde{\alpha}\tilde{R}\right)\right)$$

Where $K_{1,2}(x)$ are the modified Bessel functions of the second kind, which exponentially vanish at $x\to\infty$.

In order to calculate $C_{12}^{r}$ and $J_{12}^{r}$ we followed seminal papers [45464748]. We used the aforementioned ellipsoidal coordinates (E.10) for both electrons:

$$\mu_{i}=\frac{\tilde{r}_{ai}+\tilde{r}_{bi}}{\tilde{R}},\quad \nu_{i}=\frac{\tilde{r}_{ai}-\tilde{r}_{bi}}{\tilde{R}},\quad \tan\beta_{i}=\frac{z_{i}}{y_{i}},\quad i=1,2,$$

$$\tilde{z}_{ai}=\tilde{z}_{bi}=\tilde{z}_{i}=\frac{\tilde{R}}{2}\sqrt{\left(\mu_{i}^{2}-1\right)\left(1-\nu_{i}^{2}\right)}\sin\beta_{i},$$

$$\varphi_{ai}=\sqrt{\frac{2\tilde{\alpha}^{5}}{\pi}}\tilde{z}_{ai}\exp\left(-\tilde{\alpha}\tilde{r}_{ai}\right),\quad \varphi_{bi}=\sqrt{\frac{2\tilde{\alpha}^{5}}{\pi}}\tilde{z}_{bi}\exp\left(-\tilde{\alpha}\tilde{r}_{bi}\right), \quad \text{(E.15)}$$

$$d\tilde{\mathbf{r}}_{1}d\tilde{\mathbf{r}}_{2}=\frac{\tilde{R}^{6}}{64}\left(\mu_{1}^{2}-\nu_{1}^{2}\right)\left(\mu_{2}^{2}-\nu_{2}^{2}\right)d\mu_{1}d\mu_{2}d\nu_{1}d\nu_{2}d\beta_{1}d\beta_{2},$$

$$\int_{z_{1}>0}W\left(\tilde{\mathbf{r}}_{1},\tilde{\mathbf{r}}_{2}\right)d\tilde{\mathbf{r}}_{1}d\tilde{\mathbf{r}}_{2}=\frac{\tilde{R}^{6}}{64}\int_{1}^{\infty}d\mu_{1}\int_{1}^{\infty}d\mu_{2}\int_{-1}^{1}d\nu_{1}\int_{-1}^{1}d\nu_{2}\left(\mu_{1}^{2}-\nu_{1}^{2}\right)\left(\mu_{2}^{2}-\nu_{2}^{2}\right)\int_{0}^{\pi}d\beta_{1}\int_{0}^{\pi}d\beta_{2}W\left(\mu_{i},\nu_{i},\beta_{i}\right)$$

Expressions for distances between the electrons:

$$\tilde{r}_{12}^{\;2}=\tilde{R}^{2}\begin{pmatrix}\left(\mu_{1}\nu_{1}-\mu_{2}\nu_{2}\right)^{2}+\left(\mu_{1}^{2}-1\right)\left(1-\nu_{1}^{2}\right)+\left(\mu_{2}^{2}-1\right)\left(1-\nu_{2}^{2}\right)\\-2\sqrt{\left(\mu_{1}^{2}-1\right)\left(1-\nu_{1}^{2}\right)\left(\mu_{2}^{2}-1\right)\left(1-\nu_{2}^{2}\right)}\cos\left(\beta_{1}-\beta_{2}\right)\end{pmatrix}$$

$$=\mu_{1}^{2}+\mu_{2}^{2}+\nu_{1}^{2}+\nu_{2}^{2}-2-2\mu_{1}\nu_{1}\mu_{2}\nu_{2}-2\sqrt{\left(\mu_{1}^{2}-1\right)\left(1-\nu_{1}^{2}\right)\left(\mu_{2}^{2}-1\right)\left(1-\nu_{2}^{2}\right)}\cos\left(\beta_{1}-\beta_{2}\right), \quad \text{(E.16)}$$

$$\tilde{r}_{1'2}^{\;2}=\tilde{r}_{12'}^{\;2}=\tilde{R}^{2}\begin{pmatrix}\left(\mu_{1}\nu_{1}-\mu_{2}\nu_{2}\right)^{2}+\left(\mu_{1}^{2}-1\right)\left(1-\nu_{1}^{2}\right)+\left(\mu_{2}^{2}-1\right)\left(1-\nu_{2}^{2}\right)\\-2\sqrt{\left(\mu_{1}^{2}-1\right)\left(1-\nu_{1}^{2}\right)\left(\mu_{2}^{2}-1\right)\left(1-\nu_{2}^{2}\right)}\cos\left(\beta_{1}+\beta_{2}\right)\end{pmatrix}$$

Neumann's expansion for $\frac{1}{\tilde{r}_{12}}$ and $\frac{1}{\tilde{r}_{1'2}}$ factorization in the ellipsoidal coordinates is:

$$\frac{1}{\tilde{r}_{12}}=\frac{1}{\tilde{R}}\sum_{n=0}^{\infty}\sum_{m=0}^{\infty}D_{n}^{m}P_{n}^{m}\begin{pmatrix}\mu_{1}\\\mu_{2}\end{pmatrix}Q_{n}^{m}\begin{pmatrix}\mu_{2}\\\mu_{1}\end{pmatrix}P_{n}^{m}\left(\nu_{2}\right)P_{n}^{m}\left(\nu_{1}\right)\cos\left(m\left(\beta_{1}-\beta_{2}\right)\right) \quad \text{(E.17a)}$$

$$\frac{1}{\tilde{r}_{1'2}}=\frac{1}{\tilde{R}}\sum_{n=0}^{\infty}\sum_{m=0}^{\infty}D_{n}^{m}P_{n}^{m}\begin{pmatrix}\mu_{1}\\\mu_{2}\end{pmatrix}Q_{n}^{m}\begin{pmatrix}\mu_{2}\\\mu_{1}\end{pmatrix}P_{n}^{m}\left(\nu_{2}\right)P_{n}^{m}\left(\nu_{1}\right)\cos\left(m\left(\beta_{1}+\beta_{2}\right)\right) \quad \text{(E.17b)}$$

Where the upper variables $\mu_i$ should be used when $\mu_2 > \mu_1$ and the lower when $\mu_1 > \mu_2$. Coefficients $D_n^0 = 2n+1$ and $D_n^m = (-1)^m 2(2n+1)\left(\frac{(n-m)!}{(n+m)!}\right)^2$ for $m$ >0.

Functions $P_n^m(t)$ and $Q_n^m(t)$ are associated Legendre functions {$m$, $n$} of the first and second kind respectively, by definition:

$$P_n^m(t) = \begin{cases} (-1)^m (1-t^2)^{m/2} \dfrac{d^m}{dt^m} P_n(t), & -1 < t < 1 \\ (t^2-1)^{m/2} \dfrac{d^m}{dt^m} P_n(t), & t > 1 \end{cases} \quad \text{and } Q_n^m(t) = (t^2-1)^{m/2} \frac{d^m}{dt^m} Q_n(t), \quad t > 1.$$

Here $P_n(t) = F(-n, n+1; 1; (1-t)/2)$ are tabulated Legendre polynomials and $Q_n(t) = \frac{n!}{(2n+1)!!} t^{-n-1} F\left(\frac{n+2}{2}, \frac{n+1}{2}; \frac{2n+3}{2}; \frac{1}{t^2}\right)$, $Q_0(t) = \frac{1}{2}\ln\left(\frac{t+1}{t-1}\right)$ are the spherical functions of the second kind.

Substituting Eqs.(E.15, 17) into and we obtained that

$$\begin{aligned} J_{12}^r = \frac{\tilde{R}^{10}}{2^8} \frac{\tilde{\alpha}^{10}}{\pi^2} \int_1^\infty d\mu_1 \int_1^\infty d\mu_2 \int_{-1}^1 d\nu_1 \int_{-1}^1 d\nu_2 (\mu_1^2 - \nu_1^2)(\mu_2^2 - \nu_2^2) \exp(-\tilde{\alpha}\tilde{R}(\mu_1 + \mu_2)) \times \\ \times (\mu_1^2 - 1)(1 - \nu_1^2)(\mu_2^2 - 1)(1 - \nu_2^2) \int_0^\pi d\beta_1 \int_0^\pi d\beta_2 \cdot \left(\frac{1}{\tilde{r}_{12}} + \frac{\zeta}{\tilde{r}_{1'2}}\right) \sin^2\beta_1 \sin^2\beta_2 \end{aligned} \quad \text{(E.18)}$$

$$\begin{aligned} C_{12}^r = \frac{\tilde{R}^{10}}{2^8} \frac{\tilde{\alpha}^{10}}{\pi^2} \int_1^\infty d\mu_1 \int_1^\infty d\mu_2 \int_{-1}^1 d\nu_1 \int_{-1}^1 d\nu_2 (\mu_1^2 - \nu_1^2)(\mu_2^2 - \nu_2^2) \exp(-\tilde{\alpha}\tilde{R}(\mu_1 + \mu_2 + \nu_1 - \nu_2)) \times \\ \times (\mu_1^2 - 1)(1 - \nu_1^2)(\mu_2^2 - 1)(1 - \nu_2^2) \int_0^\pi d\beta_1 \int_0^\pi d\beta_2 \cdot \left(\frac{1}{\tilde{r}_{12}} + \frac{\zeta}{\tilde{r}_{1'2}}\right) \sin^2\beta_1 \sin^2\beta_2 \end{aligned} \quad \text{(E.19)}$$

Performing integration over the angles $\beta_i$:

$$B_m = \int_0^\pi d\beta_1 \int_0^\pi d\beta_2 \sin^2\beta_1 \sin^2\beta_2 \cos(m(\beta_1 - \beta_2)) = \begin{cases} \dfrac{\pi^2}{4} \; (m=0); \quad \dfrac{16}{9} \; (m=1); \quad \dfrac{\pi^2}{16} \; (m=2), \\ \dfrac{8(1-(-1)^m)}{m^2(m^2-4)^2}, \quad m \geq 3, \end{cases}$$

$$B_m^l = \int_0^\pi d\beta_1 \int_0^\pi d\beta_2 \sin^2\beta_1 \sin^2\beta_2 \cos(m(\beta_1 + \beta_2)) = \begin{cases} \dfrac{\pi^2}{4} \; (m=0); \quad -\dfrac{16}{9} \; (m=1); \quad \dfrac{\pi^2}{16} \; (m=2), \\ \dfrac{8(-1)^m(1-(-1)^m)}{m^2(m^2-4)^2}, \quad m \geq 3, \end{cases}$$

(E.20)

Note, that $B_m^l = (-1)^m B_m$ and coefficients vanish rapidly with $m$ increase, so the series could be cut at small $m$ with enough high accuracy.

$$J_{12}^{r}=\frac{\tilde{R}^{9}}{2^{8}}\frac{\tilde{\alpha}^{10}}{\pi^{2}}\int_{1}^{\infty}d\mu_1\int_{1}^{\infty}d\mu_2\int_{-1}^{1}d\nu_1\int_{-1}^{1}d\nu_2\left(\mu_1^2-\nu_1^2\right)\left(\mu_2^2-\nu_2^2\right)\exp\left(-\tilde{\alpha}\tilde{R}\left(\mu_1+\mu_2\right)\right)\times$$
$$\times\left(\mu_1^2-1\right)\left(1-\nu_1^2\right)\left(\mu_2^2-1\right)\left(1-\nu_2^2\right)\sum_{n=0}^{\infty}\sum_{m=0}^{\infty}D_n^m P_n^m\begin{pmatrix}\mu_1\\ \mu_2\end{pmatrix}Q_n^m\begin{pmatrix}\mu_2\\ \mu_1\end{pmatrix}P_n^m(\nu_2)P_n^m(\nu_1)\left(1+(-1)^m\zeta\right)B_m \tag{E.21}$$

$$C_{12}^{r}=\frac{\tilde{R}^{9}}{2^{8}}\frac{\tilde{\alpha}^{10}}{\pi^{2}}\int_{1}^{\infty}d\mu_1\int_{1}^{\infty}d\mu_2\int_{-1}^{1}d\nu_1\int_{-1}^{1}d\nu_2\left(\mu_1^2-\nu_1^2\right)\left(\mu_2^2-\nu_2^2\right)\exp\left(-\tilde{\alpha}\tilde{R}\left(\mu_1+\mu_2+\nu_1-\nu_2\right)\right)\times$$
$$\times\left(\mu_1^2-1\right)\left(1-\nu_1^2\right)\left(\mu_2^2-1\right)\left(1-\nu_2^2\right)\sum_{n=0}^{\infty}\sum_{m=0}^{\infty}D_n^m P_n^m\begin{pmatrix}\mu_1\\ \mu_2\end{pmatrix}Q_n^m\begin{pmatrix}\mu_2\\ \mu_1\end{pmatrix}P_n^m(\nu_2)P_n^m(\nu_1)\left(1+(-1)^m\zeta\right)B_m \tag{E.22}$$

After elementary transformations forth fold integrals Eqs.(E.21)-(E.22) can be factorized as

$$\begin{aligned}J_{12}^{r}&=\frac{\tilde{R}^{9}}{2^{8}}\frac{\tilde{\alpha}^{10}}{\pi}\sum_{n=0}^{\infty}\sum_{m=0}^{\infty}D_n^m\left(1+(-1)^m\zeta\right)B_m\times\\&\times\begin{pmatrix}\int_{1}^{\infty}d\mu_1\left(\mu_1^2-1\right)\exp\left(-\tilde{\alpha}\tilde{R}\mu_1\right)Q_n^m(\mu_1)\int_{1}^{\mu_1}d\mu_2\left(\mu_2^2-1\right)\exp\left(-\tilde{\alpha}\tilde{R}\mu_2\right)P_n^m(\mu_2)\\+\int_{1}^{\infty}d\mu_2\left(\mu_2^2-1\right)\exp\left(-\tilde{\alpha}\tilde{R}\mu_2\right)Q_n^m(\mu_2)\int_{1}^{\mu_2}d\mu_1\left(\mu_1^2-1\right)\exp\left(-\tilde{\alpha}\tilde{R}\mu_1\right)P_n^m(\mu_1)\end{pmatrix}\\&\times\int_{-1}^{1}d\nu_1\left(1-\nu_1^2\right)\left(\mu_1^2-\nu_1^2\right)P_n^m(\nu_1)\int_{-1}^{1}d\nu_2\left(1-\nu_2^2\right)\left(\mu_2^2-\nu_2^2\right)P_n^m(\nu_2)\end{aligned} \tag{E.23}$$

$$\begin{aligned}C_{12}^{r}&=\frac{\tilde{R}^{9}}{2^{8}}\frac{\tilde{\alpha}^{10}}{\pi^{2}}\sum_{n=0}^{\infty}\sum_{m=0}^{\infty}D_n^m\left(1+(-1)^m\zeta\right)B_m\times\\&\times\begin{pmatrix}\int_{1}^{\infty}d\mu_1\left(\mu_1^2-1\right)\exp\left(-\tilde{\alpha}\tilde{R}\mu_1\right)Q_n^m(\mu_1)\int_{1}^{\mu_1}d\mu_2\left(\mu_2^2-1\right)\exp\left(-\tilde{\alpha}\tilde{R}\mu_2\right)P_n^m(\mu_2)\\+\int_{1}^{\infty}d\mu_2\left(\mu_2^2-1\right)\exp\left(-\tilde{\alpha}\tilde{R}\mu_2\right)Q_n^m(\mu_2)\int_{1}^{\mu_2}d\mu_1\left(\mu_1^2-1\right)\exp\left(-\tilde{\alpha}\tilde{R}\mu_1\right)P_n^m(\mu_1)\end{pmatrix}\\&\times\int_{-1}^{1}d\nu_1\left(1-\nu_1^2\right)P_n^m(\nu_1)\left(\mu_1^2-\nu_1^2\right)\int_{-1}^{1}d\nu_2\left(1-\nu_2^2\right)\left(\mu_2^2-\nu_2^2\right)P_n^m(\nu_2)\exp\left(-\tilde{\alpha}\tilde{R}\left(\nu_1-\nu_2\right)\right)\end{aligned} \tag{E.24}$$

Using the integrals

$$I_{nm}(\mu)=\frac{(n-m)!}{(n+m)!}\int_{-1}^{1}d\nu_2\left(1-\nu_2^2\right)\left(\mu^2-\nu_2^2\right)P_n^m(\nu_2)=a_{nm}\mu^2+b_{nm}=\begin{cases}\frac{4}{3}\mu^2-\frac{4}{15}, & n=m=0,\\ -\frac{4}{15}\mu^2-\frac{4}{105}, & m=0, n=2,\\ \frac{16}{315}, & m=0, n=4,\\ 0, & m=0, n=1,3, n>4\,.......\end{cases}$$

(E.25a)

In general case $a_{nm}=I_{n0}^{m}-I_{n2}^{m}$ and $b_{nm}=I_{n4}^{m}-I_{n2}^{m}$, where

$$I_{np}^{m}=\frac{(n-m)!}{(n+m)!}\int_{-1}^{1}\nu^{p}P_{n}^{m}(\nu)d\nu=$$

$$=\frac{(-1)^{m}\sqrt{\pi}\,\Gamma\left(\frac{p+1}{2}\right)\left(1+(-1)^{n+m+p}\right)}{\Gamma\left(\frac{m+1}{2}\right)\Gamma\left(\frac{m+p+3}{2}\right)2^{2m+1}}\,{}_{3}F_{2}\left(\frac{n+m+1}{2},\frac{m-n}{2},\frac{m}{2}+1;m+1,\frac{m+p+3}{2};1\right) \qquad \text{(E.25b)}$$

Coulomb integrals

$$K_{nm}(\mu)=\int_{-1}^{1}d\nu_{2}\left(1-\nu_{2}^{2}\right)\left(\mu^{2}-\nu_{2}^{2}\right)P_{n}^{m}(\nu_{2})\exp\left(\pm\tilde{\alpha}\tilde{R}\nu_{2}\right)=c_{nm}\mu^{2}+d_{nm}=$$

$$=\begin{cases}\frac{4}{\left(\tilde{\alpha}\tilde{R}\right)^{5}}\begin{pmatrix}\mu^{2}\left(\left(\tilde{\alpha}\tilde{R}\right)^{3}\cosh\left(\tilde{\alpha}\tilde{R}\right)-\left(\tilde{\alpha}\tilde{R}\right)^{2}\sinh\left(\tilde{\alpha}\tilde{R}\right)\right)\\-\tilde{\alpha}\tilde{R}\left(12+\left(\tilde{\alpha}\tilde{R}\right)^{2}\right)\cosh\left(\tilde{\alpha}\tilde{R}\right)+\left(12+5\left(\tilde{\alpha}\tilde{R}\right)^{2}\right)\sinh\left(\tilde{\alpha}\tilde{R}\right)\end{pmatrix}, & m=n=0,\\ \ldots., \quad n=1, m=0, \\ \ldots\ldots, \quad n=2, m=0, \\ \ldots\ldots..\end{cases} \qquad \text{(E.26)}$$

Using these integrals we integrated further:

$$L_{nm}(a,b,\mu)=\int_{1}^{\mu}d\mu_{1}\left(\mu_{1}^{2}-1\right)\exp\left(-\tilde{\alpha}\tilde{R}\mu_{1}\right)P_{n}^{m}(\mu_{1})\left(a\mu_{1}^{2}+b\right)=$$

$$=\begin{cases}\begin{pmatrix}\frac{2\exp\left(-\tilde{\alpha}\tilde{R}\right)}{\left(\tilde{\alpha}\tilde{R}\right)^{5}}\left(b\left(\tilde{\alpha}\tilde{R}\right)^{2}\left(1+\tilde{\alpha}\tilde{R}\right)+a\left(2+\tilde{\alpha}\tilde{R}\right)\left(6+\tilde{\alpha}\tilde{R}\left(3+\tilde{\alpha}\tilde{R}\right)\right)\right)-\\ \frac{\exp\left(-\tilde{\alpha}\tilde{R}\mu\right)}{\left(\tilde{\alpha}\tilde{R}\right)^{5}}\begin{pmatrix}24a+2(b-a)\left(\tilde{\alpha}\tilde{R}\right)^{2}-b\left(\tilde{\alpha}\tilde{R}\right)^{4}+\\-\tilde{\alpha}\tilde{R}\left((a-b)\left(\tilde{\alpha}\tilde{R}\right)^{2}-12a\right)\left(2\mu+\mu^{2}\tilde{\alpha}\tilde{R}\right)+a\left(\tilde{\alpha}\tilde{R}\right)^{3}\left(4\mu^{3}+\mu^{4}\tilde{\alpha}\tilde{R}\right)\end{pmatrix}\end{pmatrix}, & m=n=0,\\ \ldots., \quad m=0, n=1\end{cases}$$

After elementary transformations we obtained the quadratures:

$$\begin{aligned}J_{12}^{r}=&\frac{\tilde{R}^{9}}{2^{7}}\frac{\tilde{\alpha}^{10}}{\pi^{2}}\sum_{n=0}^{\infty}\sum_{m=0}^{n}(-1)^{m}p_{m}(2n+1)\left(1+(-1)^{m}\zeta\right)B_{m}\times\\&\times\int_{1}^{\infty}d\mu_{1}\left(\mu_{1}^{2}-1\right)\exp\left(-\tilde{\alpha}\tilde{R}\mu_{1}\right)Q_{n}^{m}(\mu_{1})\left(a_{nm}\mu_{1}^{2}+b_{nm}\right)L_{nm}\left(a_{nm},b_{nm},\mu_{1}\right)\end{aligned} \qquad \text{(E.27)}$$

$$\begin{aligned}C_{12}^{r}=&\frac{\tilde{R}^{9}}{2^{7}}\frac{\tilde{\alpha}^{10}}{\pi^{2}}\sum_{n=0}^{\infty}\sum_{m=0}^{n}D_{n}^{m}\left(1+(-1)^{m}\zeta\right)B_{m}\times\\&\times\int_{1}^{\infty}d\mu_{1}\left(\mu_{1}^{2}-1\right)\exp\left(-\tilde{\alpha}\tilde{R}\mu_{1}\right)Q_{n}^{m}(\mu_{1})\left(c_{nm}\mu_{1}^{2}+d_{nm}\right)L_{nm}\left(c_{nm},d_{nm},\mu_{1}\right)\end{aligned} \qquad \text{(E.28)}$$

Coefficients $p_{0}=1$ and $p_{m}=2$ for $m>0$.

Cutting the series we obtained that:

$$J_{12}^{r00} = \tilde{\alpha}\frac{(\tilde{\alpha}\tilde{R})^9}{2^9}(1+\zeta)\int_1^\infty d\mu_1(\mu_1^2-1)\exp(-\tilde{\alpha}\tilde{R}\mu_1)Q_0(\mu_1)\left(\frac{4}{3}\mu_1^2-\frac{4}{15}\right)L_{00}(\mu_1)=$$

$$=\frac{\tilde{\alpha}}{\rho}(1+\zeta)\left(e^{-2\rho}\left(\frac{\mathbf{C}+\ln(\rho)}{2}\left(1+\rho+\frac{2}{5}\rho^2+\frac{\rho^3}{15}\right)^2-\rho\left(-\frac{93}{512}+\frac{163\rho}{256}+\frac{991\rho^2+401\rho^3}{2880}+\frac{\rho^4}{48}+\frac{\rho^5}{600}\right)\right)+\right.$$

$$\left.+\operatorname{Ei}(-2\rho)\left(-1+\frac{\rho^2}{5}-\frac{2\rho^4}{75}+\frac{\rho^6}{225}\right)+e^{2\rho}\operatorname{Ei}(-4\rho)\frac{1}{2}\left(1-\rho+\frac{2}{5}\rho^2-\frac{\rho^3}{15}\right)^2\right)$$

(E.29)

$$C_{12}^{r00} = \frac{\rho^9\tilde{\alpha}}{2^9}(1+\zeta)\int_1^\infty d\mu_1(\mu_1^2-1)\exp(-\rho\mu_1)Q_0(\mu_1)(c_{00}(\rho)\mu_1^2+d_{00}(\rho))L_{00}(c_{00}(\rho),d_{00}(\rho),\mu_1)=$$

$$=\frac{\tilde{\alpha}}{\rho}(1+\zeta)\begin{pmatrix}\frac{261}{2048\rho^5}-\frac{679}{2048\rho^3}+\frac{441}{1024\rho^2}+\frac{63}{256\rho}-\frac{125}{256}-\\ -e^{-2\rho}\left(\frac{261}{1024\rho^5}+\frac{261}{512\rho^4}-\frac{157}{1024\rho^3}-\frac{121}{512\rho^2}+\frac{107}{128\rho}+\frac{13}{128}\right)+\\ +e^{-4\rho}\left(\frac{261}{2048\rho^5}+\frac{261}{512\rho^4}+\frac{1409}{2048\rho^3}+\frac{361}{1024\rho^2}+\frac{59}{256\rho}+\frac{3}{256}\right)-\\ -\operatorname{Ei}(-2\rho)+\operatorname{Ei}(-4\rho)\frac{1}{2}+\frac{\mathbf{C}+\ln(\rho)}{2}\end{pmatrix} \quad \text{(E.30)}$$

Here $\tilde{\alpha}\tilde{R}\equiv\rho$, **C** is Euler's constant, with numerical value ≈0.577216, Ei(z) is the exponential integral function.

Using Eqs. (E.10-13) one could rewrite (E.9) as

$$\tilde{E}_{S,A}=\frac{Z}{\tilde{r}_{ab}}+\frac{2(A\pm S\,A_{ab}-C_{ab}\mp S\,J_{ab})+C_{12}\pm J_{12}}{1\pm S^2} \quad \text{(E.31)}$$

Energy levels difference is

$$\tilde{E}_S-\tilde{E}_A=\frac{4S}{1-S^4}(A_{ab}-S\,A-J_{ab}+S\,C_{ab})+2\frac{J_{12}-S^2C_{12}}{1-S^4} \quad \text{(E.32)}$$

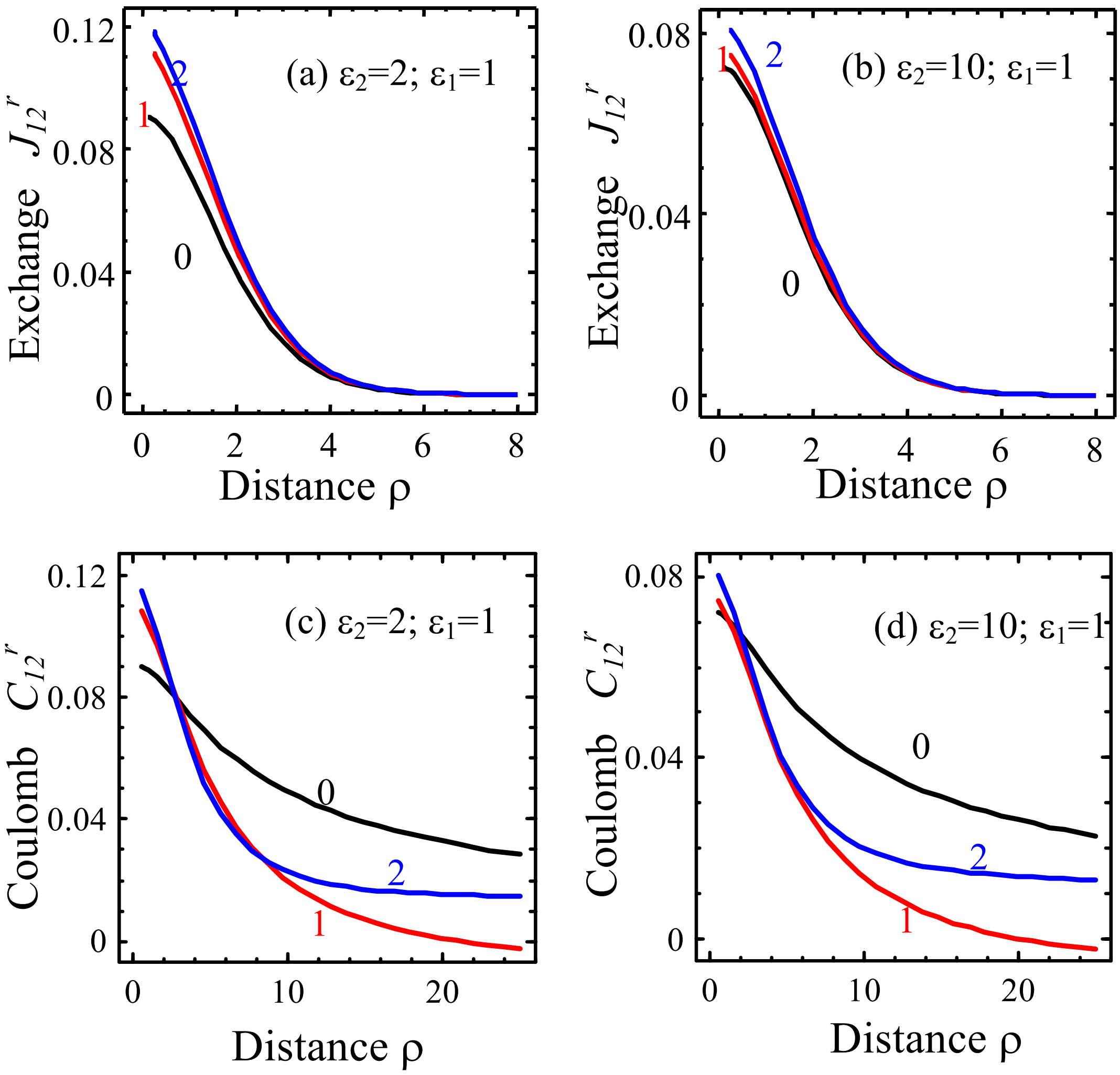

**Fig. 1E.** Figures near the curves correspond to the number of terms in the series